\documentclass[fleqn,usenatbib]{mnras}

\usepackage{newtxtext,newtxmath}

\usepackage[T1]{fontenc}

\DeclareRobustCommand{\VAN}[3]{#2}
\let\VANthebibliography\thebibliography
\def\thebibliography{\DeclareRobustCommand{\VAN}[3]{##3}\VANthebibliography}

\usepackage{graphicx}	
\usepackage{amsmath}	
\usepackage{bbm}
\usepackage{multirow}   
\usepackage[utf8]{inputenc} 
\usepackage{tabularx}
\usepackage{comment}
\usepackage{scalerel}
\usepackage{orcidlink}
\usepackage{pdflscape} 
\usepackage[table]{xcolor} 

\usepackage{caption}
\usepackage{subcaption}

\newcommand{\plotsidesize}[2]
 {\centering \leavevmode \includegraphics[width={#2\textwidth}]{#1}}

\newcommand{\nH}{n_{\rm H}}
\newcommand{\nHn}{n_{\rm H,neutral}}

\newcommand{\Teff}{T_{\rm eff}}
\newcommand{\nHrms}{n_{\rm H}^{\rm rms}}
\newcommand{\DZ}{{\rm D/Z}}

\newcommand{\Zsol}{{\rm Z}_{\sun}}
\newcommand{\Msol}{{M}_{\sun}}

\newcommand{\Mdust}{M_{\rm dust}}

\newcommand{\nanom}{\rm nm}
\newcommand{\angstrom}{\text{\AA}}
\newcommand{\cmcubed}{ {\rm cm}^{-3} }
\newcommand{\dnda}{\frac{\partial n}{\partial a}}
\newcommand{\dt}{\Delta t}
\newcommand{\dndaflat}{\partial n/\partial a}
\newcommand{\amin}{a_{\rm min}}
\newcommand{\amax}{a_{\rm max}}
\newcommand{\aic}{a_i^{\rm c}}
\newcommand{\ailower}{a_i^{\rm e}}
\newcommand{\aiupper}{a_{i+1}^{\rm e}}
\newcommand{\ajc}{a_j^{\rm c}}
\newcommand{\ajlower}{a_j^{\rm e}}
\newcommand{\ajupper}{a_{j+1}^{\rm e}}
\newcommand{\akc}{a_k^{\rm c}}
\newcommand{\aklower}{a_k^{\rm e}}
\newcommand{\akupper}{a_{k+1}^{\rm e}}
\newcommand{\vshat}{v_{\rm shat}}

\newcommand{\GIZMO}{{\small GIZMO}}
\newcommand{\Hmol}{{\rm H}_{2}}

\newcommand{\AV}{A_{\rm V}}

\defcitealias{choban_2022:GalacticDustmodellingDust}{C22}
\defcitealias{choban_2024:DustyLocaleEvolution}{C24}
\defcitealias{choban_2025:DustyDawnGalactic}{C25}

\defcitealias{hirashita_2019:RemodellingEvolutionGrain}{H19}
\defcitealias{matsumoto_2025:EvolutionGalaxyAttenuation}{M25}
\defcitealias{li_2021:OriginDustExtinction}{Li21}

\newcommand{\gizmourl}{\href{http://www.tapir.caltech.edu/~phopkins/Site/GIZMO.html}{\url{http://www.tapir.caltech.edu/~phopkins/Site/GIZMO.html}}}

\newcommand{\datastatement}[1]{\begin{small}\section*{Data Availability Statement}\end{small}{\noindent #1}\vspace{5pt}}

\title[Grain Size Evolution in the Local Group with FIRE]{Ashes of FIRE: Modeling Dust Grain Size Evolution in the Local Group with FIRE}

\author[C. R. Choban et al.]{
\parbox[t]{\textwidth}{
        Caleb R. Choban\orcidlink{0000-0001-9200-169X}$^{1}$\thanks{email: cchoban@iu.edu},
        Samir Salim\orcidlink{0000-0003-2342-7501}$^{1}$,
        Du\v{s}an Kere\v{s}\orcidlink{0000-0002-1666-7067}$^{2,3}$, Julia Roman-Duval\orcidlink{0000-0001-6326-7069}$^{4}$, Karin M. Sandstrom\orcidlink{0000-0002-4378-8534}$^{2}$
} \vspace*{4pt} \\
$^{1}$ Department of Astronomy, Indiana University, Bloomington, IN 47405, USA \\
$^{2}$ Department of Astronomy \& Astrophysics, University of California at San Diego, La Jolla, CA 92093, USA \\
$^{3}$ Department of Physics, University of California at San Diego, La Jolla, CA 92093, USA \\
$^{4}$ Space Telescope Science Institute, 3700 San Martin Drive, Baltimore, MD 21218, USA \\
}

\date{Accepted XXX. Received YYY; in original form ZZZ}

\pubyear{2025}

\begin{document}
\label{firstpage}
\pagerange{\pageref{firstpage}--\pageref{lastpage}}
\maketitle

\begin{abstract}
We introduce a new, discretized grain size evolution model, incorporated into the \GIZMO\ code and coupled with FIRE-3 stellar feedback and ISM physics, to investigate variations in dust abundance, chemical composition, and grain sizes observed in the Local Group.
This model tracks the size evolution of specific dust species, and includes stellar production of dust, dust growth through gas-phase metal accretion, dust destruction by sputtering, SNe shocks, and astration, grain-grain collisional shattering and coagulation, and turbulent dust diffusion. 
Using idealized galaxy simulations, we test the dependence of MW dust properties on variations in each dust process and find that our model uniquely predicts a bimodal grain size distribution.
This bimodality is due to our simulation's ability to resolve each dust process and where they occur in the ISM, unlike other works.
We find that Local Group dust abundances are determined by dust growth and destruction, with little dependence on coagulation or shattering, explaining why models that do not include these processes can match abundance observations. 
We also find that variations in Local Group extinction curve slopes are determined by coagulation, with inefficient coagulation leading to steeper slopes.
However, inefficient coagulation also results in stronger extinction curve bumps, which are not observed.
We also do not predict a population of very small (${<}1$ nm) carbonaceous grains, required for MIR emission features, due to their rapid growth by accretion. 
These results highlight the possible necessity of ``top-down'' PAH formation from preexisting grains as a means to inhibit carbonaceous dust growth.


\end{abstract}





\section{Introduction}

Observations of the local Universe have found interstellar dust grains are ubiquitous within all but the most metal-poor galaxies \citep[e.g.][]{galliano_2018:InterstellarDustProperties}, and recent discoveries of dusty galaxies at high-redshift ($z>4$) have further cemented that galaxies are rich in dust for most of their lives \citep[e.g.][]{schneider_2024:FormationCosmicEvolution}.
It is well known that interstellar grains play a critical role in many physical processes that shape the evolution of galaxies. They are the catalyst site for $\Hmol$ formation, indirectly forming stars \citep{hollenbach_1971:SurfaceRecombinationHydrogen}, are the dominant heat source in the warm neutral medium (WNM) and photodissociation regions via the photoelectric effect \citep{tielens_1985:PhotodissociationRegionsBasic,wolfire_2003:NeutralAtomicPhases}, and are a mediator for stellar feedback through radiation pressure, which drives stellar to galactic-scale winds \citep{thompson_2005:RadiationPressuresupportedStarburst,murray_2010:DisruptionGiantMolecular}.
Interstellar dust also affects nearly all galactic observations due to its tendency to extinguish ultraviolet (UV) and optical light through absorption and scattering, and reemit in the mid-to-far infrared (IR). 
This reprocessing of stellar light complicates the interpretation of observed stellar spectra from individual stars to stellar populations of entire galaxies \citep[e.g.][]{conroy_2013:ModelingPanchromaticSpectral, salim_2020:DustAttenuationLaw}.

To accurately account for these physical and observational effects, a detailed understanding of the amount, chemical composition, and sizes of the dust grains are needed. 
Observations of dust within the Milky Way (MW) have produced the current interstellar dust paradigm: ${\sim}40\%$ of astrophysical metals in the ISM are in dust, dust is primarily composed of silicate and carbonaceous species and extremely small (${<}10$ nm) polycyclic aromatic hydrocarbons (PAHs), and dust grains range in size from $a \simeq5\;{\angstrom} - 0.3\;\micron$, with small grains dominating by number and surface area \citep[e.g.][]{draine_2011:PhysicsInterstellarIntergalactic}.
However, observations of dust within other galaxies and even within the MW find that this paradigm is far from universal.
Gas-phase element depletion trends in the MW and Small and Large Magellanic Clouds (LMC/SMC) show that the amount of metals locked in dust increases with gas density \citep{jenkins_2009:UnifiedRepresentationGasPhase,jenkins_2017:InterstellarGasphaseElement,roman-duval_2021:METALMetalEvolution}.
Observed extinction curves slopes and bumps strength in the MW showcase a wide range of average grain sizes \citep[e.g.][]{fitzpatrick_2007:AnalysisShapesInterstellar}, while those in the LMC and SMC show an overall larger abundance of small grains (steeper slopes), but fewer carbonaceous grains (smaller bumps) compared to the MW \citep{gordon_2003:QuantitativeComparisonSmall,gordon_2024:ExpandedSampleSmall}.
These variations in dust grain size and composition are also seen in attenuation curves\footnote{Attenuation curves also depend on the relative geometry between stars and dust, making it difficult to deduce detailed information about the dust population \citep[e.g.][]{narayanan_2018:TheoryVariationDust, salim_2020:DustAttenuationLaw}.} of galactic spectral energy distributions (SEDs), depending on galactic properties and redshift \citep[e.g][]{shivaei_2025:DiversityEvolutionDust}.
Dust mass estimates from IR dust emission also find a strong correlation between galactic dust content and metallicity, with lower metallicity galaxies having less of their metals locked in dust \citep{remy-ruyer_2014:GasdustMassRatios,devis_2019:SystematicMetallicityStudy,clark_2023:QuestMissingDust}.
The James Webb Space Telescope's (JWST) unprecedented observations of mid-IR (MIR) PAH emission lines have added further scrutiny on the extremely small end of the dust grain population.
Notably, the amount of PAHs varies considerably between galaxies, depending on metallicity and specific star formation rate \citep{draine_2007:DustMassesPAH,chastenet_2024:JWSTMIRINIRCam,tarantino_2025:JWSTCapturesGrowth,lai_2025:ResolvingEmissionSmall}.
Within galaxies, PAH abundances decrease in strong radiation fields \citep{sutter_2024:FractionDustMass} and increase in the translucent ISM \citep{zhang_2025:DustextinctioncurveVariationTranslucent}, suggesting they experience unique destruction and growth processes.

\begin{table*}
    \centering
    \renewcommand{\arraystretch}{1.3}
	\begin{tabular}{ |c|ccccccccccc| }
		 \hline
		Name & $M_{\rm halo}$ & $M_{\rm *,disc}$ & $M_{\rm *,bulge}$ & $M_{\rm gas}$ & $M_{\rm bh}$ & $R_{\rm *,disc}$ & $Z_{\rm init}$ & $\delta_{\rm dust}$ & Sil-to-C & $m_{\rm gas}$ & $\epsilon^{\rm MIN}_{\rm gas}$ \\ 
        & $(10^{10}\Msol)$ & $(10^{10} \Msol)$ & $(10^{10}\Msol)$ & $(10^{10}\Msol)$ & $(10^{6}\Msol)$ & (kpc) & $(Z_{\odot})$ & & & ($\Msol$) & (pc) \\ [0.5ex] 
		\hline
            m12\_lowres & 150 & 4.7 & 0.5 & 1 & 4 & 3.2 & 0.8 & 0.5 & 2 & 7100 & 1 \\
            m12 & 150 & 4.7 & 0.5 & 1 & 4 & 3.2 & 0.8 & 0.5 & 2 & 710 & 0.1\\
            m11 & 15 & 0.29 & 0.032 & 0.3 & 0.4 & 1.6 & 0.4 & 0.3 & 4 & 710 & 0.1 \\
            m10 & 1.5  & 0.029 & 0.0032 & 0.03 & 0.04 & 0.8 & 0.2 & 0.1 & 10 & 710 & 0.1 \\
		\hline
	\end{tabular}
	\caption{Properties of idealized simulations run with our dust model. Dark matter and initial star particle resolutions and universal softening lengths are $m_{\rm DM}=100m_{\rm gas};\epsilon_{\rm DM}=100\epsilon^{\rm MIN}_{\rm gas}$ and $m_{\rm IC, star}=5 m_{\rm gas};\epsilon_{\rm IC,star}=10\epsilon^{\rm MIN}_{\rm gas}$ respectively. {\bf m12\_lowres} is rerun for testing variations in our dust model, while the rest are run only with our preferred model.
    }
	\label{tab:idealized_ICs}
\end{table*}

Despite these observations showcasing a complex evolution of dust population properties, galaxy simulations have typically treated dust as a static quantity, limiting the self-consistent investigation of the physical effects of dust and its reprocessing of light on galactic scales.
Recently, the galaxy formation community has striven to address this disregard for dust by developing and integrating dust evolution models into galaxy simulations.
These models encapsulate our current understanding of the dust life cycle which is comprised of the following processes: dust is created in stellar ejecta, dust grows via the accretion of gas-phase metals, dust is destroyed from sputtering in supernovae (SNe) shocks and hot halo gas, and dust coagulates into larger aggregates or shatters into small fragments via grain-grain collisions \citep[e.g.][]{dwek_2005:InterstellarDustWhat}.
However, how these processes are modeled depends on the treatment of grain sizes.

The first generation of galactic dust models, largely derived from semi-analytical models (SAMs) presented by \citet{dwek_1998:EvolutionElementalAbundances} and \citet{zhukovska_2008:EvolutionInterstellarDust}, assume a constant power-law size distribution ($\dndaflat \propto a^{-3.5}$; MRN  \citealt{mathis_1977:SizeDistributionInterstellar})
\citep[][]{bekki_2013:CoevolutionDustGas,bekki_2015:DustregulatedGalaxyFormation,bekki_2015:CosmicEvolutionDust,mckinnon_2016:DustFormationMilky,mckinnon_2017:SimulatingDustContent,aoyama_2018:CosmologicalSimulationDust,li_2019:DustgasDustmetalRatio,choban_2022:GalacticDustmodellingDust}.
This simplification enables each dust process to be summarized by a effective timescales and is computationally inexpensive. However, such models cannot account for the processes of coagulation or shattering and cannot make predictions for attenuation/extinction curves and PAH abundances which are grain size dependent. 
To overcome this limitation, the next generation of dust models has focused on evolving grain sizes, but the implementations vary. 
The most widely adopted grain size implementation is the ``two-size approximation'' developed by \citet{hirashita_2015:TwosizeApproximationSimple} \citep[e.g.][]{aoyama_2017:GalaxySimulationDust,aoyama_2018:CosmologicalSimulationDust,gjergo_2018:DustEvolutionGalaxy,granato_2021:DustEvolutionZoomcosmological,parente_2022:DustEvolutionMUPPI,parente_2023:1DropCosmic,dubois_2024:GalaxiesGrainsUnraveling,trayford_2026:ModellingEvolutionInfluence}. 
This model divides dust into two populations of small ($a\sim0.01\,\micron$) and large ($a\sim0.1\,\micron$) grains, evolving each separately, and accounting for the coagulation and shattering processes by transitioning grains between the two sizes.
While still computationally inexpensive, these models rely on tunable timescales which obfuscate the underlying physics.
More recently, models that self-consistently track the evolution of grain sizes by discretizing the grain size distribution have been developed (e.g. \citealt{asano_2013:WhatDeterminesGrain,mckinnon_2018:SimulatingGalacticDust,hirashita_2019:RemodellingEvolutionGrain,hirashita_2020:SelfconsistentModellingAromatic,li_2021:OriginDustExtinction,narayanan_2023:FrameworkModelingPolycyclic}, Parente el al. in prep).
These models are the most rigorous way to model and test dust processes and investigate the origin of dust variability in the Universe.
However, they are challenging to implement, have a high computational and memory overhead, and have strict time-stepping requirements. 
For these reasons, such models have typically been integrated into SAMs and low resolution galaxy simulations which do not resolve the multiphase ISM, with \citet{narayanan_2023:FrameworkModelingPolycyclic} being the one exception.

In this work, we present a new discretized grain size evolution model integrated into the magneto-hydrodynamics meshless-finite mass code \GIZMO\ coupled with the FIRE-3 (Feedback in Realistic Environments) model for stellar feedback and ISM physics.
This model tracks the separate evolution of silicate, carbonaceous, and metallic iron dust, and accounts for all major processes in the dust life cycle.
We also present a novel prescription for SNR dust destruction which accounts for the shattering of large dust grains.
Using a large suite of idealized galaxy simulations, we conduct the most rigorous tests to date on the sensitivity of MW dust abundance, composition, and sizes to variations in each dust processes.
We also investigate what causes the diversity of dust properties in Local Group galaxies.
We uniquely predict a bimodal grain size distribution, which is qualitatively different from all other works, due to FIRE's ability to resolve the multiphase ISM.
Our model reproduces the observed variations in dust abundances and extinction curve slopes seen in Local Group galaxies, caused by variable accretion and coagulation efficiencies respectively. 
However, our model cannot reproduce observed variations in extinction curve bump strengths and does not predict a population of very small carbonaceous grains responsible for MIR emission lines. 
This due to the efficient growth of small carbonaceous grains, highlighting  the possible necessity of ``top-down'' PAH formation to inhibit their growth.

This paper is organized as follows. In Section~\ref{Galaxy Simulations}, we provide a brief overview of the FIRE-3 galaxy formation model and the initial conditions of our idealized simulations. 
In Section~\ref{GSE}, we lay out the analytical framework of our grain size evolution model, compiling the full numerical implementation in Appendix~\ref{app:discretized_forms}. 
In Section~\ref{sec:fiducial_model}, we present the results of our model for a MW-mass galaxy, and in Section~\ref{sec:model_variations} we test how variations to each dust process affect predicted grain size distributions and extinction curves with a suite of simulations. 
In Section~\ref{sec:local_group_comparison}, we investigate the cause of dust abundance and extinction curve variations seen in the MW, LMC, and SMC.
We compare our predictions against other works and discuss the implications of our model on the qualitative shape of the grain size distribution, the lack of large grains in the ISM, and evolution of PAHs in Section~\ref{sec:discussion}.
Finally, we present our conclusions in Section~\ref{Conclusions}.


\section{Galaxy Simulations} \label{Galaxy Simulations}

To study the evolution of dust grain sizes in Local Group galaxies we utilize simulations of a idealized, non-cosmological MW, LMC, and SMC-mass galaxies with a dust grain size evolution model incorporated into the \GIZMO\ code base \citep{hopkins_2015:NewClassAccurate} and coupled with FIRE-3 stellar feedback and ISM physics. FIRE-3 is an update of the FIRE star-formation and stellar feedback model \citep{hopkins_2014:GalaxiesFIREFeedback,hopkins_2018:FIRE2SimulationsPhysics}, with a general overview given in Sec.~\ref{GFM}. The initial conditions used in our simulations are presented in Sec.~\ref{ICs}. The numerical approach for our grain size evolution model is explained in detail in Sec.~\ref{GSE}. 

\subsection{Galaxy Physics \&\ Feedback Mechanisms} \label{GFM}

The galaxy simulations and numerical methods used here have been extensively explained in previous works, so we only briefly summarize them here. The simulations were run with {\small GIZMO}\footnote{A public version of {\small GIZMO} is available at \href{http://www.tapir.caltech.edu/~phopkins/Site/GIZMO.html}{\url{http://www.tapir.caltech.edu/~phopkins/Site/GIZMO.html}}} \citep{hopkins_2015:NewClassAccurate}, in its meshless finite-mass MFM mode (a mesh-free finite-volume Lagrangian Godunov method). Gravity is solved with fully-adaptive Lagrangian force softening.  
The physics of cooling, star formation, and stellar feedback follows the FIRE-3 version of the Feedback in Realistic Environments (FIRE) project\footnote{\url{http://fire.northwestern.edu}} \citep{hopkins_2014:GalaxiesFIREFeedback}, described in detail in \citet{hopkins_2023:FIRE3UpdatedStellar}. Gas cooling and thermo-chemistry is followed as \citet{grudic_2021:STARFORGEComprehensiveNumerical} from $T=10-10^{10}\,$K including explicit non-equilibrium ionization/atomic/molecular chemistry as well as molecular, fine-structure, photo-electric, dust, ionization, cosmic-ray, and other heating/cooling processes (including local radiation sources and the meta-galactic background from \citealt{faucher-giguere_2020:CosmicUVXray}, with self-shielding), and opacities updated per \citet{hopkins_2024:FORGEdFIREResolving}. We follow 11 distinct abundances plus a set of tracer species accounting for turbulent metal diffusion as in \citet{escala_2018:ModellingChemicalAbundance}. Gas is converted to stars using a sink-particle prescription if and only if it is locally self-gravitating at the resolution scale \citep{hopkins_2013:MeaningConsequencesStar} with an unresolved Jeans fragmentation scale. Each star particle is then evolved as a single stellar population with IMF-averaged feedback properties calculated following the 2021 version of STARBURST99 \citep{leitherer_2014:EffectsStellarRotation} for a \citet{kroupa_2001:VariationInitialMass} IMF and its age and abundances as updated in \citet{hopkins_2023:FIRE3UpdatedStellar}. We explicitly treat mechanical feedback from SNe (Ia \&\ II) and stellar mass loss (from O/B and AGB stars) as discussed in \citet{hopkins_2018:HowModelSupernovae,hopkins_2023:FIRE3UpdatedStellar}, and radiative feedback including photo-electric and photo-ionization heating and UV/optical/IR radiation pressure with a five-band radiation scheme as discussed in \citet{hopkins_2020:RadiativeStellarFeedback}.

We highlight that all cooling and heating processes,
radiative transfer, and $\Hmol$ formation modeled in our simulations are coupled with our dust evolution model. 
Specifically, photoelectric heating\footnote{Note the photoelectric heating routine in FIRE assumes an abundance of PAHs.}, dust-gas collisional heating/cooling, radiative transfer, and equilibrium $\Hmol$ formation scale with the local D/Z ratio assuming a constant size distribution, and metal-line cooling accounts for the depletion of each element into dust. 
We also added an additional high-temperature dust cooling channel as described in Appendix~\ref{app:dust_cooling}, which scales with the local D/Z.

\subsection{Idealized Initial Conditions} \label{ICs}

We utilize idealized, non-cosmological disc galaxies, initialized with {\small MAKEDISK} \citep{springel_2005:ModellingFeedbackStars} to create a stellar disc+bulge and gaseous disc embedded in a dark matter halo, with galaxy properties and resolution details provided in Table~\ref{tab:idealized_ICs}.
For the dust model variation testing in Sec.~\ref{sec:model_variations} we utilize a low resolution MW-mass galaxy ({\bf m12\_lowres}), while for comparisons with local Group observations in Sec.~\ref{sec:local_group_comparison} we utilize high resolution MW, LMC, and SMC-mass galaxies ({\bf m12}, {\bf m11}, and {\bf m10}). The full details for our initial conditions are outlined below.

\begin{figure*}
    \plotsidesize{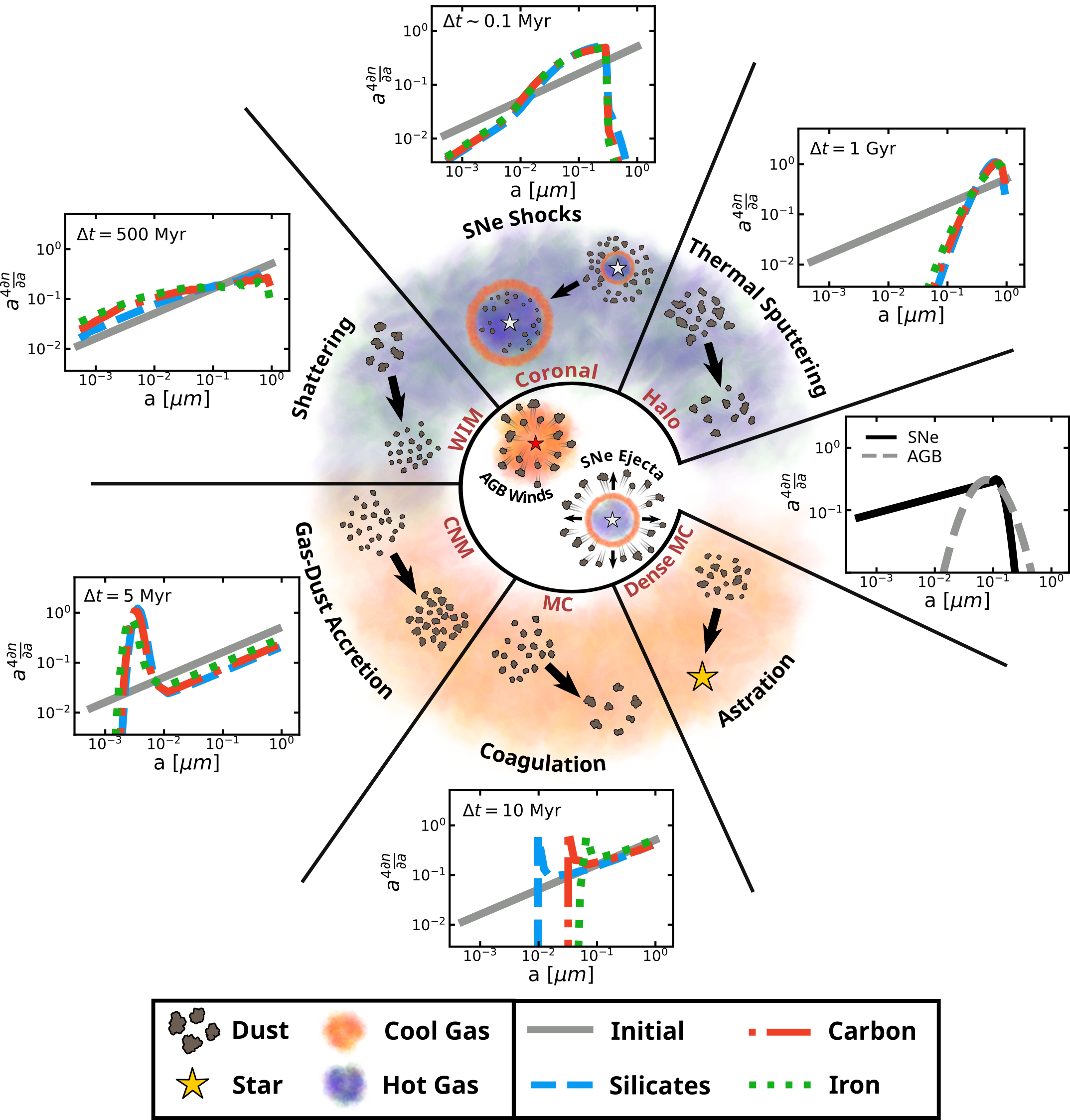}{0.99}
    \caption{Pictorial representation of each dust process included in our dust evolution model and the ISM phase where they typically dominate. 
    For each process, we include diagrams for how an initial MRN grain size distribution ({\it grey solid}) would change assuming typical gas properties and timescales for each ISM phase.
    We also show this change for silicates ({\it blue dashed}), carbonaceous ({\it red dash-dotted}), and metallic iron ({\it green dotted}) species since each process depends on dust physical properties as summarized in Table~\ref{tab:dust_quantities}.    
    {\bf Dust Creation:} Dust is initially created in the metal-rich ejecta of SNe and AGB winds, with gas-phase metals directly condensing into solid dust grains. Stardust is believed to be large $(a\gtrsim0.1\micron)$ in size but may shatter in SNe. These dust grains are injected into the ISM where they are exposed to mass and number altering processes.
    {\bf Thermal Sputtering:} Dust grains residing in hot, halo gas are eroded by collisions with energetic protons and electrons, shrinking and ultimately destroying them.
    {\bf SNe Shocks:} As SNe shocks propagate through the ISM, they shatter and sputter dust grains residing in said gas.
    {\bf Shattering:} In hot, turbulent gas, dust grains have large relative velocities which cause colliding grains to shatter into smaller grains, shifting grain mass from large to small sizes.
    {\bf Gas-Dust Accretion:} In cool, dense gas, gas-phase metals have sufficiently low energy that when they collide with a dust grain, they stick and grow the grain.
    {\bf Coagulation:} In molecular clouds, dust grains have low relative velocities allowing colliding grains to stick together, becoming one larger aggregate, shifting grain mass from large to small sizes.
    {\bf Astration:} As gas cools and collapses to form stars, the dust residing in the gas is destroyed and contributes to the stellar metallicity.}
    \label{fig:lifecycle_diagram}
\end{figure*}

For each galaxy, we set the galactic disc with gas and stellar mass of $M_{\rm disc,gas}$ and $M_{\rm disc,*}$ and utilize an exponential stellar and gas density profile $\rho(R,z)\propto e^{-R/R_{\rm d}}e^{-|z|/z_{\rm d}}$ with radial scale lengths of $R_{\rm d,*}$ and $R_{\rm d,gas} = 1.5R_{\rm d,*}$ kpc respectively and vertical scale length of $z_{\rm d}= 0.08 R_{\rm d,*}$ for both. 
We also include a stellar bulge with mass $M_{\rm bulge}$ following a Hernquist profile~\citep{hernquist_1990:AnalyticalModelSpherical} and a black hole with mass $M_{\rm bh}$. The galaxy is embedded in a NFW profile \citep{navarro_1996:StructureColdDark} dark matter halo with $M_{\rm vir}$ and halo concentration of $c=12$. 
The gas cell mass resolution is $m_{\rm gas}$  with adaptive softening lengths, achieving a minimum softening length of $\epsilon^{\rm MIN}_{\rm gas}$. 
The initial disk and bulge star particles have a resolution of $m_{\rm IC, star}=5 m_{\rm gas}$ with a universal softening length of $\epsilon_{\rm IC,star}=10\epsilon^{\rm MIN}_{\rm gas}$.
The dark matter particles have a mass resolution of $m_{\rm DM}=100m_{\rm gas}$ with a universal softening length of $\epsilon_{\rm DM}=100\epsilon^{\rm MIN}_{\rm gas}$. 
All gas cells and star particles start with an initial $Z_{\rm init}$, star particles initially have a uniform age distribution over 13.8 Gyr, and gas cells have an initial dust population assumed to be SNe II dust and whose amount is set as follows. We set an initial element depletions of $\delta_{\rm dust}$ for Si and Fe into silicate and metallic iron dust respectively, O and Mg depletion then follow from the assumed chemical composition of silicates outlined in Sec.~\ref{sec:dust_composition}. 
We determine an initial carbonaceous dust mass assuming a silicate-to-carbonaceous dust mass ratio (Sil-to-C) following estimates for the MW, LMC and SMC \citep[e.g.][]{pei_1992:InterstellarDustMilky}.
The initial grain size distribution is set to an MRN ($\dndaflat\propto a^{-3.5}$) power law for the low-resolution MW-mass galaxy\footnote{Tests assuming an initial log-normal size distribution centered at $a=0.1\micron$ exhibited no change in our {\bf Fiducial} model predictions shown in Sec.~\ref{sec:fiducial_model}.}.
For the high resolution MW, LMC, and SMC-mass simulations we use a log-normal initial size distribution, with $a_{\rm center} = 0.1\;\micron$ and $\sigma=0.6$, to imitate an initially star-dust dominated population.
To allow the gaseous disk to settle into steady star formation, we first simulate the galaxy for $0.2-0.5$ Gyr with all dust evolution processes turned off and no coupling with dust physics. The galaxy is then simulated for ${\sim}0.5$ Gyr with the dust evolution model turned on and fully coupled with all dust physics routines. We find that this timescale is long enough for our dust evolution model to reach a steady-state D/Z ratio and stable dust population chemical and source composition.

\section{Grain Size Evolution Model} \label{GSE}

\begin{table*}
    \centering
    \renewcommand{\arraystretch}{1.3}
	\begin{tabular}{| l l | l l l |}
		 \hline
		Quantity & Symbol & Silicate & Carbonaceous & Metallic Iron\\ [0.5ex] 
		\hline
            \multicolumn{5}{|c|}{Physical Properties} \\
            \hline 
            Internal density & $\rho_{\rm gr}$ (g $\cmcubed$) & 3.13 & 2.25 & 7.86 \\
            Atomic weight & $A_{\rm c}$ & 143.8 & 12.0 & 55.9 \\
            Shattering threshold & $\vshat$ (km/s) & 2.7 & 1.2 & 2.2\\
            Critical shock pressure &  $P_1$ (dyn cm$^{-2}$) & $3\times10^{11}$ & $4\times10^{10}$ & $5.5\times10^{10}$ \\
            Surface energy & $\gamma$ (erg cm$^{-2}$) & 25 & 75 & 3000 \\
            Young's modulus & $E$ (dyn cm$^{-2}$) & $5.4\times10^{11}$ & $1.0\times10^{11}$ & $2.1\times10^{12}$ \\
            Poisson's ratio & $\nu$ & 0.17 & 0.32 & 0.27 \\
            Small grain Coulomb enh. &  $D_{\rm small}$ & 10 & 3 & 20 \\
            Large grain Coulomb enh.  &  $D_{\rm large}$ & 0.5 & 0 & 1 \\
   		\hline
            \multicolumn{5}{|c|}{Stardust Creation} \\
            \hline          
            SNe II dust creation & $M^{\rm SNe\;II}_{\rm dust}$ ($\Msol$) & 0.03 & 0.06 & 0.008 \\
            SNe Ia dust creation & $M^{\rm SNe\;Ia}_{\rm dust}$ ($\Msol$) & 0 & 0 & 0.003 \\
            AGB dust creation & $M^{\rm AGB}_{\rm dust}$ ($\Msol$) & \multicolumn{3}{c|}{Following \citet{zhukovska_2008:EvolutionInterstellarDust}} \\
       	\hline
            \multicolumn{5}{|c|}{SNe Destruction} \\
            \hline      
            SNe sputtering efficiency &
            $\delta^{\rm SN}_{\rm sput}$ & 0.3 & 0.2 & 0.24 \\
            SNe shattering efficiency &
            $\delta^{\rm SN}_{\rm shat}$ & 0.1 & 0.13 & 0.15 \\
            Total destruction eff. & $\epsilon_{\rm dest}$ & 0.4 & 0.35 & 0.41 \\
       	\hline
            \multicolumn{5}{|c|}{Other Parameters} \\
            \hline
            Accretion turnoff temp. & $T_{\rm cutoff}\;(K)$ &\multicolumn{3}{c|}{300 } \\
            Coag. density enhancement & $C_{\rm coag}$ &\multicolumn{3}{c|}{see Eq.~\ref{eq:coag_enhancement}} \\
		\hline
	\end{tabular}
	\caption{Table summarizing physical properties and parameters for each dust species we evolve in our simulations.}
	\label{tab:dust_quantities}
\end{table*}


Our grain size evolution model includes the creation of dust by SNe (Sec~\ref{sec:SNe_dust}) and AGB stars (Sec~\ref{sec:AGB_dust}), the growth of preexisting dust grains via gas-dust accretion (Sec~\ref{sec:accretion}), the destruction of dust via thermal sputtering (Sec~\ref{sec:sputtering}), SNe shocks (Sec~\ref{sec:SNe_shocks}), and astration (Sec~\ref{sec:astration}), the shattering (Sec~\ref{sec:shattering}) and coagulation (Sec~\ref{sec:coagulation}) of colliding grains, and incorporates sub-resolution turbulent dust diffusion in the same manner as the turbulent metal diffusion in FIRE-3.
This model also tracks the separate evolution of silicate, carbonaceous, and metallic iron dust species (Sec~\ref{sec:dust_composition}), affecting how each dust process is modeled due to their dependence on numerous dust species properties, which are listed in Table~\ref{tab:dust_quantities}.
We provide a general overview of our model in 
Fig.~\ref{fig:lifecycle_diagram}, which presents pictorial representations of each dust process, the typical phase of the ISM they occur in, and how they change an MRN grain size distribution for silicates, carbonaceous, and metallic iron species.
Our model considers minimum and maximum grain sizes of  $\amin=5 \, \angstrom$ and $\amax=1 \, \micron$, divided into $N_{\rm bin}=16$ logarithmically-spaced bins with convergence tests provided in Appendix~\ref{app:convergence_tests}.

\subsection{Dust Species Chemical Composition} \label{sec:dust_composition}

As shown in \citetalias{choban_2022:GalacticDustmodellingDust}, \citetalias{choban_2024:DustyLocaleEvolution}, and \citetalias{choban_2025:DustyDawnGalactic} accounting for and evolving individual dust species with set chemical compositions has broad impacts on the inferred dust population evolution. Individual species have different key element abundances, grain charging, shattering thresholds, sputtering yields, and material `softness', shown in Table~\ref{tab:dust_quantities}, which are inputs into grain size evolution models. 
Therefore, we track the evolution of individual dust species, similar to \citetalias[][]{choban_2022:GalacticDustmodellingDust}. In particular, we model the evolution of silicates ($\rm Mg_{1.06} Fe_{0.57} Si O_{3.63}+O_2$), carbonaceous (pure C), and metallic iron (pure Fe), tracking the evolution of individual grain size distributions for each.
Note the addition of $\rm O_2$ to silicates is to account for the observed excess in depletion of O that cannot be explained by silicates alone \citep{whittet_2010:OxygenDepletionInterstellar} but for which no clear carrier has been found \citep{poteet_2015:CompositionInterstellarGrains,wang_2015:InterstellarOxygenCrisis,jones_2019:EssentialElementsDust}. This extra O depletion is especially important for total dust masses due to it being the most abundant metal. 

\subsection{Sub-resolved Clumping of Gas}

\begin{figure*}
    \plotsidesize{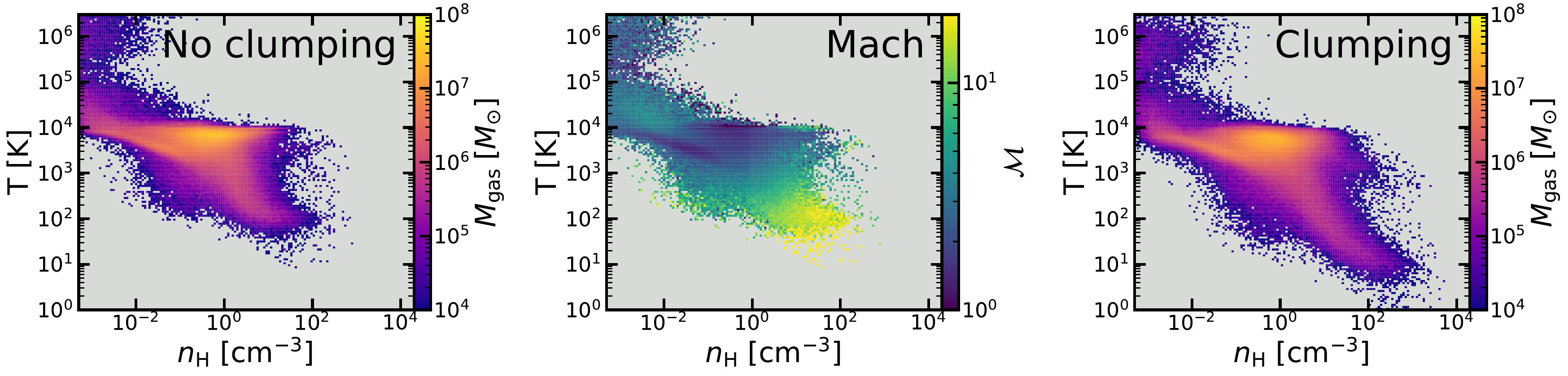}{0.99}
    \caption{Gas phase diagram for all gas in our {\bf m12\_lowres} simulation at simulation end. (\textit{left}) Phase-space diagram using volume-averaged density and temperature tracked in the simulation. (\textit{middle}) Median sonic Mach number tracked in the simulation. (\textit{right}) Phase-space diagram using rms density and effective temperature, factoring in sub-resolved clumping of gas following Appendix~\ref{app:gas_clumping}. $\mathcal{M}$ rapidly increases in cool, dense gas, increasing the rms density and decreasing the effective temperatures beyond those tracked by the volume-averaged quantities in our simulation.}
    \label{fig:gas_phase_clumping}
\end{figure*}

Turbulence drives the clumping of gas and dust on scales smaller than typically resolved by our simulations. 
This clumping increases the rate of gas-gas, dust-gas, and dust-dust interactions. Thus, using only the gas cell volume-averaged density for a given cell $i$ ($\bar{n} \equiv \langle n \rangle_{i} = N_{i,} / V_{i}$) will underestimate these interaction rates and introduce a resolution dependence. 
To account for sub-resolution clumping, we use the scheme outlined in Appendix~\ref{app:gas_clumping} and briefly described below. Using the local sonic Mach number $\mathcal{M}$ tracked for each gas cell and an assumed gas density probability density function, we calculate second-order clumping factors $C_2=1+b^2\mathcal{M}^2$, where $b=0.5$, used to enhance all gas-dust and dust-dust interaction rates.
For gas-dust accretion specifically, we use a modified clumping factor $C_{2}^{\rm acc}$ which accounts for the ``turn off'' of accretion past a maximum density due to CO or ice formation depending on the dust species. 
For the process of gas-dust accretion and coagulation, we also use the local root-mean-squared number densities ($\nH^{\rm rms}=\sqrt{C_2}\bar{n}$) and effective temperature ($\Teff=\bar{T}/\sqrt{C_2}$) instead of the volume-averaged equivalents.
We demonstrate the typical mach numbers predicted in our simulation and differences between the $\bar{n}-\bar{T}$ and $\nH^{\rm rms}-\Teff$ gas-phase diagram in Fig.~\ref{fig:gas_phase_clumping}.

\subsection{Discretized Grain Size Distribution} \label{sec:GSD}

The discretized formulation of our dust grain size evolution model largely follows the second-order piecewise linear method developed in \citet{mckinnon_2018:SimulatingGalacticDust}.
For brevity, we only summarize the discretized formulation below and focus on analytical equations in the succeeding sections. The full discretized forms of all equations are compiled in Appendix~\ref{app:discretized_forms} and referenced in the main text as appropriate. We also provide additional timestep sub-cycling criteria enforced in our simulations due to dust evolution processes in Appendix~\ref{app:time_stepping}.

Each gas cell in our simulation tracks a grain size distribution for each dust species discretized into $N$ logarithmically spaced bins between minimum and maximum grain sizes $\amin=5 \, {\rm nm}$ and $\amax=1 \, \micron$ with logarithmic bin width
\begin{equation}
    \log \delta = \frac{\log\amax - \log\amin}{N_{\rm bin}}.
\end{equation}
The edges of the $N_{\rm bin}$ bins are $(a_0^{\rm e},a_1^{\rm e},\ldots,a_{N_{\rm bin}}^{\rm e}$), where $\ailower\equiv\delta^i\amin$. 
For values below/above the min/max grain sizes, we define the edges $a_{-1}=0$ and $a_{N_{\rm bin}+1}=\infty$ respectively.
We assume that the grain size distribution within each bin $i$ at time $t$ takes the linear form
\begin{equation} \label{eq:dnda_discretized}
    \dnda = \frac{N_i(t)}{\aiupper-\ailower} + s_i(t)(a-\aic),
\end{equation}
where $\aic=(\ailower+\aiupper)/2$ is the linear mid-point of the bin, $s_i(t)$ is the slope, and $N_i(t)$ is the total number of grains in the bin. With this form, each bin inherently tracks the total number of grains of size $\ailower<a<\aiupper$
\begin{equation} \label{eq:N_from_dnda}
      N_i(t) = \int_{\ailower}^{\aiupper}\dnda \, da,
\end{equation}
and $s_i(t)$ is set such that each bin also tracks the total mass of grains of size $\ailower<a<\aiupper$. Assuming spherical grains, the mass of grains in bin $i$ is
\begin{equation} \label{eq:M_from_dnda}
         M_i(t) = \int_{\ailower}^{\aiupper} m(a) \dnda \, da = \int_{\ailower}^{\aiupper} \frac{4 \pi \rho_{\rm gr}}{3}  a^3 \dnda \, da,
\end{equation}
where $\rho_{\rm gr}$ is the internal density of the given dust species. $s_i(t)$ is then determined such that
\begin{equation} \label{eq:slope_from_M}
     M_i(t) =  \frac{4 \pi \rho_c}{3} \left[ \frac{N_i a^4}{4 (\aiupper-\ailower)} + s_i(t) \left( \frac{a^5}{5} - \frac{\aic a^4}{4} \right) \right]^{a=\aiupper}_{a=\ailower}.
\end{equation}


After a given time step, $\dt$, the number, $N_i(t+\dt)$, and mass, $M_i(t+\dt)$, of grains for a bin can be modified due to various physical processes which we describe in later sections. Given the new $N_i(t+\dt)$ and $M_i(t+\dt)$ a new bin slope $s_i(t+\dt)$ is determined using Eq.~\ref{eq:slope_from_M}. With each update, slopes are also limited following Appendix~\ref{app:slope_limiting} as needed to avoid unphysical values, such as negative $\dndaflat$.

\subsection{Initial Grain Size Distribution and Stardust Creation}

The first seeds of a dust population are formed in the metal-rich ejecta and winds of stars. SNe and AGB stars are believed to produce the majority of this so-called stardust in present-day galaxies (although more exotic stars could be dominant stardust sources in earlier times, such as Wolf-Rayet or Pop III stars).
Below, we briefly describe how grain size bins are updated due to stardust injection from an arbitrary source and provide specifics for the amount and size of dust grains prescribed for SNe and AGB in Sec.~\ref{sec:SNe_dust} and~\ref{sec:AGB_dust} respectively.

Over a given time step $\dt$, star particles can inject dust created in SNe ejecta or AGB winds into nearby gas parcels, with the total dust mass injected labelled as $M_{\rm total}^{\rm inj}$.
Then, given an assumed $\dndaflat$ for each creation process, we determine the number and mass of grains to be injected into bin $i$, $N_i^{\rm inj}$ and $M_i^{\rm inj}$, using Eq.~\ref{eq:N_from_dnda} and~\ref{eq:M_from_dnda} respectively. The resulting dust grain number and mass after the time step are then $N_i(t+\dt)=N_i(t)+N_i^{\rm inj}$ and $M_i(t+\dt)=M_i(t)+M_i^{\rm inj}$ respectively. We then determine the new slope $s_i(t+\dt)$ for each bin using Eq.~\ref{eq:slope_from_M} and limit slopes as needed.


\subsubsection{Dust Creation by SNe} \label{sec:SNe_dust}

Understanding the amount and size of grains injected into the ISM by SNe first requires constraining the amount and size of dust initially created within the SNe.
Observations find large quantities of dust can form within cool clumps of Type II SNe ejecta with masses ranging from ${\sim}10^{-5} - {\sim}0.5 \Msol$ and tentatively increasing with SNe age \citep[e.g.][]{shahbandeh_2024:JWSTMIRIObservations,sarangi_2025:TwoDecadesDust}. 
On the other hand, Type Ia SNe are generally believed to produce little if any dust \citep{nozawa_2011:FormationDustEjecta, gomez_2012:DustHistoricalGalactic}, but recent JWST observations of one SNe Ia suggest the exact opposite
\citep{wang_2024:NewlyFormedDust}.
In regards to grain sizes, there is strong observational evidence that this newly formed dust is preferentially large, $a\sim0.1 \, \micron$ \citep{gall_2014:RapidFormationLarge,wesson_2015:TimingLocationDust,bevan_2016:ModellingSupernovaLine,priestley_2020:DustMassesGrain}, but the exact size distribution is unknown and varies between theoretical predictions \citep{nozawa_2003:DustEarlyUniverse,sarangi_2015:CondensationDustEjecta,sluder_2018:MolecularNucleationTheory}.

To be incorporated into the ISM, these newly formed grains must be subsequently propelled out of the SNR. 
During this period, grains are processed by the reverse shock of the SNe,  drastically changing the final grain size distribution.
Numerous analytical and hydrodynamic models have been developed to determine the final size distribution of these surviving grains \citep[e.g.][]{nozawa_2007:EvolutionDustPrimordial,bocchio_2016:DustGrainsHeart,slavin_2020:DynamicsDestructionSurvival,kirchschlager_2019:DustSurvivalRates,kirchschlager_2020:SilicateGrainGrowth,kirchschlager_2023:DustSurvivalRates,kirchschlager_2024:TotalDestructionComplete}, but they vary in their assumed dust physics, such as grain-grain collisions, grain charging, ion trapping, and assumed initial size distributions, leading to large variations in the final mass produced. 
However, certain results are shared among them: (1) silicate grains are more easily destroyed compared to carbonaceous grains due to increased sputtering yields and (2) mainly large grains ($a\gtrsim0.05\;\micron$) survive. 
This ultimately leads to a final size distribution similar to the initial distribution of large grains with a possible added tail toward smaller grains due the shattering of grains via grain-grain collisions. 
These results are also tentatively supported by observations of Cas A \citep{priestley_2022:DustDestructionSurvival} which indicate differing grain size populations in components of the SNe; with large grains $(a\sim0.1\,\micron)$ in the diffuse, X-ray emitting shocked ejecta and small grains $(a\sim5-10\, \nanom)$ in dense clumps which have passed through the reverse shock.

Given the lack of consensus on SNe II dust yields, we assume SNe II are moderate dust producers, with a constant 20\% fraction of the Si, C, and Fe SNe II metal yields locked in silicates, carbonaceous, and metallic iron dust, respectively ($\delta^{\rm SNe\,II}_{\rm all}=20\%$).
This results in ${\sim}0.1\;\Msol$ of dust produced per SNe II (see Table~\ref{tab:dust_quantities} for the yields of each species), which falls in the middle of observed estimates \citep[e.g.][]{shahbandeh_2024:JWSTMIRIObservations}.
For the final grain size distribution, we follow the results of \citet{kirchschlager_2019:DustSurvivalRates,kirchschlager_2020:SilicateGrainGrowth} due to their more in-depth dust physics compared to other works. \citet{kirchschlager_2019:DustSurvivalRates,kirchschlager_2020:SilicateGrainGrowth} predicts a two component grain size distribution: (1) a log-normal distribution of large grains similar to an assumed initial distribution formed within the SNe and (2) a power-law distribution of small grains produced by the shattering of large grains via grain-grain collisions. Thus we employ a functional form similar in shape to the results in \citet[][]{kirchschlager_2020:SilicateGrainGrowth} (see Fig.~\ref{fig:lifecycle_diagram}) such that
\begin{equation} \label{SNe dn/da}
\dnda =
\begin{cases}
      \frac{C_1}{a}\exp\left(-\frac{\ln^2(a/a_{\rm SNe})}{2 \sigma_{\rm SNe}^2}\right) & a \geq a_{\rm cut} \\
      C_2\,a^{-\gamma} & a_{\rm min} \leq a < a_{\rm cut}, \\
\end{cases} 
\end{equation}

where we take $a_{\rm SNe}=0.1\,\micron$, $\sigma_{\rm SNe}=0.2$, $a_{\rm cut}=a_{\rm SNe}$, $\gamma=3.5$, $a_{\rm min}{\sim}1\,\nanom$, and $C_1$ and $C_2$ are normalization factors such that $\int_{0}^{\infty} m(a) \dndaflat \, da = M_{\rm dust}$ and $\dndaflat$ is continuous. 
We highlight that \citet{kirchschlager_2019:DustSurvivalRates,kirchschlager_2020:SilicateGrainGrowth} solely model dusty clumps passing through the reverse shock and not the slowing of the grains until they eventually merge with the surrounding ISM. 
During this time grains are subject to destruction by thermal and kinetic sputtering which preferentially destroys small grains \citep{slavin_2020:DynamicsDestructionSurvival} which could reduce $\gamma$ and increase $a_{\rm min}$ substantially.
Due to the lack of literature on dust mass yields or grain size distributions from SNe Ia, we assume a small amount of metallic iron dust is produced in SNe Ia ($\delta^{\rm SNe\,Ia}_{\rm iron}=0.5\%$) and assume the same size distribution as SNe II.

\subsubsection{Dust Creation by AGB Stars} \label{sec:AGB_dust}

Large amounts of dust can form in the metal-rich winds of AGB stars, with the amount and type of grain species produced depending on the stellar mass and metallicity which is seen in both observations \citep[e.g.][]{mcdonald_2010:RustyOldStars,marini_2019:EvolvedStarsLMC,maercker_2022:InvestigatingDustProperties,kraemer_2024:DustiestGalacticStars} and theoretical models 
\citep[e.g.][]{rosenfield_2014:EvolutionThermallyPulsing,ventura_2018:GasDustSolar,ventura_2020:GasDustMetalrich,ventura_2021:GasDustExtremely}.
However, there is little literature on the grain size distribution for the dust they produce. \citet{yasuda_2012:FormationSiCGrains} investigated the formation of SiC dust around a $1 \Msol$ C-rich pulsating AGB star utilizing hydrodynamical simulations. 
Their results suggest dust produced by AGB stars has a log-normal mass distribution per logarithmic radius $(a^4 \dndaflat)$ with a peak at $a\sim0.2-0.3 \, \micron$. 
Observations of dust in a sample of C-rich AGB wind-ISM interaction regions by \citet{maercker_2022:InvestigatingDustProperties} exhibit even larger carbonaceous dust grains $(a\gtrsim2 \micron)$, suggesting grains continue to grow outside of the circumstellar envelope. 
Furthermore, in situ studies of pre-solar graphite grains from meteorites, which likely originate from AGB stars, have also found sizes of $a\sim1\; \micron$ \citep[e.g.][]{bernatowicz_1996:ConstraintsStellarGrain,xu_2016:FirstDiscoveryPresolar}. 

For the amount of each dust species produced by AGB stars, we follow the prescription laid out in \citetalias{choban_2022:GalacticDustmodellingDust}, using the AGB dust production results from \citet{zhukovska_2008:EvolutionInterstellarDust}, convolved over a \citet{kroupa_2002:InitialMassFunction} IMF. 
For the grain sizes, we assume AGB dust has a log-normal mass  distribution ($a^4 \dndaflat$) following results from \citet{yasuda_2012:FormationSiCGrains} (see Fig.~\ref{fig:lifecycle_diagram}) such that

\begin{equation}
    \dnda = \frac{C}{a^5}\exp\left(-\frac{\ln^2(a/a_{\rm AGB})}{2 \sigma_{\rm AGB}^2}\right)
\end{equation}
where $a_{\rm AGB}=0.1$ $\micron$, $\sigma_{\rm AGB}=0.47$, and $C$ is a normalization constant such that $\int_{0}^{\infty} m(a) \dndaflat \, da = M_{\rm dust}$ $\left( C=\sqrt{\frac{3 a_{\rm AGB} M_{\rm dust}}{2^{5/2}\pi^{3/2}\rho_{\rm gr}\sigma_{\rm AGB}}}  e^{-\sigma_{\rm AGB}/4}\right)$, and $M_{\rm dust}$ is the total dust mass returned for a given stellar wind event.

\subsection{Number Conserving Processes} \label{sec:number_conserving_processes}

Processes such as gas-dust accretion and thermal sputtering alter the size of dust grains, conserving the total number of grains but altering the total mass.
Over a timestep $\dt$, these processes will change a given grain's radius by $\dot{a}(a,t) \times \Delta t$. 
Given this, we update the number and slope of each bin following the procedure laid out in Appendix~\ref{app:update_number_conserving}. 
Below we describe the gas-dust accretion and thermal sputtering routines and how $\dot{a}(a,t)$ is determined for each.

\subsubsection{Gas-Dust Accretion} \label{sec:accretion}

In cool environments, dust grains can grow via gas-dust accretion whereby gas-phase metals collide and stick to the surface of preexisting dust grains. 
Similar to \citetalias{choban_2022:GalacticDustmodellingDust}, we limit the accretion rate for each dust species by the key element\footnote{Here key element refers to the element for which $N/i$ has the lowest value, where $N$ is the number density of the element and $i$ is the number of atoms of the element in one formula unit of the dust species under consideration.} for that species and assume that all gas-phase metals are in atomic form.
Thus, the change in mass for a single grain of dust species $k$ due to gas-dust accretion is
\begin{equation} \label{eq:mass_accretion}
    \frac{d m_{{\rm gr},k}}{dt} = \left( \sigma^{\rm eff}_{{\rm gr},k} \, \varv_{\rm key,th} \, n_{\rm key} \right) \, \left(\xi_{\rm key} \, m_{\rm added} \right),
\end{equation}
where the first term is the interaction rate between the dust grain and key element, and the second term is the mass added to the dust grain after each interaction. The variables are as follows: $\sigma^{\rm eff}_{{\rm gr},k}$ is the effective interaction cross-sectional area between the dust grain and atomic metals, $\varv_{\rm key, th}=\sqrt{\frac{8\, k\, T }{\pi A_{{\rm key},k}m_{\rm H}}}$ is the thermal velocity of the key element, $n_{\rm key}$ is the gas-phase number density of the key element (i.e. amount not locked up in dust), $\xi_{\rm key}$ is the sticking efficiency for each collision, $m_{\rm added}=A_{{\rm gr},k} \, m_{\rm H} / \alpha_{{\rm g},k}$ is the mass added to the dust grain with each collision, $A_{{\rm gr},k}$ and $A_{{\rm key},k}$ are the atomic weight of one formula unit of the dust material under consideration and of the key element respectively, and $\alpha_{{\rm gr},k}$ is the number of atoms of the key element contained in one formula unit of the dust material under consideration.

Assuming spherical dust grains ($m_{{\rm gr},k}=4/3\pi\rho_{{\rm gr},k} a_k^3$) and purely hard-sphere type encounters ($\sigma^{\rm eff}_{{\rm gr},k}=\pi a_k^2$), we can express the change in grain size due to accretion as 
\begin{equation} \label{eq:size_accretion1}
    \frac{da_k}{dt} = \frac{\varv_{{\rm key,th}} \, n_{\rm key} \, \xi_{\rm key} \, m_{\rm added}}{4 \rho_{{\rm gr},k}}.
\end{equation}
For the sticking efficiency, we follow \citet{zhukovska_2016:ModelingDustEvolution} and take a simple step function with $\xi_{\rm key} = 1$ for $\Teff<T_{\rm cutoff}$ and $\xi_{\rm key} = 0$ for $\Teff>T_{\rm cutoff}$, with the fiducial value of $T_{\rm cutoff}$ provided in Table~\ref{tab:dust_quantities}. 
This is roughly in line with detailed molecular dynamic simulations which find sticking efficiencies drop with increasing temperature \citep{bossion_2024:AccurateStickingCoefficient,hansson_2025:BindingEnergiesAtoms}.

We also consider two other factors: (1) sub-resolved clumping and (2) Coulomb enhancement.

{\bf (1) Sub-resolved clumping: }
Turbulence drives gas clumping on sub-resolution scales, which can enhance gas-dust interaction rates.
However, beyond a certain density, gas-dust accretion is halted due to either the freeze-out of molecules onto dust grain surfaces or the conversion of all gas-phase key elements into molecules that do not accrete onto dust grains, such C into CO. 
To account for this, we calculate an effective change in grain size as
\begin{equation}
    \left(\frac{da_k}{dt}\right)_{\rm eff} = C_{2,k}^{\rm acc} \frac{da_k}{dt}.
\end{equation}
where $C_{2,k}^{\rm acc}$ is a 2nd-order clumping factor for dust species $k$ that accounts for sub-resolved gas clumping up to a maximum density above which accretion effectively turns off, provided in Appendix~\ref{app:gas_clumping}.

{\bf (2) Coulomb Enhancement:} The interaction cross-sectional area between ionized gas-phase metals and dust grains can vary depending on the grain charge distribution. 
In the CNM, small dust grains can be negatively charged while large grains are positively charged, greatly enhancing/decreasing their effective interaction area respectively \citep{weingartner_1999:InterstellarDepletionVery,draine_2011:PhysicsInterstellarIntergalactic,hensley_2017:ThermodynamicsChargingInterstellar,ibanez-mejia_2019:DustChargeDistribution}.
To account for this Coulomb enhancement, the effective cross-sectional area can be written as 
$\sigma^{\rm eff}_{{\rm gr},k}=\pi a_k^2 D_k(a)$ where $D_k(a)$ is the grain size dependent electrostatic enhancement for dust species $k$.
We utilize a piecewise function prescription for $D_k(a)$, following findings that small grains ($a\lesssim0.01\micron$) have a noticeable fraction of negatively charged grains and large grains ($a\gtrsim0.01\micron$) are largely positively charged in CNM environments \citep{draine_2011:PhysicsInterstellarIntergalactic,ibanez-mejia_2019:DustChargeDistribution}.
Thus,
\begin{equation}
    D_k(a) = 
\begin{cases}
      D_{{\rm small},k} & a\leq a_{\rm min} \\
      \frac{(D_{{\rm large},k} - D_{{\rm small},k})}{\log(a_{\rm mid}/a_{\rm min})} \log\left(\frac{a}{a_{\rm min}} \right) +  D_{{\rm small},k} & a_{\rm min} < a \leq a_{\rm mid} \\
      D_{{\rm large},k} & a>a_{\rm mid},\\
\end{cases} 
\end{equation}
where $D_{{\rm small},k}$ and $D_{{\rm large},k}$ are the small and large grain Coulomb enhancement factors provided in Table~\ref{tab:dust_quantities}, $a_{\rm min}=0.001\;\micron$, and $a_{\rm mid}=0.01\;\micron$.
The chosen values for $D_{{\rm small},k}$ and $D_{{\rm large},k}$ are derived from general trends in \citet{weingartner_1999:InterstellarDepletionVery} for silicates and carbonaceous dust. For metallic iron, no enhancement factors are readily available, but it should to be more negatively charged than silicates and have higher enhancement factors \citep{hensley_2017:ThermodynamicsChargingInterstellar, zhukovska_2018:IronSilicateDust}. We therefore assume similar values to silicates increased by an arbitrary factor of 2.

Incorporating the above factors, Eq.~\ref{eq:size_accretion1} becomes
\begin{equation} \label{eq:dadat_acc}
\begin{aligned}
     \left(\frac{da_k}{dt}\right)_{\rm eff} 
     & = C_{2,k}^{\rm acc} \sqrt{\frac{k m_{\rm H}}{2\pi}} \frac{\xi_{\rm key} \, D_k(a) \, A_{{\rm gr},k}}{\alpha_{{\rm gr},k} \, A_{{\rm key},k}^{1/2}} \\
     & \;\;\;\;\; \times \left( \frac{1  \; {\rm g \; cm}^{-3}}{\rho_{{\rm gr},k}} \right) \left( \frac{n_{\rm key}}{1 \; {\rm cm}^{-3}} \right) \left( \frac{\Teff}{1 \; {\rm K}} \right)^{1/2}\; {\rm cm/s} \\
     & = \left( 0.110 \; {\rm \frac{\micron}{Gyr}} \right)  C_{ 2,k}^{\rm acc} \frac{\xi_{\rm key} \, D_k(a) \, A_{{\rm gr},k}}{\alpha_{{\rm gr},k} \, A_{{\rm key},k}^{1/2}} \\
     & \;\;\;\;\; \times \left( \frac{3  \; {\rm g \; cm}^{-3}}{\rho_{{\rm gr},k}} \right) \left( \frac{n_{\rm key}}{10^{-2} \; {\rm cm}^{-3}} \right) \left( \frac{\Teff}{300 \; {\rm K}} \right)^{1/2}.
\end{aligned}
\end{equation}

Given $(da_k/dt)_{\rm eff}$, we update the number and mass of grains in each bin as outlined in Appendix~\ref{app:update_number_conserving}. In the case where the resulting total mass of dust species $k$ after accretion ($M_{\rm dust, key}$) exceeds the available mass of gas-phase elements ($M_{\rm total, key}$), we scale down all grain size bin masses by $M_{\rm total, key}/M_{\rm dust, key}$ and recalculate the bin slopes (i.e. we shift all grains down in size).

\subsubsection{Thermal Sputtering} \label{sec:sputtering} 

\begin{table}
    \centering
    \begin{tabular}{|ccccccc|}
        \hline	
        & $a_0$ & $a_1$ & $a_2$ & $a_3$ & $a_4$ & $a_5$ \\
        \hline	
        $y_{\rm sil}$     & -226.95 & 127.94   & -29.920 & 3.5354 & -0.21055 & 0.0050362 \\
        $y_{\rm carb}$    & -226.85  & 133.44   & -32.573 & 4.0057 & -0.24747 & 0.0061212 \\
        $y_{\rm iron}$    & -156.88 & 82.110   & -18.238 & 2.0692 & -0.11933 & 0.0027788 \\
        \hline
    \end{tabular}
    \caption{Polynomial coefficients for our adopted erosion rate fits from data in Fig. 2 of \citet{nozawa_2006:DustDestructionHighVelocity}, where $y = \sum_{i=0}^{5} a_i x^i$, $x = \log_{10} (T/{\rm K})$, and $y = \log_{10} (Y / {\rm \mu m \ yr^{-1}cm^3})$. Note large number of significant digits needed to reproduce the curve.}
    \label{tab:sputtering_rates}
\end{table}


Dust grains residing in hot ($T\gtrsim10^6$ K) gas undergo thermal sputtering, by which atoms are ejected from the grain surface due to the bombarded of energetic protons and ions, eroding grain sizes over time. \citep{draine_1979:DestructionMechanismsInterstellar,tielens_1994:PhysicsGrainGrainCollisions,tsai_1995:InterstellarGrainsElliptical}. 
The rate of erosion differs between dust species primarily due to physical properties such as surface binding energies \citep{nozawa_2006:DustDestructionHighVelocity}, with silicate dust generally eroding faster than carbonaceous and metallic iron.
We follow the prescription for thermal sputtering from \citet{hu_2019:ThermalNonthermalDust}\footnote{We caution that the fit parameters given in \citet{hu_2019:ThermalNonthermalDust} do not reproduce the sputtering erosion rates due to a lack of significant figures.}, denoting the effective decrease in size of a grain of species $k$ due to sputtering as
\begin{equation}
    \left(\frac{da_k}{dt}\right)_{\rm eff} = C_2 \nH Y_k(T)
\end{equation}
where $C_2$ is the gas-dust clumping factor as described previously and in Appendix~\ref{app:gas_clumping}, $\nH$ is the hydrogen number density of the gas the dust resides in, and $Y_k(T)$ is the dust species erosion rate from gas with temperature $T$. For $Y_k(T)$, we utilize polynomial fits of thermal sputtering erosion rates determined by \citet{nozawa_2006:DustDestructionHighVelocity} (see their Fig. 2) for various dust species residing in $Z=10^{-4} \Zsol$ gas. For carbonaceous, silicates, and metallic iron dust, we adopt their rates for MgSiO$_3$, C, and Fe respectively, and provide their fit parameters in Table~\ref{tab:sputtering_rates}.
We highlight that this approximation does not take into account Coloumb forces from charged grains and size-dependant sputtering due to finite grain size which can affect the sputtering yield \citep[e.g.][]{kirchschlager_2019:DustSurvivalRates}.
Given $(da_k/dt)_{\rm eff}$, we update the number and mass of grains in each bin as outlined in Appendix~\ref{app:update_number_conserving}.

\subsection{Mass Conserving Processes} \label{sec:mass_conserving_processes}

\begin{figure}
    \centering
    \includegraphics[width=\linewidth]{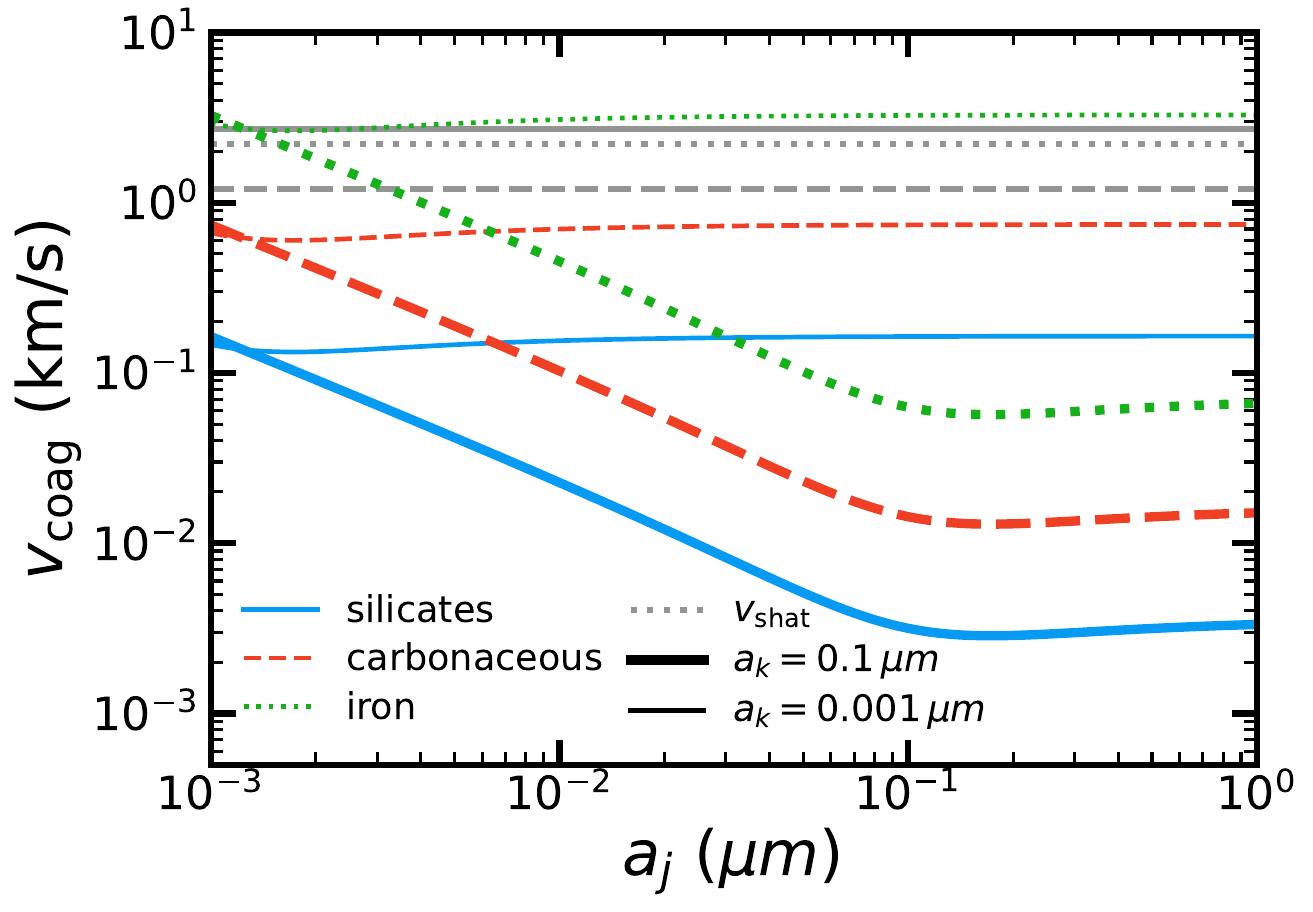}
    \caption{Predicted coagulation velocity thresholds from \citet{chokshi_1993:DustCoagulation,dominik_1997:PhysicsDustCoagulation,yan_2004:DustDynamicsCompressible} for interactions between grains of size $a_k$ and $a_j$ of the same dust species. We show the thresholds for silicate ({\it blue line}), carbon ({\it red dashed}), and metallic iron ({\it green dotted}) species with corresponding shattering velocity thresholds ({\it grey}). $a_k$ is set to 0.1 ({\it thick}) and 0.001 $\micron$ ({\it thin}). Larger grains have lower coagulation thresholds, and carbon and metallic iron have higher coagulation thresholds than silicate.}
    \label{fig:coag_velocity}
\end{figure}

\begin{figure*}
    \plotsidesize{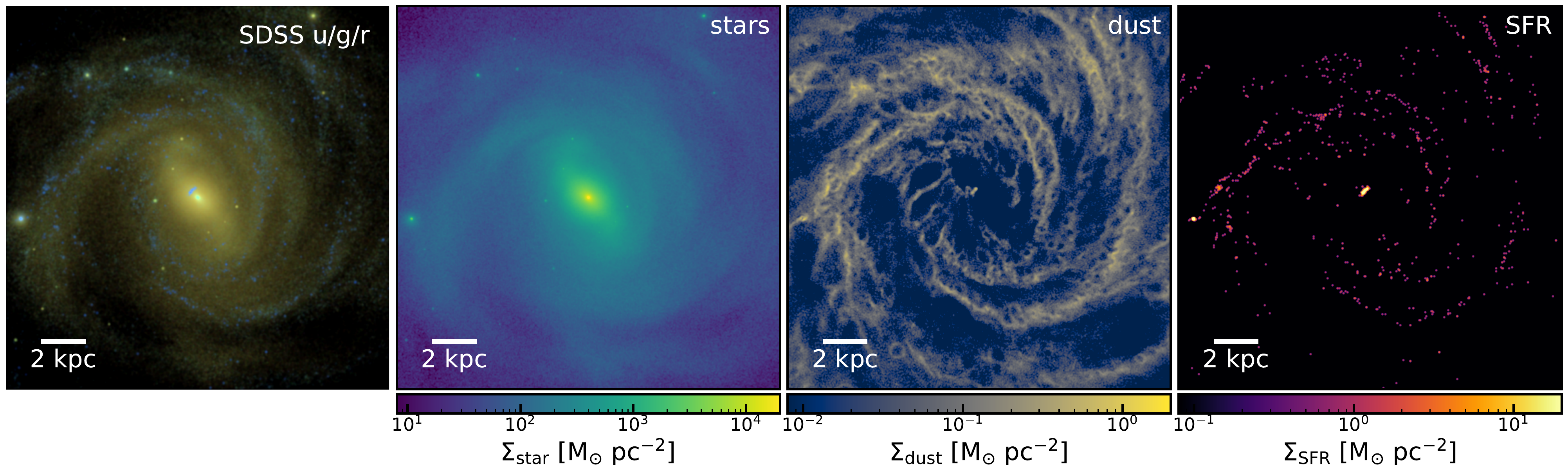}{0.95}
    \caption{Image and projections of our idealized MW-mass spiral galaxy. ({\it left}) Mock SDSS u/g/r image following the \citet{lupton_2004:PreparingRedGreenBlueImages} color algorithm created with SKIRT utilizing the local D/Z produced by our model, assuming MW dust opacities. ({\it left middle}) Stellar surface density projection. ({\it right middle}) Dust surface density projection. ({\it right}) Stellar surface density projection for stars formed in the last 10 Myr.}
    \label{fig:mock_image}
\end{figure*}

Dust grains not only change size by losing or gaining mass but also by sticking to one another (coagulation) or colliding with such force as to break apart into smaller grains (shattering). These two processes are unique from the previous dust processes mentioned because they conserve total grain mass but not total grain number. 
Given this, it is useful to discuss these processes in terms of the volume-averaged mass density of grains in each grain bin 
\begin{equation}
    \frac{\partial \bar{\rho} }{\partial a} = \left( \frac{m(a)}{V_{\rm cell}}\right) \dnda
\end{equation}
where $V_{\rm cell}\equiv m_{\rm cell}/\bar{\rho}_{\rm cell}$ is the volume of the gas cell the grains reside, $\bar{\rho}_{\rm cell}$ and $m_{\rm cell}$ are the volume-averaged mass density and total mass of the cell respectively, and $m(a)$ is the mass of a grain of size $a$. For our discussion of coagulation and shattering below, we will use this formalism instead of $\dndaflat$. 
Over a timestep $\dt$, these processes will change the mass of dust grains within a given bin $i$ by $dM_i/dt \times \Delta t$. Given this, we follow the procedure laid out in Appendix~\ref{app:update_mass_conserving} to update the number and slope of each grain size bin.

\subsubsection{Grain Shattering}  \label{sec:shattering}
Turbulence in the ISM accelerates dust grains depending on their size, which can lead to large relative velocities between dust grains of different sizes.
When two dust grains collide, if their relative velocity is above a critical threshold, the grains will shatter and fragment into smaller grains. 
Following \citet{hirashita_2009:ShatteringCoagulationDust}, \citet{asano_2013:WhatDeterminesGrain}, and \citet{hirashita_2023:SubmillimetreGalaxiesLaboratories}, shattering changes the mass density of grains of size $a$ such that
\begin{equation} \label{eq:shat_analytic}
\begin{aligned}
    \frac{1}{C_{2}}\frac{\partial}{\partial t} \left[ \frac{\partial \bar{\rho} }{\partial a}  \right] &= -m(a) \frac{\partial \bar{\rho} }{\partial a} \int_{\amin}^{\amax} \alpha(a,a_1) \frac{\partial \bar{\rho}}{\partial a_1} da_1 \\
    &+ \int_{\amin}^{\amax} \int_{\amin}^{\amax}  \Bigg[ \alpha(a_1,a_2) \frac{\partial \bar{\rho}}{\partial a_1} \frac{\partial \bar{\rho}}{\partial a_2} \\
    &\times m_{\rm shat} (a,a_1,a_2) \Bigg]  da_1 da_2
\end{aligned}
\end{equation}
where the first term on the right-hand side is the removal rate of grains of size $a$ due to shattering collisions with grains of size $a_1$ and the second term is the injection rate of grain fragments of size $a$ due to the shattering collisions between grains of size $a_1$ and $a_2$.
The variables are as follows: $C_2$ is the dust-dust clumping factor which accounts for sub-resolution clumping of dust as described in Appendix~\ref{app:gas_clumping}.
\begin{equation}
\alpha(a_1,a_2) = \frac{\pi (a_1+a_2)^2 v_{\rm rel}(a_1,a_2)}{m(a_1)m(a_2)} \mathbbm{1}_{v_{\rm rel} > \vshat}(a_1,a_2), \\
\end{equation}
where
\begin{equation}
     \mathbbm{1}_{v_{\rm rel} > \vshat}(a_1,a_2) =
    \begin{cases}
     1, \; & {\rm if } \, v_{\rm rel}(a_1,a_2)>\vshat \\
     0, \; & {\rm else} \\
    \end{cases}    
\end{equation}
is a term arising from the mean free time between collisions of dust grains of size $a_1$ and $a_2$. 
This depends on the effective cross-sections of each grain and their relative velocity $v_{\rm rel}(a_1,a_2)$ and only considers collisions with relative velocities above the shattering threshold $\vshat$. 
$m_{\rm shat}(a,a_1,a_2)$ is the mass of grains of size $a$ produced by the shattering of grains of size $a_1$ due to collisions with grains of size $a_2$\footnote{Note that this $m_{\rm shat}$ definition accounts for the mass produced from each of the colliding grains separately (i.e. from either $a_1$ or $a_2$). Therefore, we do not include an extra factor of 1/2 compared to other $m_{\rm shat}$ prescriptions \citep[e.g.][]{mckinnon_2018:SimulatingGalacticDust}.}. 
Using simplifying assumptions, Eq.~\ref{eq:shat_analytic} can be discretized into the form $d M_i / dt$, as outlined in Appendix~\ref{app:discretized_shat}, allowing us to track the movement of grain mass between grain size bins.

For $\vshat$, we take the values for silicates, graphite, and iron presented in \citet{jones_1996:GrainShatteringShocks} and shown in Table~\ref{tab:dust_quantities}, with carbonaceous dust shattering at lower velocities than silicates or metallic iron.
To determine $v_{\rm rel}$, we must first determine the velocity of grains at a given size $v(a)$. 
Following \citet{hirashita_2019:RemodellingEvolutionGrain,hirashita_2023:SubmillimetreGalaxiesLaboratories, li_2021:OriginDustExtinction}, we determine the grain velocities by assuming grains are coupled to local gas turbulence via drag forces. 
Assuming a Kolmogorov power spectrum of turbulence, the functional form is 
\begin{equation}
\begin{aligned}
      v_{\rm gr}(a) &= 0.32\ \left(\frac{\mathcal{M}}{3}\right) \left( \frac{a}{1\,\micron} \right)^{1/2} \left( \frac{\Teff}{100\,{\rm K}} \right)^{1/4} \left( \frac{\nH^{\rm rms}}{10^3\,\cmcubed} \right)^{-1/4} \\
      &\times \left( \frac{\rho_{\rm gr}}{3.5\,{\rm g \, \cmcubed}} \right)^{1/2} \; {\rm km \, s^{-1}}, 
\end{aligned}
\end{equation}
where $\mathcal{M}$ is the local Mach number and $\nH^{\rm rms}=\sqrt{C_2} \bar{n}_{\rm H}$ and $\Teff=\bar{T} / \sqrt{C_2} $ are the root-mean-squared density and effective temperature of the gas cell as described in Appendix~\ref{app:gas_clumping}.

Given $v_{\rm gr}(a)$, we then determine the relative velocity between colliding grains following \citet{hirashita_2009:ShatteringCoagulationDust} with
\begin{equation}
    v_{\rm rel}(a_1,a_2)=\sqrt{v(a_1)^2+v(a_2)^2-2v(a_2)v(a_1)\cos\theta},
\end{equation}
where $\cos \theta$ is the impact angle between grains which is randomly sampled from the interval $[-1,1]$ at each calculation of $\alpha(a_1,a_2)$ (i.e. each timestep) to account for stochastic variations in relative velocities.

Determining $m_{\rm shat}(a,a_1,a_2)$ depends on the total mass ejected from grain $a_1$ ($m_{\rm ej}$) and the size distribution of the ejected fragments ($\partial n_{\rm frag} / \partial a$) such that 
\begin{equation}
\begin{aligned}
    m_{\rm shat}(a,a_1,a_2) &= m(a) \frac{\partial n_{\rm frag}}{\partial a} (a, a_1, a_2) \\
    &+ (m(a_1)-m_{\rm ej}) \delta (a-a_{\rm rem}),
\end{aligned}
\end{equation}
where the first term is the mass of the fragments and the second term is the mass of the remnant, $\delta (a-a_{\rm rem})$ is a delta function, and $a_{\rm rem}=(a_1^3-m_{\rm ej}/(4/3\pi \rho_{\rm gr}))^{1/3}$ is the size of the remnant. $m_{\rm ej}$ has a complex dependence on numerous dust material properties, but \citet{hirashita_2013:EvolutionDustGrain} argue that the most important parameter is the velocity threshold for catastrophic fragmentation (when more than half of grain is fragmented). We, therefore, follow the prescription from \citet{kobayashi_2010:FragmentationModelDependence}, \citet{hirashita_2013:EvolutionDustGrain}, and \citet{hirashita_2019:RemodellingEvolutionGrain} with
\begin{equation}
    m_{\rm ej} = \frac{\phi}{1+\phi} m(a_1),
\end{equation}
where 
\begin{equation}
    \phi = \frac{E_{\rm imp}}{m(a_1) Q_D^*},
\end{equation}
\begin{equation}
    E_{\rm imp} = \frac{1}{2} \frac{m(a_1)m(a_2)}{m(a_1)+m(a_2)} v_{\rm rel} (a_1,a_2)^2,
\end{equation}
and $Q_D^*$ is the specific impact energy that causes catastrophic disruption of the grain. This formalism satisfies the two extreme cases for weak collisions ($\phi\ll1 ; \, m_{\rm ej}\sim E_{\rm imp}/Q_D^*$) and strong collisions ($\phi\gg1 ; \, m_{\rm ej}\sim m(a_1)$).
$Q_D^*$ is estimated as $Q_D^*=P_1/(2 \rho_{\rm gr})$, where $P_1$ is the critical shock pressure for each dust species taken from \citet{jones_1996:GrainShatteringShocks} and listed in Table~\ref{tab:dust_quantities}. This ultimately results in metallic iron grains ejecting more mass than carbonaceous or silicate dust during shattering.
For the size distribution of the fragments, we adopt a power-law
\begin{equation} \label{eq:frag_dnda}
   \frac{\partial n_{\rm frag}}{\partial a} = C_{\rm frag} a^{-\alpha_{\rm frag}},
\end{equation}
where $\alpha_{\rm frag}=3.3$ \citep{hellyer_1970:FragmentationAsteroids,jones_1996:GrainShatteringShocks}\footnote{The exact value of $\alpha_{\rm shat}$ is not important as long as $\alpha_{\rm shat}<4$ \citep{hirashita_2013:EvolutionDustGrain}.},
$C_{\rm frag}$ is a normalization constant such that $\int_{a_{\rm frag,min}}^{a_{\rm frag,max}} m(a) \partial n_{\rm frag}/\partial a \, da = m_{\rm ej}$, and
$a_{\rm frag,max}=(0.02 m_{\rm ej}/m(a_1))^{1/3} a_1$ and $a_{\rm frag,min}=0.01 a_{\rm frag,max}$ are the maximum and minimum sizes of the fragments \citep{hirashita_2019:RemodellingEvolutionGrain}.

\subsubsection{Grain Coagulation} \label{sec:coagulation}

While grains with high relative velocities can collide and shatter, grains with low relative velocities, below the coagulation velocity threshold $v_{\rm coag}$, can collide and stick together, forming one larger aggregate.  
$v_{\rm coag}$ is determined by the relative sizes and physical properties of the colliding grains and presence of ice mantels, with theoretical and experimental results predicting a range of $v_{\rm coag}$ on order of ${\sim}1-80$ m/s \citep{chokshi_1993:DustCoagulation,dominik_1997:PhysicsDustCoagulation,blum_2000:LaboratoryExperimentsPreplanetary,yan_2004:DustDynamicsCompressible,wada_2009:CollisionalGrowthConditions,wada_2013:GrowthEfficiencyDust}. 
Furthermore, voids are created in coagulated grains leading to grains with varying porosity, which can nominally affect their evolution  observable quantities \citep{hirashita_2021:EvolutionDustPorosity,hirashita_2022:EvolutionDustGrain}.
For our purposes, we neglect the effects of grain porosity, assuming aggregate grains are solid spheres, and assume coagulation occurs before ice mantels can form.

Luckily, the analytical formalism for coagulation closely follows that of shattering so we can reuse Eq.~\ref{eq:shat_analytic} with some minor changes. 
In particular, we replace $m_{\rm shat}(a, a_1, a_2)$ with 
\begin{equation}
    m_{\rm coag}(a, a_1, a_2) = \frac{m(a_1)+m(a_2)}{2}\; \delta \left( a-\left( \frac{m(a_1)+m(a_2)}{4 \pi \rho_{\rm gr}/3} \right)^{1/3} \right)
\end{equation}
which represents the mass of coagulated grains of size $a$ due to coagulating collisions between grains of size $a_1$ and $a_2$. Note the $1/2$ factor arises from the form of Eq.~\ref{eq:shat_analytic}, which counts the collision twice.
We also replace $\mathbbm{1}_{v_{\rm rel} > \vshat}(a_1,a_2) $ with
\begin{equation}
    \mathbbm{1}_{v_{\rm rel} < v_{\rm coag}}(a_1,a_2)  = 
\begin{cases}
    1, \; & {\rm if} \; v_{\rm rel}(a_1,a_2) < v_{\rm coag}(a_1,a_2), \\
    0, \; & {\rm else,} 
\end{cases}
\end{equation}
where $v_{\rm coag}(a_1,a_2)$ is the velocity threshold for coagulation for grains of size $a_1$ and $a_2$. This limits coagulation to only collisions between grains with relative velocities below the coagulation threshold. 

We determine $v_{\rm coag}(a_1,a_2)$ following the prescription from \citet{chokshi_1993:DustCoagulation}, \citet{dominik_1997:PhysicsDustCoagulation}, and \citet{yan_2004:DustDynamicsCompressible} with
\begin{equation} \label{eq:vcoag}
     v_{\rm coag}(a_1,a_2) = 2.14 F_{\rm s} \left[ \frac{a_1^3 + a_2^3}{(a_1 + a_2)^3} \right]^{1/2} \frac{\gamma^{5/6}}{(\mathcal{E}^*)^{1/3}R_{\rm 1,2}^{5/6} \rho_{\rm gr}^{1/2}} \,{\rm cm/s}
\end{equation}
where a $F_{\rm s}=10$ is a fudge factor from \citet{yan_2004:DustDynamicsCompressible} to match experimental work, $\gamma$ is the grain surface energy per unit area, $R_{\rm 1,2}=a_1 a_2 /(a_1+a_2)$ is the reduced radius of the colliding grains, and $\mathcal{E}^*$ is the reduced elastic modulus related to each grain's Poisson's ratios ($\nu_1, \nu_2$) and Young's modulus ($E_1, E_2$) by $(\mathcal{E}^*)^{-1}=(1-\nu_1)^2/E_1+(1-\nu_2)^2/E_2$.
The values for $\gamma$, $\nu$, and $E$ for each dust species are taken from \citet{chokshi_1993:DustCoagulation} and provided in Table~\ref{tab:dust_quantities}. 
We show the relation between $v_{\rm coag}$ and grain sizes for each dust species in Fig.~\ref{fig:coag_velocity}. 
For metallic iron $v_{\rm coag}$ can be greater than $v_{\rm shat}$, in this case we limit $v_{\rm coag}=v_{\rm shat}$.

We highlight that the coagulation process is efficient in molecular cloud environments with $\nH\sim10^4 \,\cmcubed$, which our simulations do not reach even when accounting for sub-resolved clumping. 
Furthermore, resonant-drag instabilities in molecular clouds will further enhance the clumping of large grains, increasing the coagulation rate \citep[e.g.][]{hopkins_2022:DustWindResonant}.
Initial testing of our model using only the aforementioned gas clumping resulted in only grains with $a\lesssim 0.01\;\micron$ being below the coagulation threshold in our simulations.
To overcome this limitation, we include a coagulation enhancement scheme as follows.
For gas with $\Teff \leq T_{\rm cutoff}$ we set $\mathcal{M}=1$ when calculating $v_{\rm grain}(a)$ and 
include a density enhancement factor, $C_{\rm coag}$, only used for the coagulation process. 
For $C_{\rm coag}$, we utilize a density-dependence similar to 
\citet{trayford_2026:ModellingEvolutionInfluence} such that
\begin{equation} \label{eq:coag_enhancement}
    C_{\rm coag} = 
\begin{cases}
      1 & \nH \leq n_{\rm H,min} \\
      (\nH/n_{\rm H,min})^\beta & n_{\rm H,min} < \nH \leq n_{\rm H,max} \\
      C_{\rm caog}^{\rm max} & \nH > n_{\rm H,max}\\
\end{cases} 
\end{equation}
where $n_{\rm H,min}=0.1\, \cmcubed$, $n_{\rm H,max}=100\, \cmcubed$, $C_{\rm coag}^{\rm max}$ is the maximum enhancement factor, and $\beta=\log(C_{\rm coag}^{\rm max})/ \log(n_{\rm H,max}/n_{\rm H,min})=2/3$.
In Sec~\ref{sec:model_variations}, we explore tuning $C_{\rm coag}^{\rm max}$ to set the mass fraction of large grains.

\subsection{Dust Destruction in SNR} \label{sec:SNe_shocks}

As SNe remnants (SNR) propagate through the ISM, they shock heat gas, shattering dust grains and destroying them via thermal and non-thermal sputtering.  
The amount of gas shocked to high enough velocities to destroy dust (typically ${\gtrsim} 100$ km/s) is determined by the local density, metallicity, and gas inhomogeneity due to turbulence \citep{cioffi_1988:DynamicsRadiativeSupernova, mckee_1989:DustDestructionInterstellar,kirchschlager_2022:SupernovaInducedProcessing,scheffler_2025:DustDestructionSupernova}.
Meanwhile, the total amount/fraction of dust destroyed and the resulting grain size distribution depends on the initial amount and size distribution of dust and dust-gas coupling \citep{kirchschlager_2022:SupernovaInducedProcessing, kirchschlager_2024:SupernovaDustDestruction}.
The resolution needed to resolve the aforementioned processes is well beyond the resolution of typical FIRE simulations, and we use a sub-resolution prescription. 
Below, we describe our prescription for determining the amount of gas shocked by an SNe and how the dust size distribution in the shocked gas is processed.



\begin{figure*}
    \plotsidesize{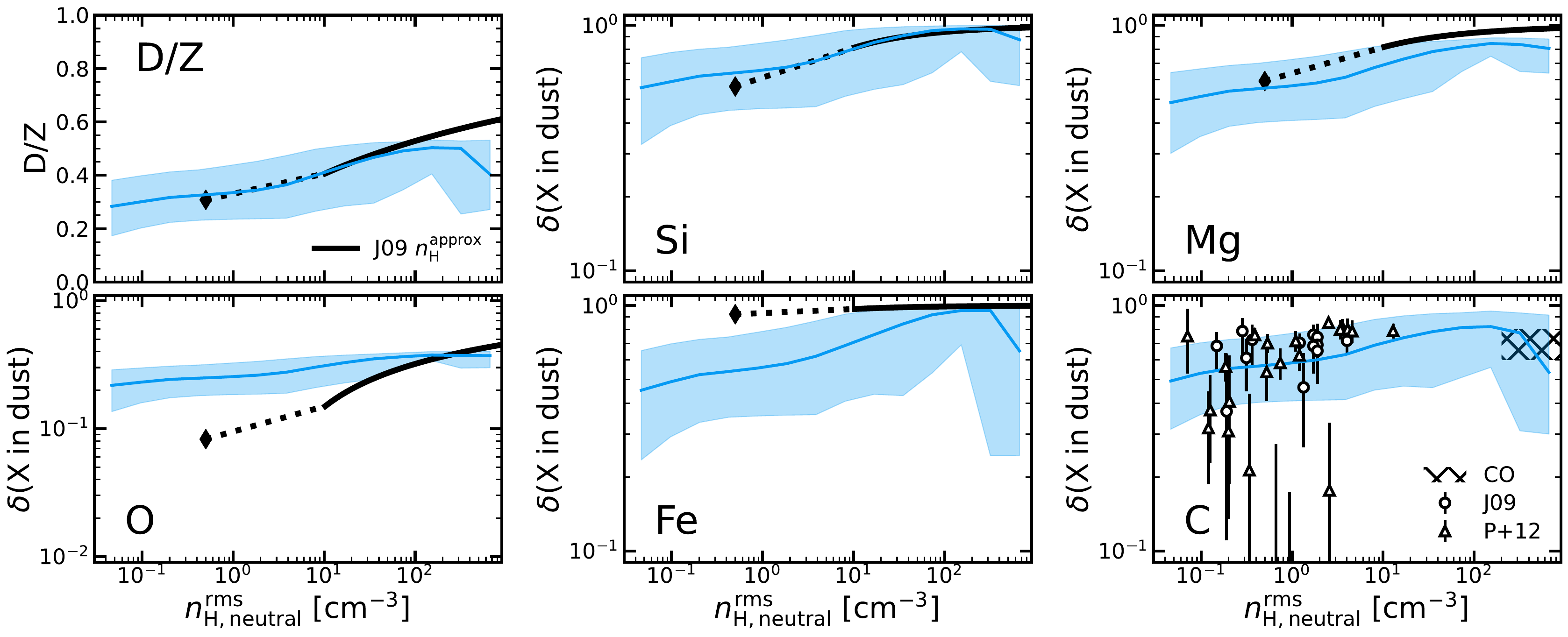}{0.95}
    \caption{Resulting median and 16/84th percentiles of D/Z and element depletion into dust trends with neutral gas density for our {\bf Fiducial} model. 
    We include fits to sight line depletions from \citet{jenkins_2009:UnifiedRepresentationGasPhase} ({\it thick black}), estimates for the WNM depletions ({\it diamond}), and an interpolation to the Jenkins' relation ({\it black-dotted}). These trends are aggregated for D/Z. Due to the scarcity of C depletion observations, we show individual sight lines from \citet{jenkins_2009:UnifiedRepresentationGasPhase} and \citet{parvathi_2012:ProbingRoleCarbon} as an upper limit and a range of expected C depletions in dense gas from observations of C in CO \citep[e.g.][]{irvine_1987:ChemicalAbundancesMolecular, vandishoeck_1993:ChemicalEvolutionProtostellar, vandishoeck_1998:ChemicalEvolutionStarForming,lacy_1994:DetectionAbsorptionH2}.
    Overall the {\bf Fiducial} model reproduces observed element depletions and aggregated D/Z trends, but deviates from observed O and Fe depletions due to only considering silicates as O-bearing dust and possibly underpredicting Fe-bearing dust accretion rates respectively.}
    \label{fig:depletion_relation}
\end{figure*}


To determine the mass of gas shocked to ${\sim}100$ km/s by a single SNe for a neighboring gas $i$ cell ($M_{{\rm shock},i}$) we utilize results from \citet{yamasawa_2011:RoleDustEarly}\footnote{We note models vary in their adopted shocked mass and local gas property scaling relations with typical values of $M_{\rm shock}\sim1500-6000\Msol$ \citep[e.g.][]{mckee_1989:DustDestructionInterstellar,hu_2019:ThermalNonthermalDust}.} with
\begin{equation}
    M_{{\rm shoc}k,i} =  1535 \, \tilde{\omega}_i \, E_{51} \left( \frac{n_{{\rm gas},i}}{1 \, \cmcubed} \right)^{-0.202} \left( \frac{Z_i}{\Zsol} + 0.039 \right)^{-0.298} \Msol,
\end{equation}
where $\tilde{\omega}_i$ is the gas cell weight determined by the \citet{hopkins_2018:HowModelSupernovae} mechanical SNe feedback routine, $E_{\rm 51}$ is the energy released in a typical supernova in units of $10^{51}$ erg, and $n_{{\rm gas},i}$ and $Z_i$ are the gas cell number density and metallicity respectively.
Note that \citet{yamasawa_2011:RoleDustEarly} only considers $E_{\rm 51}=1$, so we have added a linear scaling with $E_{51}$ \citep[e.g.][]{mckee_1989:DustDestructionInterstellar}. 
We note that this assumes the dust is thoroughly mixed in each gas cell so that equal proportions of each dust species are shocked. 

To determine how the grain size distribution is processed in the shocked gas, we must consider the grain processes at play. 
Simulations from \citet{kirchschlager_2022:SupernovaInducedProcessing, kirchschlager_2024:SupernovaDustDestruction} find that grain shattering and subsequent destruction by sputtering are synergistic, affecting the amount of dust destroyed and the resulting grain size distribution compared to sputtering alone.
However, the most widely used prescriptions for SNe dust processing only account for dust destruction via sputtering \citep[e.g.][]{nozawa_2007:EvolutionDustPrimordial,hirashita_2019:RemodellingEvolutionGrain}.
Therefore, we introduce a new analytical prescription that accounts for shattering and sputtering, outlined in Appendix~\ref{app:SNe_dust_processing}.
Assuming an initial MRN size distribution, this prescription is tuned to match an expected mass fraction of dust destroyed of $\epsilon_i\sim0.4$ and produces a final grain size distribution of similar qualitative shape to the results from \citet{kirchschlager_2024:SupernovaDustDestruction}, such that there is a substantial reduction in the number of large ($a\gtrsim0.3\micron$) grains and survival of some small grains due to shattering.
As we discuss in Sec.~\ref{sec:large_grains}, this shattering channel is necessary to reproduce the observed cutoff of large grains.

To avoid double counting of shattering and thermal sputtering due to our separate treatment of these processes and SNe destruction, we prevent thermal sputtering and shattering from occurring in gas which has been affected by an SNe event in the past 0.3 Myr, which is the typical time it takes for all dust destruction process to cease after a single SNe event \citep{hu_2019:ThermalNonthermalDust}.

\subsection{Dust Destruction by Star Formation} \label{sec:astration}

As gas cools and collapses to form stars, dust residing in said gas is destroyed and contributes to the stellar metallicity. 
This process is called astration, and is intrinsically tracked in our simulations. 
As star particles form from gas cells or fractions thereof, the corresponding (cell-averaged) mass of dust is removed (added to the stellar metallicity), and the shape of the grain size distribution is assumed to remain unchanged.
However, we stress that little work has been done investigating the astration process. Astration may be less efficient for more massive stars \citep{soliman_2024:DustEvacuatedZonesMassive} and it has been theorized that it is inefficient at high-z \citep[e.g.][]{mattsson_2021:MinimalAstrationHypothesisa}.








\section{Results} \label{Results}

We first present the results of our dust model as previously described in Sec.~\ref{GSE}, which we call {\bf Fiducial}, and compare its successes and shortcomings at reproducing observed MW depletion trends and extinction curves in Sec.~\ref{sec:fiducial_model}. 
We then investigate the effects of model variations on these results in Sec.~\ref{sec:model_variations}.
We then settle on a preferred model which matches both MW depletion trends and extinction curves, present the results of this model for Local Group analogs, and compare to MW, LMC, and SMC observations in Sec.~\ref{sec:local_group_comparison}.
For all calculations below, we only consider gas within the galactic disk for each galaxy ($r< 2R_{\rm d,gas}$; $|z|<2z_{\rm d}$). 
We also use $\nHrms$ and $\Teff$ when showcasing dust trends, but the results are similar when using the volume averaged $\nH$ and $T$ tracked in our simulations.

\subsection{Fiducial Model: Successes and Shortcomings} \label{sec:fiducial_model}

We first showcase a mock SDSS {\it ugr} composite image along with stellar and dust surface density projections for our {\bf Fiducial} MW-mass simulation in Fig.~\ref{fig:mock_image} to highlight the dusty spiral arms with embedded star formation. 
These images were created using the radiative transfer code SKIRT \citep{camps_2015:SKIRTAdvancedDust} with BPASS \citep{eldridge_2017:BinaryPopulationSpectral} to compute the stellar spectra for each star particle given their age and metallicity. We use the tracked dust mass for each gas cell produced by our dust model and assume a MW dust composition and size distribution to calculate attenuation \citep{weingartner_2001:DustGrainSizeDistributions}.

\begin{figure*}
    \plotsidesize{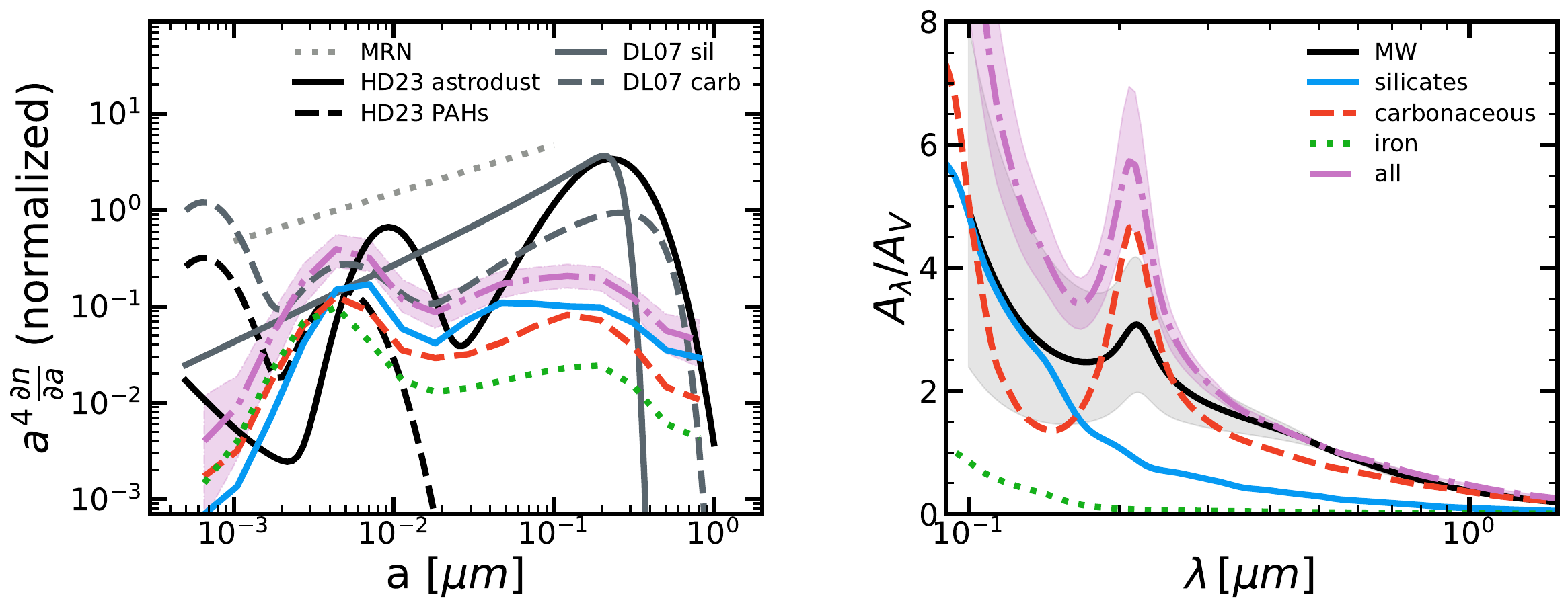}{0.95}
    \caption{Resulting dust mass-weighted median and 16/84th percentile grain size distributions and extinction curves for our {\bf Fiducial} model. {\it Left:} Grain size distribution for ({\it blue}) silicates, ({\it red}) carbonaceous, ({\it green}) metallic iron, and ({\it lavender}) all species normalized by total dust mass. MW-derived size distributions are also shown, including ({\it thick light grey}) \citet{mathis_1977:SizeDistributionInterstellar} power-law distribution and parametric two species distributions from \citet{draine_2007:InfraredEmissionInterstellar} ({\it thick dark grey}) and \citet{hensley_2023:Astrodust+PAHModelUnified} ({\it thick black}).
    The {\bf Fiducial} model produces a bimodal distribution, with a peak at $a\sim5 \; \nanom$ and $a\sim0.1 \; \micron$ similar to \citet{draine_2007:InfraredEmissionInterstellar} and \citet{hensley_2023:Astrodust+PAHModelUnified}, but unpredicts the abundance of large grains.
    The model also does not produce the peak of small carbonaceous/PAH grains at ${\sim}0.7$ nm, although we do not explicitly track the formation/evolution of PAHs. The variation in size distributions between dust species is relatively small, with the small/large grain peak being slightly larger/smaller for silicates compared to the other dust species.
    {\it Right:} Extinction curves normalized by $\AV$ for each dust species treating each gas cell as a sight line. ({\it thick black}) Observed range of extinction curves in the MW from \citet{fitzpatrick_2007:AnalysisShapesInterstellar} and compiled by \citet{salim_2020:DustAttenuationLaw} are also included.
    The {\bf Fiducial} model predicts a steeper extinction curve and stronger $2175\;\angstrom$ bump than observed in the MW due to the underpredicted mass fraction of large silicate and carbonaceous grains. Tuning the coagulation density enhancement parameter, as shown in Sec.~\ref{sec:model_variations}, fixes this discrepency.}
    \label{fig:median_size_AV} 
\end{figure*}

\subsubsection{Dust Abundance and Composition}

The detailed observations of dust abundances and its elemental constituents within the MW provide the first set of constraints for our model.
Notably, gas-phase element abundances in the Local Group, derived from spectral UV absorption along sight lines to O/B-type stars, show that the fraction of refractory elements locked in dust (i.e. not in the gas-phase) increases with neutral hydrogen column density \citep{jenkins_2009:UnifiedRepresentationGasPhase,jenkins_2017:InterstellarGasphaseElement,roman-duval_2022:METALMetalEvolutiona}.
Spatially resolved dust-to-metals and dust-to-gas ratios derived from IR dust emission show a similar increase with local gas surface density \citep{chiang_2021:ResolvingDustMetalsRatio,clark_2023:QuestMissingDust}.

Looking at the building blocks of dust, in Fig.~\ref{fig:depletion_relation} we show the resulting relation between the fraction of major refractory elements (C, O, Si, Mg, and Fe) along with the total fraction of metals (D/Z) locked into dust, and the neutral gas density $\nHn^{\rm rms}$. 
We specifically show $\delta({\rm X \; in \; dust}) = 1-n_{\rm X,gas}/n_{\rm X,total}$, where $n_{\rm X,gas}$ and $n_{\rm X,total}$ are the gas-phase number density and total (gas+dust) number density of element X.
We also include fits to the sight line depletion trends observed in the MW from \citet{jenkins_2009:UnifiedRepresentationGasPhase} using \citet{zhukovska_2016:ModelingDustEvolution, zhukovska_2018:IronSilicateDust} mean sight line density to physical density fit (see \citet{choban_2022:GalacticDustmodellingDust} Sec.\ 3.2.2 for details). 
Due to the limited observations of C depletion, we show the individual observations from \citet{jenkins_2009:UnifiedRepresentationGasPhase} and \citet{parvathi_2012:ProbingRoleCarbon} as an upper limit and a range of expected maximum C depletions in dense environments based on observations of 20\%\ to 40\%\ of C in CO in the Milky Way \citep[e.g.][]{irvine_1987:ChemicalAbundancesMolecular, vandishoeck_1993:ChemicalEvolutionProtostellar, vandishoeck_1998:ChemicalEvolutionStarForming,lacy_1994:DetectionAbsorptionH2}. 
For D/Z, we include an observed relation aggregated from \citet{jenkins_2009:UnifiedRepresentationGasPhase} depletion trends.

Overall, the {\bf Fiducial} model reproduces the observed depletion trends of Si, Mg, and C, but not O or Fe.
For O, our model overpredicts its depletion in all but the most dense gas.
This is due to our model assuming silicates are the only species to sequester oxygen which cannot explain observed O depletion trends, and an additional unknown O-bearing dust species is needed \citep[e.g.][]{whittet_2010:OxygenDepletionInterstellar}.
For Fe, our model underpredicts its depletion in diffuse gas.
This suggests that the metallic iron species is particularly robust to destruction processes, as posited by \citet{zhukovska_2018:IronSilicateDust}, or it has higher accretion rates due to some unaccounted for physical process.
The resulting D/Z trend matches well with observations, but does not reach the expected $\DZ=0.6$ in the densest gas, due to the max D/Z being set by our choice of dust chemical compositions.
Similar to \citetalias{choban_2024:DustyLocaleEvolution}, the variations between C, Si, and Fe depletions, which are the key elements of each dust species, is due to variable accretion rates for each species due to elemental abundances, Coulomb enhancement, and the sequestration of gas-phase C into CO in dense gas which we show in Sec.~\ref{sec:acc_dest_variations}.

The broad agreement of the {\bf Fiducial} model with dust abundance trends is reassuring but not surprising. Notably, previous versions of this model which did not include grain size evolution produced similar agreement with MW depletion observations \citepalias{choban_2022:GalacticDustmodellingDust,choban_2024:DustyLocaleEvolution}.
As we highlight in Sec.~\ref{sec:model_variations}, the accretion and SNe dust destruction processes are the main determinator of dust abundances as long as a steady supply of small grains is produced by large grain shattering.
In the case of models which do not track the evolution of grains, the typical assumption of an MRN size distribution means small grains always exist.

\begin{table*}
        \renewcommand{\arraystretch}{1.15}
	\centering
	\begin{tabular}{l|cccc|ccccc} 
		\hline
        & \multicolumn{4}{c|}{Dust Process Variations} & \\
		Sim. Name & Accretion & SNe Dest. & Shattering & Coagulation & D/Z & Si/C & STL & S & B \\
           
        \hline
        Fiducial & \textemdash & \textemdash & \textemdash & \textemdash & 0.34 & 2.1 & 1.1 & 3.8 & 0.51 \\
        \hline
             &  &  &  &  & \multicolumn{5}{c}{Logarithmic Difference from Fiducial} \\
            \hline
        ConstSpecAcc & $n_{\rm max} = \infty$ \& $D_{\rm k}=1$ & \textemdash & \textemdash & \textemdash & \cellcolor{red!14} -0.1 & \cellcolor{red!70} -0.2 & \cellcolor{red!15} -0.2 & \cellcolor{red!16} -0.1 & \cellcolor{green!0}  0 \\
        4xAcc & 4 x increase &  \textemdash & \textemdash & \textemdash & \cellcolor{green!8} 0.1 & \cellcolor{red!13} -0 & \cellcolor{green!3}  0 & \cellcolor{green!0}  0 & \cellcolor{green!2}  0 \\
        1/4xAcc & 4 x decrease & \textemdash & \textemdash & \textemdash & \cellcolor{red!26} -0.2 & \cellcolor{green!44} 0.1 & \cellcolor{red!12} -0.1 & \cellcolor{red!7} -0 & \cellcolor{red!23} -0.1 \\
        4xSNRDest & \textemdash & 4 x increase & \textemdash & \textemdash & \cellcolor{red!70} -0.5 & \cellcolor{green!7}  0 & \cellcolor{green!11} 0.1 & \cellcolor{green!12}  0 & \cellcolor{green!3}  0 \\
        1/4xSNRDest & \textemdash & 4 x decrease & \textemdash & \textemdash & \cellcolor{green!17} 0.1 & \cellcolor{red!4} -0 & \cellcolor{red!2} -0 & \cellcolor{green!0}  0 & \cellcolor{green!2}  0 \\
        Tcutoff1000K & $T_{\rm cut}=1000$K & \textemdash & \textemdash & \textemdash & \cellcolor{green!3}  0 & \cellcolor{red!7} -0 & \cellcolor{red!1} -0 & \cellcolor{red!4} -0 & \cellcolor{red!1} -0 \\
        Tcutoff100K & $T_{\rm cut}=100$K & \textemdash & \textemdash & \textemdash & \cellcolor{red!8} -0.1 & \cellcolor{green!10}  0 & \cellcolor{red!6} -0.1 & \cellcolor{red!2} -0 & \cellcolor{red!4} -0 \\
        NoISMShat  & \textemdash & \textemdash & only in SNR & \textemdash & \cellcolor{red!16} -0.1 & \cellcolor{green!31} 0.1 & \cellcolor{red!23} -0.2 & \cellcolor{red!15} -0.1 & \cellcolor{red!31} -0.1 \\
        10xISMShat  & \textemdash & \textemdash & 10 x increase & \textemdash & \cellcolor{green!1}  0 & \cellcolor{red!7} -0 & \cellcolor{green!25} 0.3 & \cellcolor{green!29} 0.1 & \cellcolor{green!15} 0.1 \\
        NoSNRShat  & \textemdash & only sputtering$^{\it a}$ & \textemdash & \textemdash & \cellcolor{red!6} -0 & \cellcolor{red!0} -0 & \cellcolor{green!12} 0.1 & \cellcolor{green!13} 0.1 & \cellcolor{green!13}  0 \\
        WeakSNRShat  & \textemdash & 4 x reduced shat. eff.$^{\it a}$ & \textemdash & \textemdash & \cellcolor{red!5} -0 & \cellcolor{green!3}  0 & \cellcolor{green!1}  0 & \cellcolor{red!0} -0 & \cellcolor{green!4}  0 \\
        LargerSNRFrag  & \textemdash & larger $a^{\rm SN}_{\rm frag}=0.6\;\micron^{\it a}$ & \textemdash & \textemdash & \cellcolor{red!7} -0 & \cellcolor{green!3}  0 & \cellcolor{red!2} -0 & \cellcolor{red!5} -0 & \cellcolor{green!2}  0 \\
        NoCoag  & \textemdash & \textemdash & \textemdash & none & \cellcolor{green!1}  0 & \cellcolor{red!5} -0 & \cellcolor{green!42} 0.4 & \cellcolor{green!5}  0 & \cellcolor{green!28} 0.1 \\
        10xCCoag & \textemdash & \textemdash & \textemdash & 10 x $C_{\rm coag}$ & \cellcolor{red!1} -0 & \cellcolor{green!7}  0 & \cellcolor{red!39} -0.4 & \cellcolor{red!33} -0.1 & \cellcolor{red!33} -0.1 \\
        20xCCoag & \textemdash & \textemdash & \textemdash & 20 x $C_{\rm coag}$ & \cellcolor{red!3} -0 & \cellcolor{green!11}  0 & \cellcolor{red!50} -0.5 & \cellcolor{red!46} -0.2 & \cellcolor{red!43} -0.2 \\
        100xCCoag & \textemdash & \textemdash & \textemdash & 100 x $C_{\rm coag}$  & \cellcolor{red!10} -0.1 & \cellcolor{green!21} 0.1 & \cellcolor{red!70} -0.7 & \cellcolor{red!70} -0.3 & \cellcolor{red!70} -0.2 \\
        10xVCoag  & \textemdash & \textemdash & \textemdash & 10 x $v_{\rm coag}$ & \cellcolor{red!4} -0 & \cellcolor{green!2}  0 & \cellcolor{red!6} -0.1 & \cellcolor{red!25} -0.1 & \cellcolor{red!1} -0 \\
        NoVCoag  & \textemdash & \textemdash & \textemdash & $v_{\rm coag} = v_{\rm shat}$ & \cellcolor{red!1} -0 & \cellcolor{green!2}  0 & \cellcolor{red!7} -0.1 & \cellcolor{red!21} -0.1 & \cellcolor{red!0} -0 \\
            \hline
	   \hline
	\end{tabular}
\caption{Dust model parameter variations for the {\bf  m12\_lowres} simulation.
    We show the median dust abundance, composition, sizes, and resulting extinction parameterizations weighted by dust mass for the {\bf Fiducial} model. For all other simulations we show the logarithmic differences of each value from {\bf Fiducial}. 
    Color gradients are used to highlight positive ({\it green}) or negative ({\it red}) differences with the intensity of the color scaled to the maximum for each column. 
    Variations in dust destruction by thermal sputtering are not considered since the idealized galaxy lacks a gaseous halo where this process would occur.
    \textbf{(1)} Name of simulation.
    \textbf{(2)} Accretion routine modification.
    \textbf{(3)} SNe destruction routine modification.
    \textbf{(4)} Shattering routine modification.
    \textbf{(5)} Coagulation routine modification.
    \textbf{(6)} Dust-to-metals ratio.
    \textbf{(7)} Silicate-to-carbonaceous dust mass ratio.
    \textbf{(8)} Small-to-large grain mass ratio ($a_{\rm small} \leq 0.02\;\micron$).
    \textbf{(9)} Extinction UV-optical slope.
    \textbf{(10)} Extinction UV bump strength.
    Generally, coagulation and shattering variations affect the STL ratio, extinction slope, and bump strength but have little effect on the total dust abundance/composition as long as some ISM shattering occurs and coagulation is not too strong such that it competes with the accretion process.
    Gas-dust accretion and SNe destruction rates set the median D/Z. Si/C is determined by the relative accretion rates of the different dust species.\\
    $^{\it a}$SNe sputtering efficiencies are modified such that the total dust destruction efficiency is unchanged.}
    \label{tab:sim_suite}
\end{table*}

\subsubsection{Grain Sizes and Extinction Curves}

The observed wavelength-dependent extinction and reddening of stellar spectra encodes information on the size and chemical composition of dust along the line of sight. 
In broad strokes, the steepness of the extinction curve provides information on the mass fraction of small silicate grains, while the 2175 $\angstrom$ bump strength provides information on the mass fraction of small carbonaceous grains.
Parametric dust populations, specifying dust species composition and size distributions, can be derived from averaged extinction curves along with emission spectra and wavelength-dependent polarization \citep{mathis_1977:SizeDistributionInterstellar,weingartner_2001:DustGrainSizeDistributions,draine_2007:InfraredEmissionInterstellar,hensley_2023:Astrodust+PAHModelUnified}.  
These observed extinction curves and derived size distributions provide the next set of constraints for our model.

To compute extinction curves in our simulation we approximate each gas cell as a sight line, calculating a total extinction curve from each dust species' grain size distribution as outlined in Appendix~\ref{app:calc_extinction}.
We caution that this is not an apples-to-apples comparison with observations since extinction sight lines extend up to ${\sim}1$ kpc \citep[e.g.][]{fitzpatrick_2007:AnalysisShapesInterstellar}, probing multiple ISM phases,
while each gas cell in our simulation represents one ISM phase due to their high resolution.\footnote{For reference, only a few $\%$ of gas cells have $\AV>0.1$, while extinction observations typically probe $\AV\gtrsim1$.}
Nevertheless, this comparison is useful for understanding how changes in the average size distribution affect typical extinction curves.
In Fig.~\ref{fig:median_size_AV}, we show the resulting normalized, grain mass distribution per log grain size $\left( \frac{\partial m}{\partial \ln a}  \propto a^4 \dnda \right)$ weighted by gas mass for each dust species and corresponding extinction curves normalized by $\AV$, the extinction in the V band ($\lambda=0.547\;\micron$).
We include an idealized \citet{mathis_1977:SizeDistributionInterstellar} power-law size distribution $ \left( \dnda \propto a^{-3.5} \right)$ and parametric grain size distributions derived for the MW from \citet{draine_2007:InfraredEmissionInterstellar}, which considers seperate silicate and carbonaceous+PAHs, and \citet{hensley_2023:Astrodust+PAHModelUnified}, which considers conglomerate astrodust and PAHs.

Focusing first on the successes of the {\bf Fiducial} model, it produces a bimodal (two-peak) mass distribution at $a\sim5 \; \nanom$ and $a\sim0.1 \; \micron$ similar in shape to carbonaceous dust in \citet{draine_2007:InfraredEmissionInterstellar} and astrodust in \citet{hensley_2023:Astrodust+PAHModelUnified}.
We stress that observationally derived size distributions are parametric fits using assumed functional forms (i.e. power-law, log-normal) and so a qualitatively similar shape is encouraging.
As we show in Sec.~\ref{sec:model_variations}, the $a\sim5 \; \nanom$ peak is due to the rapid growth of small grains via accretion, while
the $a\sim0.1 \; \micron$ peak is due to the coagulation of accretion-grown grains.
The sharp downturn in $a\gtrsim0.3\;\micron$ grains is due to the shattering of grains in SNR.
The size distributions are relatively similar between silicate, carbonaceous, and metallic iron, although the silicate small/large grain peaks are slightly larger/smaller compared to the other dust species.
The shift in small grains is due to a higher accretion rate which grows grains to larger sizes, while the shift in large grains is due to lower coagulation rates and coagulation velocity thresholds for silicates as seen in Fig.~\ref{fig:lifecycle_diagram}.

Focusing now on the shortcomings of the {\bf Fiducial} model, the mass fraction of small grains is overpredicted compared to observations.
This overabundance of small silicate and carbonaceous grains produces extinction curves which are steeper and have stronger bumps than those observed in the MW as can be seen in Fig.~\ref{fig:median_size_AV}.
As we show in Sec~\ref{sec:model_variations}, this is due to inefficient coagulation in our {\bf Fiducial} model. 
Tuning the coagulation density enhancement parameter ($C_{\rm coag}$) largely fixes this discrepancy while having only minor effects on dust abundances, highlighting the dependence of our predicted extinction curves on sub-resolved coagulation.
The {\bf Fiducial} model also does not produce a second peak of small carbonaceous grains at ${\sim}0.7$ nm as as seen in the MW.
These small grains are responsible for a majority of the mid-IR emission lines produce by aromatic carbonaceous dust (e.g. PAHs), with $a\sim 5$ nm aromatic grains being inefficient line emitters \citep[e.g.][]{hensley_2023:Astrodust+PAHModelUnified}.
We discuss our models implications for PAH evolution in Sec.~\ref{sec:very_small_grains}.

\subsection{Testing Model Variations} \label{sec:model_variations}

\begin{figure*}
    \plotsidesize{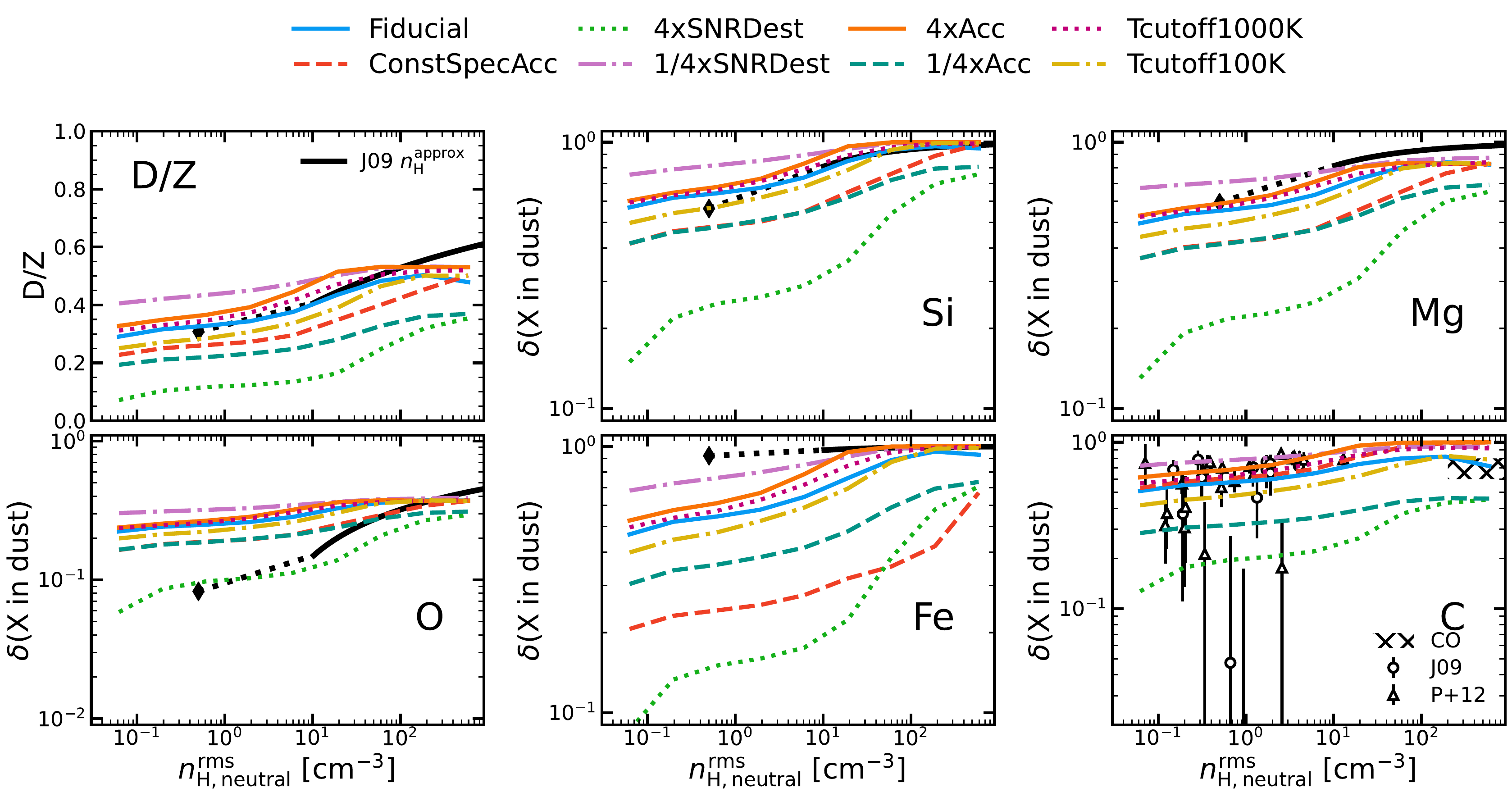}{0.95}
    \caption{Same as Fig.~\ref{fig:depletion_relation} for variations in accretion and SNe destruction rates. 
    The relative accretion and SNe destruction rates set the overall offset of the D/Z and depletion relations. Increased SNe destruction rates ({\bf 4xSNeDest}) or decreased accretion rates ({\bf 1/4xAcc}) reduce the D/Z and depletion trends overall and vis versa for increased accretion rates ({\bf 4xAcc}) or decreased SNe destruction rates ({\bf 1/4xSNeDest}).
    Assuming no species dependent Coulomb enhancement or accretion ``turn off'' ({\bf ConstSpecAcc}) leads to weaker depletions for all elements besides C. 
    This is due to carbonaceous dust having higher accretion rates than other species due to the high abundance of C compared to other elements.
    Varying the accretion cutoff temperature to 100 K and 1000 K ({\bf Tcutoff100K} and {\bf Tcutoff1000K}) slightly offsets the D/Z and depletion relations, highlighting that gas-dust accretion only needs to be appreciable for $T\lesssim 100\;K$ gas.
    }
    \label{fig:acc_dest_variations_DZ} 
\end{figure*}

\begin{figure*}
    \plotsidesize{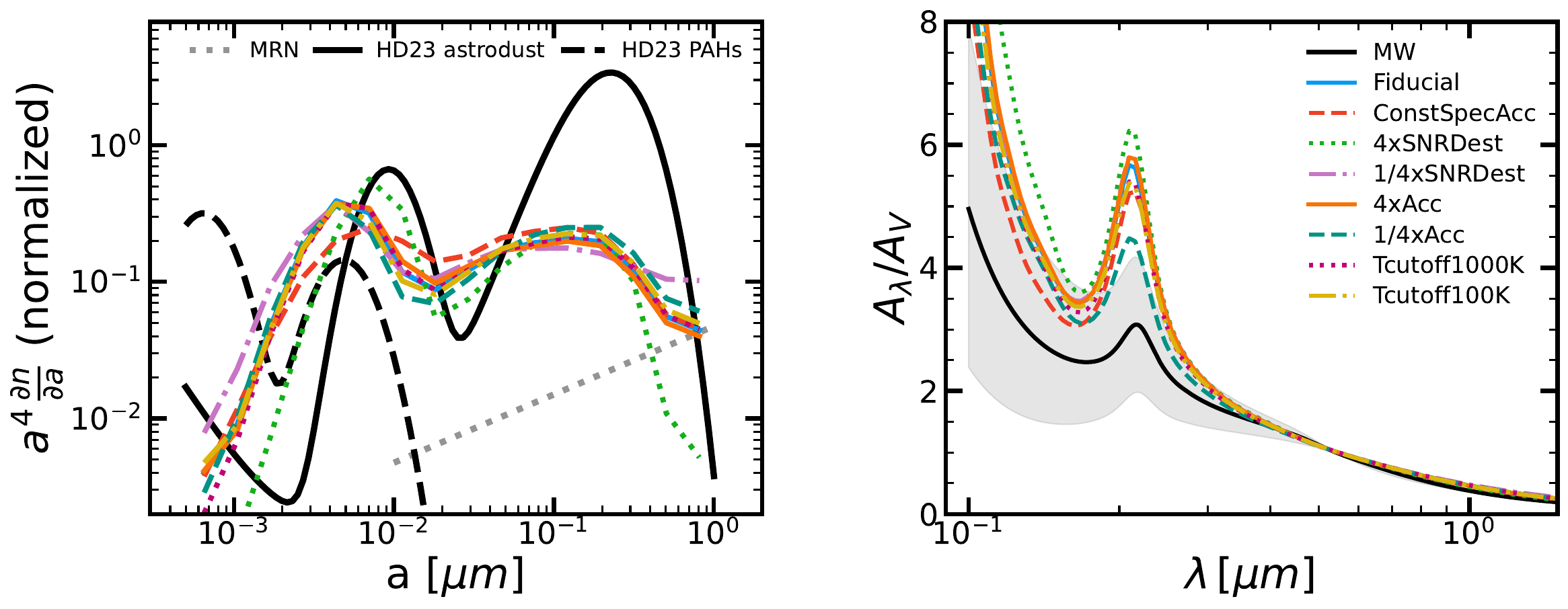}{0.95}
    \caption{Same as Fig.~\ref{fig:median_size_AV} for variations in accretion and SNe destruction rates. 
    Increasing/decreasing the SNe destruction rate ({\bf 4xSNeDest} and {\bf 1/4xSNeDest}) determines the abundance of extremely large ($a>0.3\;\micron$) grains due to SNe shattering. This, in turn, affects the abundance of small grains, with more/less SNe shattering producing more/less small grains.
    Increasing/decreasing the accretion rate ({\bf 4xAcc} and {\bf 1/4xAcc}) increases/decreases the mass in small grains. Only {\bf 4xSNeDest} and {\bf 1/4xAcc} noticeably steepen or flatten the extinction curve, respectively.
    Varying the accretion cutoff temperature to 100 K and 1000 K ({\bf Tcutoff100K} and {\bf Tcutoff1000K}) negligibly changes the size distribution and extinction curve.
    }
    \label{fig:acc_dest_variations_dmdloga} 
\end{figure*}

To understand how variations in each dust process affect the abundance, composition, and sizes of dust grains and the resulting extinction curve, we ran a suite of simulation reruns varying each dust process as detailed in Table~\ref{tab:sim_suite}.
For straightforward qualitative comparisons between each simulation, we provide the median D/Z ratio, silicate-to-carbonaceous dust mass ratio (Si/C), small-to-large grain mass ratio (STL) where small grains are $<0.02\;\micron$, and the UV-optical slope (S) and UV bump strength (B) of the median extinction curve as defined by Eq. 2 and 4 in \citet{salim_2020:DustAttenuationLaw}.

\subsubsection{Accretion and Destruction} \label{sec:acc_dest_variations}

Gas-dust accretion and SNe shocks are the primary source of dust growth and destruction in local galaxies \citep[e.g][]{mckee_1989:DustDestructionInterstellar} but various details of each process are uncertain, such as the accretion sticking efficiency and mass of gas shocked per SNe, and depend on the structure of the ISM predicted by our simulations. 
To determine the sensitivity of our results to variations in accretion and SNe destruction, we reran simulations individually increasing/decreasing by a factor of 4 the accretion rates ({\bf 4xAcc} and {\bf 1/4xAcc}) and the mass shocked by a single SNR ({\bf 4xSNRDest} and {\bf 1/4xSNRDest}), removing species-dependent accretion parameters such as dense gas accretion turn off and Coulomb enhancement ({\bf ConstSpecAcc}), and set the temperature cutoff for accretion sticking efficiency and coagulation density enhancement to $T_{\rm cutoff} =$ 100 K and 1000 K ({\bf Tcutoff100K} and {\bf Tcutoff1000K}).
The resulting D/Z and depletion trends, grain size distributions, and extinction curves compared to the {\bf Fiducial} model are shown in Fig.~\ref{fig:acc_dest_variations_DZ} and~\ref{fig:acc_dest_variations_dmdloga}.

Overall, the median and slope of the D/Z and depletion relations are set by the relative accretion and SNe destruction rates.
Weaker accretion and stronger SNe destruction decreases the median due to a lower steady-state dust abundance being reached between the two processes.
Stronger accretion or stronger destruction also increases the slope of the relation since more dust is grown or destroyed as gas cycles between cool and hot phases. 
The inverse of these effects are readily seen, producing increased medians and shallower slopes.
The relative offset in depletion trends between C, Si, and Fe are due to the differing accretion rates between dust species, since these are the key elements for each species.
When considering only the differing element abundances and atomic masses, carbonaceous accretion is higher than the other species, leading to C having the strongest depletions, since it is one of the most abundant and lightest refractory elements.
Meanwhile the accretion turn off due to CO or ice formation limits depletions in the densest gas. 
Without this turn off, C is overdepleted, leaving little gas-phase C to produce CO.
Accounting for Coulomb enhancement of silicate and metallic iron dust increases the depletions for their constituent elements (O, Si, Mg, and Fe).

The abundance of extremely large ($a>0.3\;\micron$) grains is set by the destruction rate due to SNR shattering.
This also moderately affects the mass fraction of small grains as more/less SNR shattering produces more/less small grains.
However, these results are degenerate with the shattering efficiency assumed in SNR as shown in Sec.~\ref{sec:shattering_variations}.
Changes to the accretion rate also moderately affect the mass fraction of small grains.
In regards to the resulting extinction curves, increased SNe destruction increases the slope while decreased accretion weakens the bump strength, but their D/Z and depletion trends fall well below observations.
Notably, all of our results are relatively insensitive to changes in $T_{\rm cutoff}$, highlighting that the sticking efficiency of gas-phase metals only needs to be appreciable for $T\lesssim100$ K gas. This is in line with detailed molecular simulations find non-negligible sticking efficiencies well above $T\sim100$ K \citep{bossion_2024:AccurateStickingCoefficient}.

\subsubsection{Shattering} \label{sec:shattering_variations}

\begin{figure}
    \centering
    \includegraphics[width=\linewidth]{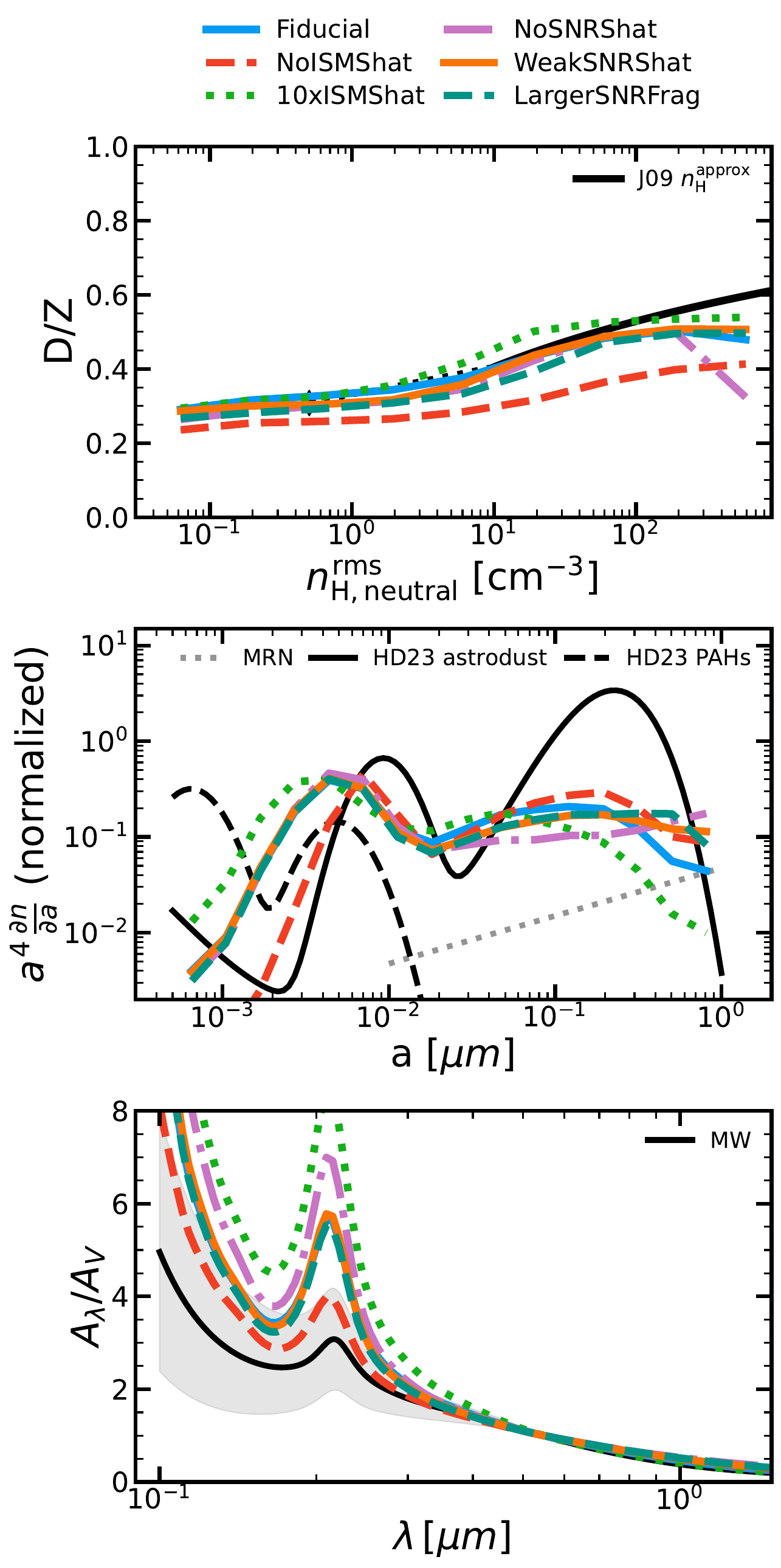}
    \caption{Resulting ({\it top}) D/Z relation, ({\it middle}) grain size distribution, and ({\it bottom}) extinction curves resulting from variations in ISM and SNR shattering compared to the {\bf Fiducial} model.
    Included observations are the same as  Fig.~\ref{fig:depletion_relation} and~\ref{fig:median_size_AV}.
    Turning off shattering in the ISM ({\bf NoISMShat}) reduces D/Z due to a decrease in the abundance of small grains which can efficiently grow via accretion.
    This results in more large grains and decreased extinction curve slope and bump strength. 
    Changes to grain shattering in SNR have lesser effects on D/Z, highlighting that the turbulent ISM is the main producer of small grains.
    Assuming no shattering in SNR ({\bf NoSNRShat}) reduces the abundance of large grains due to the increased sputtering needed to keep the same SNR destruction efficiency. This results in an extinction curve with a steeper slope and stronger bump. 
    However, grains larger than $a>0.2\;\micron$ survive, in contrast to observations, since they are not efficiently shattered in the ISM.
    The typical grain size and strength of the large grain cutoff are set by the maximum size of grains to survive shattering in SNR ({\bf LargerSNRFrag}) and SNR shattering efficiency ({\bf WeakSNRShat}) respectively, but the changes to the extinction curve are minimal.
    }
    \label{fig:shat_variations_DZ_GSD_AV}
\end{figure}

Shattering of dust grains is the primary means by which small grains are produced. However, most size evolution models only accounted for shattering in the WIM and not in SNRs as highlighted in Appendix~\ref{app:SNe_dust_processing}.
To determine the relative importance of shattering in the WIM and SNR and test free parameters in our novel SNR dust destruction routine, we reran simulations with no shattering in the ISM ({\bf NoISMShat}), 10x increased ISM shattering rates ({\bf 10xISMShat}), no shattering in SNR ({\bf NoSNRShat}), reduced shattering in SNR ({\bf WeakSNRShat}), and increased the size of grains to survive shattering in SNR ({\bf LargerSNRFrag})\footnote{Note that for all changes to shattering efficiency in the SNR destruction routine we also modify the sputtering efficiency so that the total destruction efficiency is unchanged.}.
The resulting D/Z trends, grain size distributions, and extinction curves compared to the {\bf Fiducial} model are shown in Fig.~\ref{fig:shat_variations_DZ_GSD_AV}.

Assuming no shattering in the ISM or SNR decreases D/Z by ${\sim}20\%$ and ${\sim}10\%$ respectively.
Without shattering, fewer small grains are produced which eventually grow via accretion, leading to an overall reduction in D/Z.
On the other hand, increased ISM shattering slightly increases D/Z due to more small grains being produced which can then grow.
As seen in the grain size distribution, ISM shattering is the dominant shattering source for $a\gtrsim0.08\;\micron$ grains, while SNR shattering sets the cut off in abundance of $a>0.3\;\micron$ grains\footnote{We note that the large grain peak is still apparent and offset for silicates and carbonaceous species similar to Fig.~\ref{fig:median_size_AV}, but when averaged together appears as a power-law.} seen in observations.
The maximum size of grain fragments to survive shattering in SNR and the efficiency of SNR shattering determine the typical grain size and strength of the cutoff respectively, and marginally affect the overall D/Z.
Assuming no ISM shattering also produces a lower small grain mass fraction, resulting in a shallower slope and weaker bump strength, and vice versa for increased ISM shattering.
Counterintuitively, assuming no SNR shattering results in a higher small grain mass fraction due to sputtering-only SNR destruction models destroying a large amount of $a\sim0.1 \, \micron$ grains as seen in Fig.~\ref{fig:SNe_dust}.

\subsubsection{Coagulation}

\begin{figure}
    \centering
    \includegraphics[width=\linewidth]{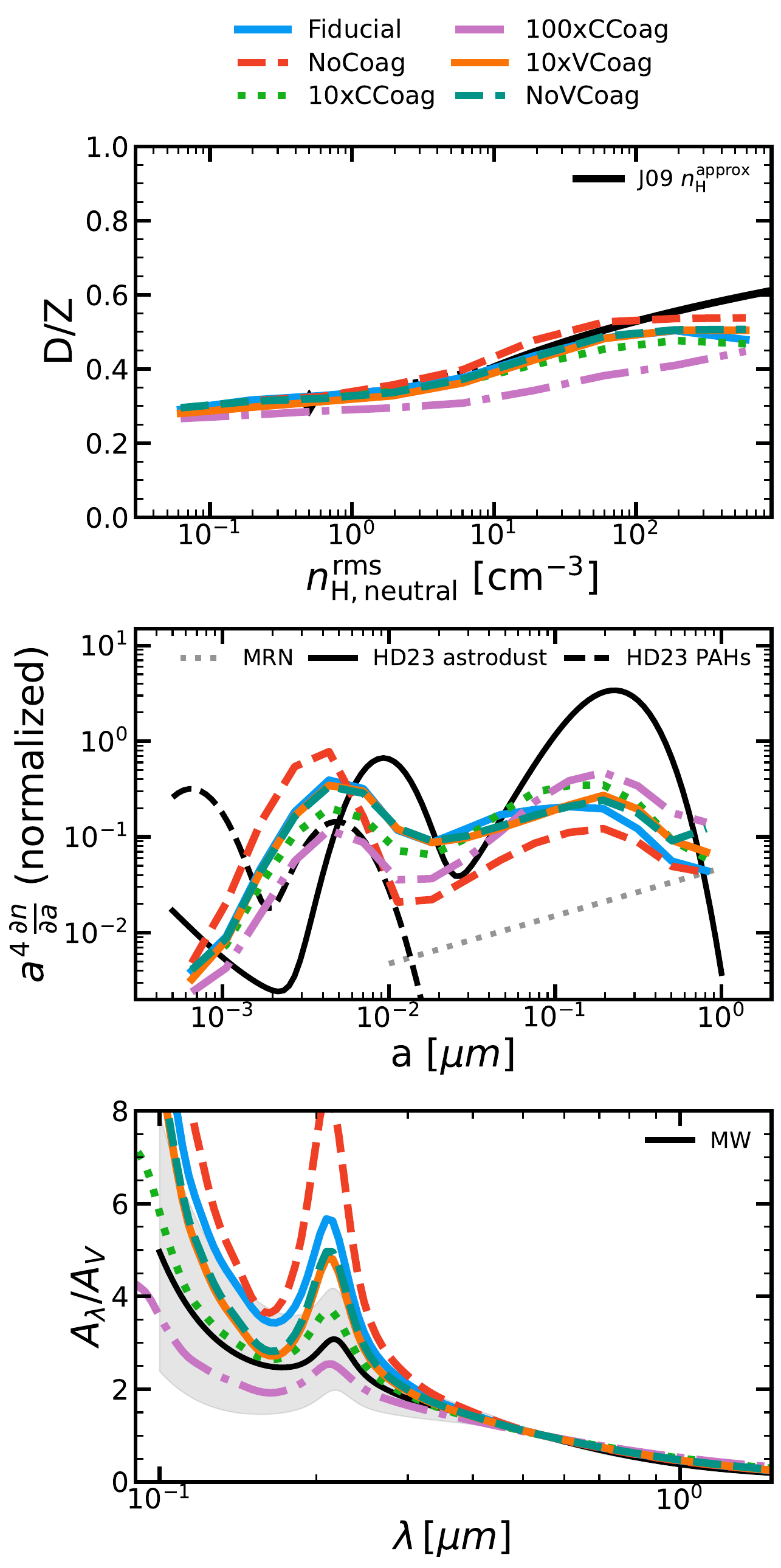}
    \caption{Same as Fig.~\ref{fig:shat_variations_DZ_GSD_AV} for variations in coagulation. 
    The D/Z relation is offset relative to the strength of coagulation due to coagulation removing a fraction of small grains which would otherwise grow via accretion. 
    This effect is relatively weak with D/Z changing by ${\sim}20\%$ from no coagulation ({\bf NoCoag}) to 2 dex increases in coagulation enhancement factors ({\bf 10xCCoag} and {\bf 100xCCoag}).
    The abundance of large grains is largely determined by the strength of coagulation showing that SNe and AGB are minor producers of large grains.
    Stronger coagulation leads to more large grains and flatter extinction slopes and bump strengths.
    Assuming an increased or no coagulation velocity threshold ({\bf 10xVCoag} and {\bf NoVCoag}) allows silicates, which have the lowest velocity threshold, to grow to larger sizes, shifting the peak size of large grains and only decreasing the extinction slope.
    }
    \label{fig:coag_variations_DZ_DSG_AV}
\end{figure}

Self-consistent modeling of grain-grain coagulation in molecular clouds is beyond the resolution of current galaxy simulations.
Given this fact, the treatment of coagulation requires a sub-resolution scheme that varies between works, including subresolved 'dense gas' prescriptions, density clumping factors, and usage or removal of coagulation velocity thresholds.
To determine the effects of coagulation and key parameters in the coagulation routine, we reran simulations with no coagulation ({\bf NoCoag}), $10\times$ increased coagulation velocity thresholds ({\bf 10xVCoag}), thresholds set to the shattering threshold ({\bf NoVCoag}), and $10 \times$ and $100 \times$ increased coagulation density enhancement ({\bf 10xCCoag} and {\bf 100xCCoag}).
The resulting grain size distributions and extinction curves compared to the {\bf Fiducial} model are shown in Fig.~\ref{fig:coag_variations_DZ_DSG_AV}.

The resulting D/Z trend scales weakly with the strength of coagulation, decreasing by ${\sim}20\%$ between no coagulation and a ${>}2$ dex increase in coagulation rates. 
This is due to the competition between the accretion and coagulation of small grains.
As the coagulation rate increases, efficient coagulation in lower density gas reduces the abundance of small grains, which reduces the effective accretion rate. 
However, this will typically be subdominant given the lower collisional rates between small grains compared to gas-phase metals and small grains.
The qualitative bimodal shape of the grain size distribution does not vary with coagulation, but the mass fraction of large grains increases with the coagulation rate.
When assuming no coagulation, large grains are only produced in SNe and AGB ejecta, which accounts for ${<}10\%$ of the dust mass in our simulations.
This heavily suppresses the abundance of large grains, increasing the slope and bump strength of the extinction curve.
Increasing the coagulation rate shifts more mass into large grains, decreasing the slope and bump strength, and bringing the model predictions into agreement with MW observations.
Increasing the coagulation velocity threshold increases the peak size of large grains and moderately decreases only the extinction curve slope.
This change is due to silicate dust coagulating to larger sizes, since it has a lower coagulation velocity threshold than the other dust species (as seen in Fig.~\ref{fig:coag_velocity}).

\subsection{Local Group Dust Abundance and Extinction Curves} \label{sec:local_group_comparison}

\begin{figure*}
    \plotsidesize{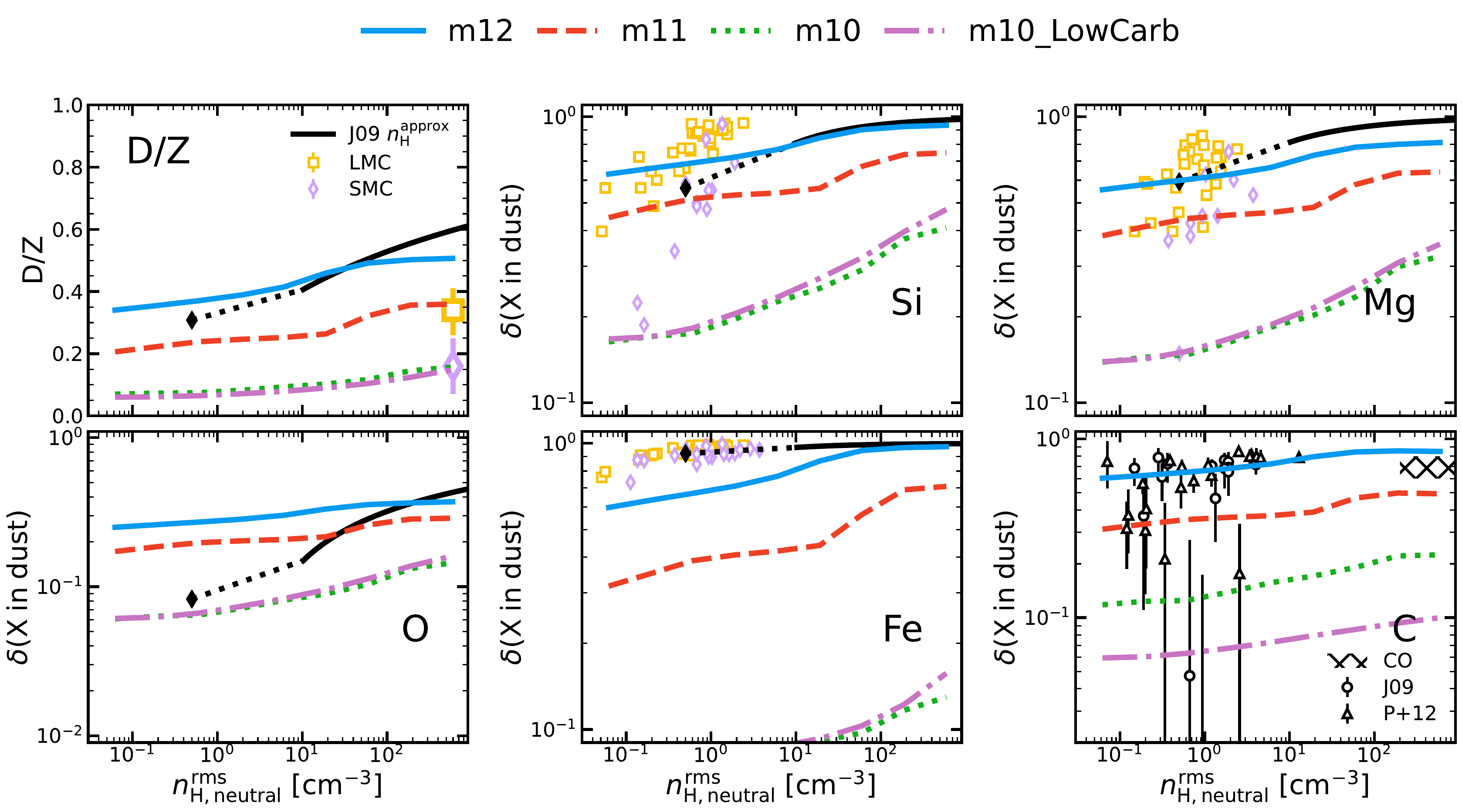}{0.95}
    \caption{Same as Fig.~\ref{fig:depletion_relation} for simulated MW, LMC, and SMC-mass galaxies ({\bf m12}, {\bf m11}, {\bf m10}). We include measured LMC ({\it gold diamond}) and SMC ({\it lilac square}) Si, Mg, and Fe element depletions and estimates of the total D/Z  with error bars showing the observed range~\citep{jenkins_2017:InterstellarGasphaseElement,roman-duval_2021:METALMetalEvolution,roman-duval_2022:METALMetalEvolutiona}.
    The LMC and SMC depletion trends should be treated as upper limits, since we assume a fixed sight line length and physical density, and are used to show the range of depletions observed in each galaxy.
    The resulting D/Z and depletion trends are similar in shape but offset to lower values for {\bf m11} and {\bf m10} compared to {\bf m12}.
    This agrees with the range of observed D/Z, Si, and Mg depletions for the MW, LMC and SMC. 
    However, too little Fe is depleted into dust for {\bf m11} and {\bf m10}, highlighting that our model underpredicts the accretion rate of Fe. Reruns of the SMC-mass galaxy with lower carbonaceous accretion rates ({\bf m10\_LowCarb}) decrease C depletions but have little effect on the total D/Z.
    }
    \label{fig:local_group_DZ} 
\end{figure*}

Observations in the MW offer a limited window for our understanding of dust evolution, but the MW's satellites provide a glimpse of dust in low metallicity environments.
Notably, the LMC and SMC gas-phase depletions are lower than the MW, with less metals being locked into dust. 
LMC and SMC extinction curves are also steeper and have a weaker bump strength than the MW, suggesting a larger/smaller contribution of small silicate/carbonaceous grains to the total dust mass respectively.
To investigate the cause of these variations in dust amount and composition, we ran three simulations of idealized galaxies with similar stellar mass and metallicity as MW, LMC, and SMC called {\bf m12}, {\bf m11}, {\bf m10} (see Table~\ref{tab:idealized_ICs}).
To ensure the multiphase ISM of {\bf m10} is resolved, and to avoid any effects of varying resolution, these runs have 10x higher resolution than the simulations presented in Sec.~\ref{sec:model_variations}. 
Given our {\bf Fiducial} model underpredicts coagulation and these simulations are higher resolution than {\bf m12i\_lowres}, we use our dust model with $C_{\rm coag}=200$\footnote{The {\bf 10xCCoag} model is the closest analog to the dust model used in these high-resolution simulations.} and assume an initial log-normal grain size distribution with $a_{\rm center} = 0.1\;\micron$ and $\sigma=0.6$.
We also assume an initial log-normal grain size distribution with $a_{\rm center} = 0.1\;\micron$ and $\sigma=0.6$.
We summarize the median dust properties and extinction curve parameterizations for each simulated galaxy and those observed for the MW, LMC, and SMC in Table~\ref{tab:local_group}.

\begin{table}
        \renewcommand{\arraystretch}{1.15}
	\centering
	\begin{tabular}{|l|ccccc|} 
		\hline
		Name & D/Z & Si/C & STL & S & B \\
           
        \hline 
        \multicolumn{6}{|c|}{Simulations} \\
        \hline
        m12 & 0.39 & 2.08 & 0.253 & 2.39 & 0.335 \\
        m11 & 0.22 & 2.77 & 0.644 & 3.92 & 0.434 \\
        m10 & 0.086 & 2.29 & 1.4 & 4.8 & 0.526 \\
        m10\_LowCarb & 0.072 & 4.85 & 1.02 & 5.51 & 0.228 \\
       \hline 
       \multicolumn{6}{|c|}{Observations} \\
       \hline
        MW & $0.40^{\it a}$ & ${\sim}3^{\it b}$ & $0.25^{\it c}$ & 2.61$^{\it d}$ & 0.33$^{\it d}$ \\
        LMC & $0.34^{\it a}$ & \textemdash & \textemdash & 2.79 & 0.32 \\
        LMC 30 Dor & \textemdash & \textemdash & \textemdash & 3.4 & 0.19 \\
        SMC &  $0.16^{\it a}$ & \textemdash & \textemdash & 4.8 & 0.098 \\     
       \hline
	\end{tabular}
	\caption{
    Median dust abundance, composition, sizes, and extinction curve slope and bump strength for our simulated Local Group analogs and those derived from MW, LMC, and SMC observations. \\
    $^{\it a}$ Average from \citet{roman-duval_2022:METALMetalEvolutiona} element depletions observations. \\
    $^{\it b}$ Average between \citet{roman-duval_2022:METALMetalEvolutiona} element depletions and \citet{draine_2007:DustMassesPAH} grain size distributions. \\
    $^{\it c}$Average between \citet{draine_2007:DustMassesPAH} and \citet{hensley_2023:Astrodust+PAHModelUnified} grain size distributions. \\
    $^{\it d}$Derived from $R_V=3.1$.\\}
    \label{tab:local_group}
\end{table}

The resulting D/Z and depletion trends for each galaxy are shown in Fig.~\ref{fig:local_group_DZ}.
We also include measured LMC and SMC Si, Mg and Fe depletions along with estimates of the total D/Z from these depletions \citep{jenkins_2017:InterstellarGasphaseElement,roman-duval_2021:METALMetalEvolution,roman-duval_2022:METALMetalEvolutiona}.
Note, the LMC and SMC depletions are used to showcase the range of depletions observed and should be treated as upper limits\footnote{We approximate the physical density of LMC and SMC sight line depletions by assuming a constant density along a line of sight of ${\sim}1$ kpc, roughly consistent with the depths of the LMC and SMC \citep{subramanian_2009:DepthEstimationLarge}.}.
The predicted D/Z and depletion trends are overall similar in shape for each galaxy, with more metals locked into dust as density increases. 
However, these trends are offset, with less metals locked in dust for {\bf m11} and {\bf m10} compared to {\bf m12}.
This is due to accretion being the dominate producer of dust mass for each species, but the accretion rates are lower for {\bf m11} and {\bf m10} due to their lower metallicity, leading to lower depletions on average.
These results agree with the range of D/Z, Si, and Mg depletions observed in the MW, LMC, and SMC.
However, Fe depletions are heavily underpredicted compared to observations for all galaxies.
This suggests that our model underpredicts Fe accretion either into metallic iron dust or some other Fe-bearing dust.

\begin{figure*}
    \plotsidesize{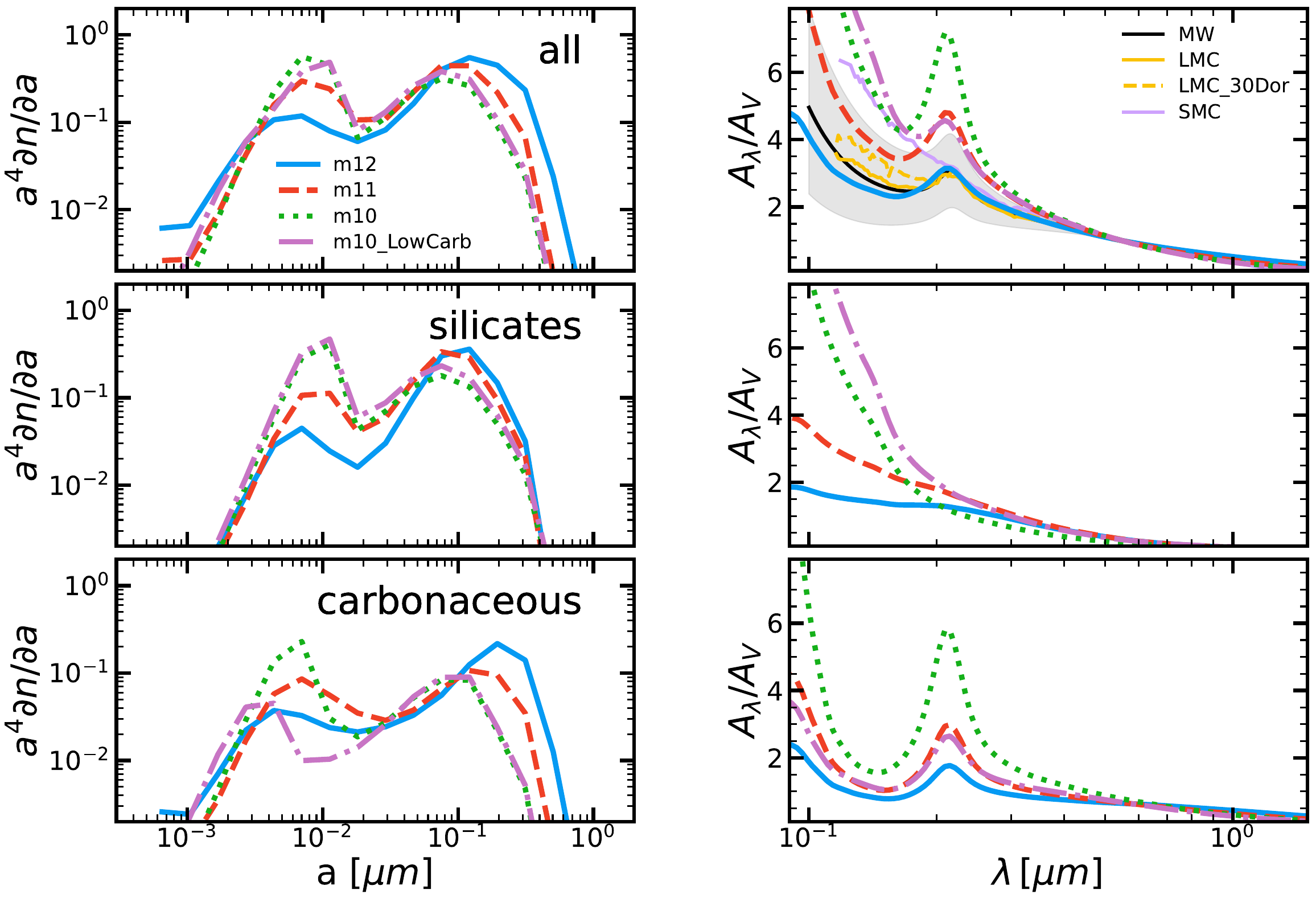}{0.95}
    \caption{Resulting median grain size distributions ({\it left}) and extinction curves ({\it right}) for simulated MW, LMC, and SMC-mass galaxies ({\bf m12}, {\bf m11}, {\bf m10}). We show the results for all dust species ({\it top}) and individual contributions from silicates ({\it middle}) and carbonaceous ({\it bottom}) species. We include observed extinction curves from the MW, average LMC and 30 Doradus star-forming complex, and SMC \citep{salim_2020:DustAttenuationLaw,gordon_2003:QuantitativeComparisonSmall}.
    Our model predicts a similar bimodal size distribution for all galaxies and dust species, but small grains constitute a larger fraction of the dust mass and the average size of the small/large grain peak increases/decreases for {\bf m11} and {\bf m10} compared to {\bf m12}.
    This is due to accretion efficiently growing small grains in all galaxies, but coagulation is less efficient in the LMC and SMC due to overall lower gas densities. 
    Compared to observations, the predicted extinction curves reproduce the steeper slopes of the LMC and SMC due to the increased mass fraction of small silicate grains. However, the increasing bump strengths are in stark disagreement with observations, which is caused by the increased mass fraction of small carbonaceous grains. 
    Reruns of {\bf m10} with carbonaceous accretion rates decreased by a factor of 4 ({\bf m10\_LowCarb}) reduces the bump strength, but not to the degree observed.
    }
    \label{fig:local_group_gsd_AV} 
\end{figure*}

The resulting grain size distributions and extinction curves for each galaxy are shown in Fig.~\ref{fig:local_group_gsd_AV}.
We specifically show the results from all dust species and the individual contributions from silicates and carbonaceous dust.
We also present observed extinction curves from the LMC, including the average and the 30 Doradus star-forming complex, and SMC from \citet{gordon_2003:QuantitativeComparisonSmall}.
Our model predicts a bimodal size distribution for all dust species in all galaxies, with higher small grain mass fractions and smaller/larger average sizes for large/small grains for {\bf m11} and {\bf m10} compared to {\bf m12}. 
These variations are ultimately due to {\bf m11} and {\bf m10} having lower gas densities, along with ${\sim}1$ dex lower maximum densities, and metallicities leading to lower dust densities.
Lower densities reduce coagulation rates and increase grain velocities, which in turn reduces the maximum grain size due to the coagulation velocity threshold.
Reduced coagulation rates also means accretion is able to grow small grains to larger sizes.

The increase in mass fraction of small silicates grains in {\bf m11} and {\bf m10} results in increased extinction curve slopes, similar to what is seen in the LMC and SMC.
However, the increase in the mass fraction of small carbonaceous grains leads to stronger bump strengths for {\bf m11} and {\bf m10}, while the LMC and SMC exhibit weaker, or non-existent bump strengths.
One possible reason for this disagreement is that our model underestimates the coagulation rate of carbonaceous dust, however this is unlikely given any increase would also affect silicates, decreasing the slope.
A more likely reason is our model overestimates carbonaceous accretion rates.
To investigate this, we reran {\bf m10} with the carbonaceous accretion rate decreased by a factor of 4 ({\bf m10\_LowCarb}), which is included in Fig.~\ref{fig:local_group_DZ} and~\ref{fig:local_group_gsd_AV}.
While decreasing carbonaceous accretion does decreases the bump strength, it is not enough to match observations of the SMC.
Furthermore, assuming this decrease for all galaxies would lead to {\bf m12} underpredicting C depletions.
This underscores that small carbonaceous dust grains grow too efficiently in the ISM to explain the low bump strengths seen in the LMC and SMC.
This points to a possible missing process for carbonaceous dust in our model, such as PAH evolution, which we discuss in Sec.~\ref{sec:very_small_grains}.

\section{Discussion} \label{sec:discussion} 

\subsection{Comparison with Other Works} \label{sec:comparison_with_others}

\begin{figure}
    \centering
    \includegraphics[width=\linewidth]{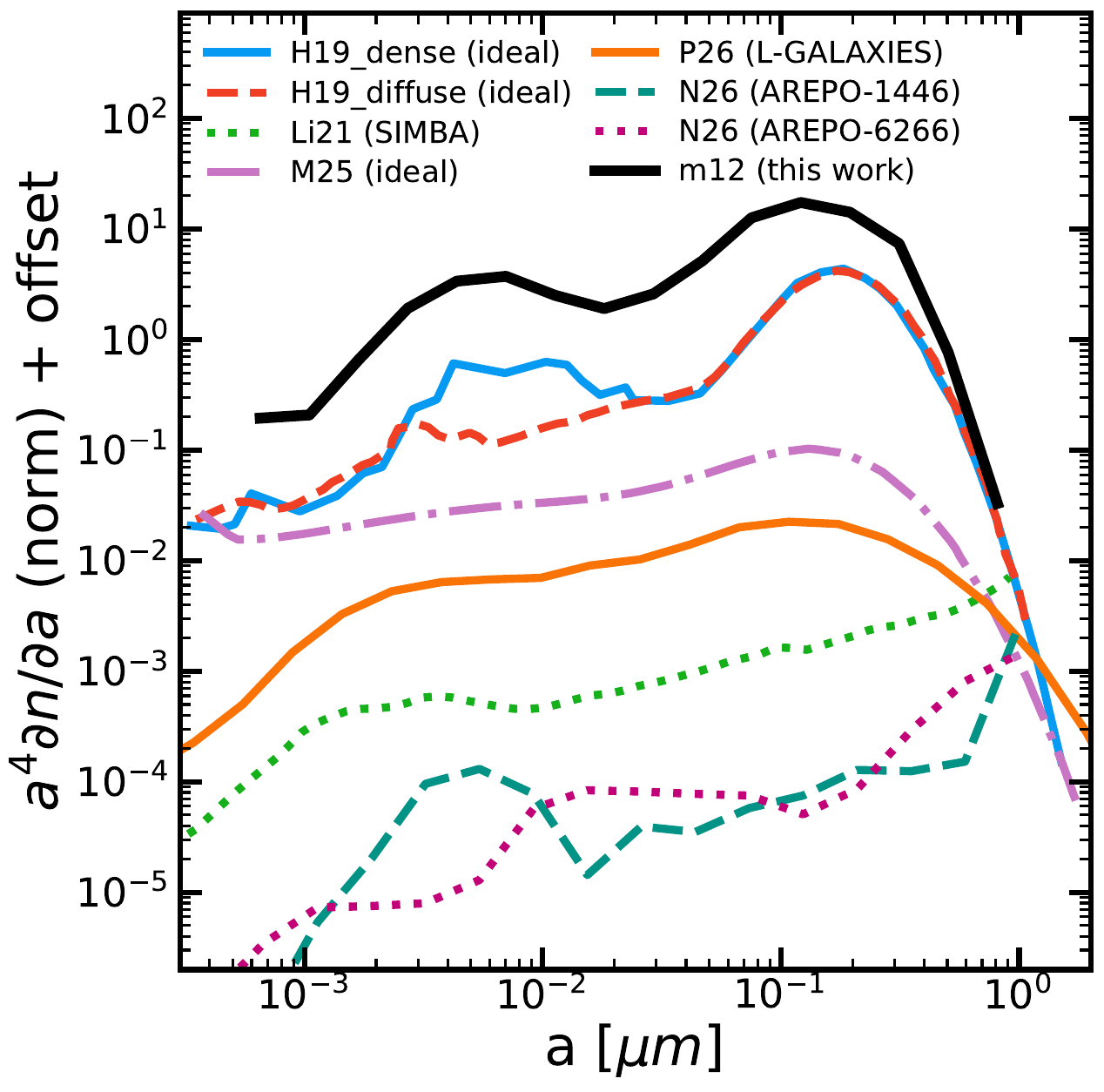}
    \caption{
    Comparison between the median grain size distributions from our model ({\bf m12}) and other works for MW-analogs, all of which reproduce the median MW extinction curve. We show results for idealized simulations \citepalias{hirashita_2019:RemodellingEvolutionGrain,matsumoto_2025:EvolutionGalaxyAttenuation}, L-GALAXIES SAM (P26), cosmological SIMBA simulations \citepalias{li_2021:OriginDustExtinction}, and two zoom-in AREPO simulations (N26). 
    Our model predicts a bimodal distribution, arising from our simulation's ability to track the accretion and coagulation processes in different ISM phases. 
    Most other model's predict a near MRN power-law distribution likely due to accretion and coagulation being subresolved.
    This also leads to a large abundance of $a<1$ nm grains due to shattering and accretion occurring in the same environment.
    Meanwhile, our model predicts such small grains quickly grow which is see to a lesser extent in lower resolution simulations (\citetalias{li_2021:OriginDustExtinction}; P26; N26). 
    Our model also uniquely predicts the cutoff of large grains due to SNR shattering, while other models predict this is caused by limits on coagulation (\citetalias{hirashita_2019:RemodellingEvolutionGrain,matsumoto_2025:EvolutionGalaxyAttenuation}) or do not produce any cutoff (\citetalias{li_2021:OriginDustExtinction};  N26). 
    }
    \label{fig:gsd_sim_comparison}
\end{figure}

A number of grain size evolution models have been developed and integrated into galaxy simulations, varying in their numerical treatment of grain sizes, dust species composition, exact details of each dust process, and incorporated star formation models and ISM physics (e.g.~\citealt{hirashita_2015:TwosizeApproximationSimple,hirashita_2019:RemodellingEvolutionGrain,aoyama_2020:GalaxySimulationEvolution,li_2021:OriginDustExtinction, granato_2021:DustEvolutionZoomcosmological,narayanan_2023:FrameworkModelingPolycyclic,dubois_2024:GalaxiesGrainsUnraveling,matsumoto_2024:ObservationalSignaturesDust, matsumoto_2025:EvolutionGalaxyAttenuation,trayford_2026:ModellingEvolutionInfluence}, Parente et al. in prep.).
Despite these differences, nearly all models reproduce observed dust abundances in the Local Universe due to the competing processes of accretion and SNR destruction, similar to our findings.
Similarly, the few models which track the evolution of separate dust species also find that differing silicate and carbonaceous accretion rates produces observed Si and C depletion trends.
However, these trends are also reproduced by simpler models which assume a constant MRN grain size distribution \citep{hou_2019:DustScalingRelations,li_2019:DustgasDustmetalRatio,graziani_2020:AssemblyDustyGalaxies,choban_2024:DustyLocaleEvolution}.
This agreement is not surprising, given all models predict that efficient accretion is the main determinator of galactic dust abundances. 
While grain size models add the additional constraint that small grain production by shattering must also be efficient, which is a baked-in assumption for an MRN size distribution.

Comparing against the predicted grain size distributions and extinction curves offer a more robust comparison. 
We showcase the predicted grain size distributions of other discretized models against our model in Fig.~\ref{fig:gsd_sim_comparison}.
All simulation results are the median for MW-analog galaxies and reproduce MW extinction curves, including the L-GALAXIES SAM from Parente et al. (in prep) (P26), idealized simulations from \citet{hirashita_2019:RemodellingEvolutionGrain} (\citetalias{hirashita_2019:RemodellingEvolutionGrain})\footnote{Notably, the dust model presented in \citetalias{hirashita_2019:RemodellingEvolutionGrain} is the template for numerous other works~\citep[e.g.][]{rau_2019:ModellingEvolutionPAH,aoyama_2020:GalaxySimulationEvolution,matsumoto_2024:ObservationalSignaturesDust, matsumoto_2025:EvolutionGalaxyAttenuation}.} and \citet{matsumoto_2025:EvolutionGalaxyAttenuation} (\citetalias{matsumoto_2025:EvolutionGalaxyAttenuation}), cosmological SIMBA simulations from \citet{li_2021:OriginDustExtinction} (\citetalias{li_2021:OriginDustExtinction}), and two zoom-in AREPO simulations from Narayanan et al. (in prep) (N26)\footnote{We do not include \citet{mckinnon_2018:SimulatingGalacticDust} since they utilize galaxy simulations without feedback, complicating any comparison.}.
We stress that while we only show the median MW-analog distributions from other works for ease of comparison, many predict large deviations between galaxies with time but their qualitative shapes remain largely unchanged.
Notably, our model differs in qualitative shape (bimodal versus power-law) and abundance of large ($a>0.3\;\micron$) and very small ($a<2$ nm) grains compared to the other works presented.
We discuss the various implications of our results in comparison with other models in the sections below.

We also highlight that a number of dust models have been implemented using the two-size approximation developed by~\citet{hirashita_2015:TwosizeApproximationSimple} \citep[e.g.][]{granato_2021:DustEvolutionZoomcosmological,parente_2022:DustEvolutionMUPPI,dubois_2024:GalaxiesGrainsUnraveling,trayford_2026:ModellingEvolutionInfluence}.
These models assume two populations of small and large grains at fixed sizes, approximating each dust processes as adding or removing mass or transfer mass between the two sizes. 
While we cannot directly compare grain size distributions, our predicted bimodal distribution naturally agrees with the two-size approximation, although the peak sizes of small and large grains changes based on the local environment as seen in Fig.~\ref{fig:local_group_gsd_AV}.

\subsection{Bimodal or Power-Law Size Distribution?}

Our model predicts a bimodal grain size distribution with a peak of small $a\sim6$ nm and large $a\sim0.1\;\micron$ grains and a dearth of $a<1$ nm grains.
In contrast, most other models
predict a near MRN power-law distribution down to  $a\sim1$ nm or lower.
Observed extinction curves can be readily reproduced by either of these qualitative distributions, raising the question: \textit{What causes a model to predict a bimodal versus power-law size distribution and how can we constrain them?}

The resolution-dependent treatment of shattering, accretion, and coagulation is the likely culprit producing a bimodal or power-law distribution. 
Our simulations resolve the multiphase ISM, allowing shattering, accretion, and coagulation to occur in different ISM phases. 
As shown in Sec.~\ref{sec:model_variations}, shattering in the WIM produces small grains, accretion of these grains in the CNM produces the small grain peak, and their succeeding coagulation in molecular gas produces the large grain peak.
In contrast \citetalias{hirashita_2019:RemodellingEvolutionGrain}, \citetalias{matsumoto_2025:EvolutionGalaxyAttenuation}, and P26 utilize sub-resolution treatments for accretion and coagulation which forces all three processes to compete in the same environment, leading to a power-law distribution.
\citetalias{hirashita_2019:RemodellingEvolutionGrain}, \citetalias{matsumoto_2025:EvolutionGalaxyAttenuation}, and P26 use a `dense cloud' prescription which sets a fraction of a gas particle to a fixed density and temperature (e.g. $\nH=10^3\;\cmcubed;T=50$ K). 
In contrast, the simulations by \citetalias{li_2021:OriginDustExtinction} do not resolve the multiphase ISM, instead using clumping factors and more efficient coagulation than other models\footnote{\citetalias{li_2021:OriginDustExtinction} adopts lower grain velocities by assuming a supersonic turbulence power spectrum.}, leading to a similar competition between the three processes.
The notably different bimodal versus power-law results of 
\citetalias{hirashita_2019:RemodellingEvolutionGrain} and \citetalias{matsumoto_2025:EvolutionGalaxyAttenuation}, which use similar model frameworks but apply the `dense gas' prescription to only `dense' or all gas particles respectively, bolsters this conclusion.

Resolution dependent treatment of shattering and accretion is also responsible for the variations in the abundance of very small $a < 1$ nm grains.
Our model predicts these grains quickly grow via accretion resulting in a sharp drop off in their abundance.
In contrast, other models which allow accretion and shattering to occur in the same environment predict a persistent population of very small grains due to their constant replenishment by shattering as
seen for \citetalias{hirashita_2019:RemodellingEvolutionGrain}, \citetalias{matsumoto_2025:EvolutionGalaxyAttenuation}, \citetalias{li_2021:OriginDustExtinction}, and P26.
This resolution dependence can readily be seen when contrasting \citetalias{li_2021:OriginDustExtinction} and NZ26 which use a similar dust model but simulations that have ${\sim}2$ dex differing resolutions.
\citetalias{li_2021:OriginDustExtinction} predicts a drop off around $a\sim 1$ nm, which shifts to $a\sim 10$ nm for NZ26. 
This dearth of a very small grain population has major implications for PAH evolution, which we discuss in Sec.~\ref{sec:very_small_grains}.

\begin{figure*}
    \plotsidesize{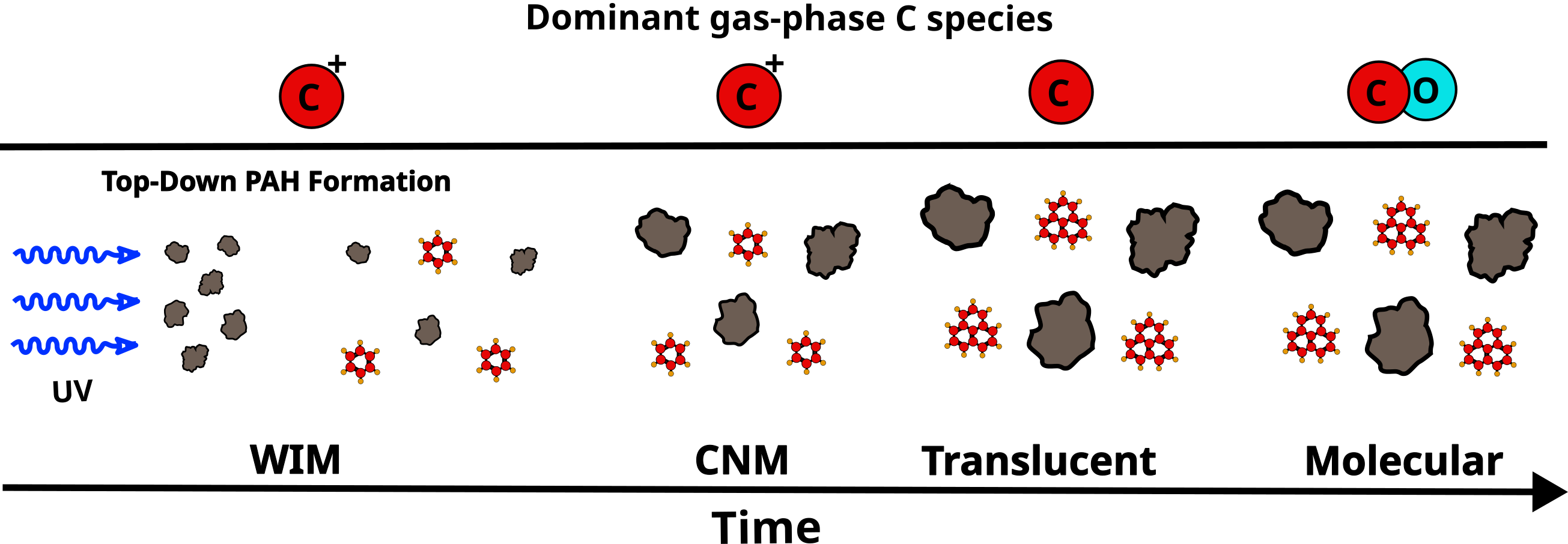}{0.95}
    \caption{A pictographic diagram for how ``top-down'' PAH formation and limited PAH growth could suppress the effective growth of C-based dust and lead to a persistent population of PAHs in the diffuse ISM.
    In the WIM, a fraction of very small ($a<1$ nm) carbonaceous grains created by shattering are converted into PAHs via UV radiation (``top-down'' formation). 
    While carbonaceous dust can grow in the CNM and dense ISM, observations suggest PAHs only grow in the dense ISM.
    Additionally, the rapid conversion of C into CO in the molecular ISM ``turns-off'' the accretion growth of C-based dust.
    Therefore, PAH grains have a limited window for growth which reduces the effective mass they can gain via accretion compared to carbonaceous dust. 
    }
    \label{fig:PAH_model} 
\end{figure*}

This comparison between models highlights that MW extinction curves are ill-suited for constraining the exact shape of the grain size distribution. 
The expected emission spectra and wavelength-dependent dust polarization provide additional observational constraints which may break this degeneracy.
Notably, the \citet{hensley_2023:Astrodust+PAHModelUnified} astrodust+PAH dust model suggest a trimodal size distribution (see Fig.~\ref{fig:median_size_AV}), with a population of large astrodust grains needed to reproduce observed dust polarization and a population of very small ($a\sim0.8$ nm) aromatic carbonaceous grains to reproduce mid-IR emission lines (we discuss the full implications of this in Sec~\ref{sec:very_small_grains}).
However, it is unclear if these observations are also degenerate with a power-law size distribution.
An investigation into the dependence of dust emission spectra and polarization on bimodal and power-law size distributions are therefore needed.

\subsection{The Cutoff of Large Grains in the ISM} \label{sec:large_grains}

Our model predicts a sharp cutoff of large grains due to their efficient shattering in SNR, and that the typical size of large grains varies between Local Group galaxies due to varying coagulation efficiencies limiting the maximum aggregate size.
In contrast, other models either do not predict this cutoff or find it is produced by a different mechanism, and all models do not predict a varying size cutoff.
Observed extinction curves in the MW, LMC, and SMC require a sharp cutoff of large grains at $a\sim0.3\;\micron$ \citep{weingartner_2001:DustGrainSizeDistributions}, but this may not be the case in other galaxies, raising the question: \textit{What determines the cutoff of large grains in the ISM and how may this change between galaxies?} Two equally plausible mechanisms could explain this.

1) Large grains cannot escape molecular clouds.
Grain coagulation in molecular clouds and protostellar disks produce grain sizes far in excess of $0.3\;\micron$ \citep[e.g.][]{lebreuilly_2023:ProtostellarCollapseSimulations}.
However, large grains typically are not coupled to gas or important feedback processes (e.g. radiation pressure) compared to smaller grains due to their lower surface area to mass ratio.
This could cause them to be preferentially `trapped' in molecular clouds, contributing to protostellar disks, while smaller grains are able to escape and be recycled back into the ISM.
The models of \citetalias{hirashita_2019:RemodellingEvolutionGrain}, \citetalias{matsumoto_2025:EvolutionGalaxyAttenuation}, and P26 implicitly include this assumption in their `dense cloud' prescription, which fixes the maximum size of coagulated grains (see Sec. 3 in \citetalias{hirashita_2019:RemodellingEvolutionGrain}). 
This results in a similar cutoff of large grains as our model, but the typical size of these grains does not change between environments.
Meanwhile, \citetalias{li_2021:OriginDustExtinction} and NZ26 do not include such a limitation on coagulation, resulting in an plethora of large grains up to $a\sim1\;\micron$.
To our knowledge no study has been made on grain evolution or retainment in star formation simulations, and so the typical size of grains retained in molecular clouds, and its dependence on cloud properties, is unknown.
The closest analogs are STARFORGE simulations with gas-dust dynamics by 
\citet{soliman_2024:DustEvacuatedZonesMassive,soliman_2024:ThermodynamicsGiantMolecular,soliman_2024:AreStarsReally}, which find that varying grain sizes has large effects on star formation efficiency and dust-to-gas ratios near stars.

2) Large grains are efficiently shattered in SNRs.
Detailed SNR dust destruction simulations by \citet{scheffler_2025:DustDestructionSupernova} find grain shattering can occur on short timescales in the dense forward shock of SNR. However, they also find this is only efficient if the SNR travels through a high-density ($\nH \sim 100\;\cmcubed$) ISM, suggesting SNR are major sources of grain shattering only in dense, star-forming galaxies.
No dust model, besides our own, accounts for SNR grain shattering, and results from \citetalias{li_2021:OriginDustExtinction} and our simulations assuming no SNR shattering ({\bf NoSNRShat}) showcase that grain shattering in the turbulent ISM cannot produce a large grain cutoff.
Therefore, if large dust grains can escape molecular clouds, SNR must efficiently shatter them.

Ultimately, the cutoff of large grains in the ISM are degenerate with assumptions of SNR dust shattering and dust coagulation in galactic dust evolution models.
Therefore, detailed simulations of these environments incorporating dust evolution and gas-dust dynamics are needed to provide robust constraints.

\subsection{Implications for PAH Evolution and the 2175 $\angstrom$ Bump} \label{sec:very_small_grains}

MW observations suggest there are two populations of large ($a\sim 7$ nm) and small ($a\sim 0.8$ nm) PAH+graphite\footnote{While these two populations are labeled as PAHs in the literature, they utilize a grain size dependent PAH and graphitic cross section to reproduce MIR continuum. 
Notably only the the $a\sim 0.8$ nm population is completely PAHs while the $a\sim 7$ nm population is roughly half PAHs and half graphite.} grains as shown in Fig.~\ref{fig:median_size_AV}.
The large grains account for ${\sim}2/3$ of the total mass, but the small grains are responsible for nearly all MIR emission features \citep[][]{draine_2007:DustMassesPAH,hensley_2023:Astrodust+PAHModelUnified}. 
While we do not explicitly model the evolution of PAHs, we can consider its implications on PAH evolution through the perspective of "top-down" PAH formation.
This formation pathway posits preexisting carbonaceous grains are converted into PAHs by UV radiation in the diffuse ISM (aromatization).
PAHs may also be converted back to carbonaceous dust through the accretion of H and/or C (aliphatization) \citep[e.g.][]{murga_2019:SHIVADustDestruction}.
Currently, this is the only PAH formation mechanism incorporated into galactic dust evolution models \citep{hirashita_2020:SelfconsistentModellingAromatic,narayanan_2023:FrameworkModelingPolycyclic} and other works which do not explicitly evolve PAHs have post-processed PAH populations from carbonaceous dust to approximate their evolution \citep{matsumoto_2024:ObservationalSignaturesDust, matsumoto_2025:EvolutionGalaxyAttenuation}.

With ``top-down'' PAH formation in mind, the inability of our model to produce a sizable population of $a<1$ nm grains highlights a major issue, small carbonaceous grains quickly grow via accretion in the CNM\footnote{For reference, Eq.~\ref{eq:dadat_acc} yields a timescale of ${<}1$ Myr to accrete 1 nm to a $a<1$ nm carbonaceous grain in a MW CNM environment.} even when the overall accretion rate is low (e.g. {\bf m10\_LowCarb}).
This leads to our model overpredicting the $2175 \; \angstrom$ bump strength for the LMC and SMC due to the large mass in small carbonaceous grains.
Therefore, PAH growth must be severely limited compared to other dust species, possibly due to low sticking efficiencies or PAH charging causing Coulomb repulsion of ionized gas-phase C, and their formation could limit the growth of C-based dust as summarized in Fig.~\ref{fig:PAH_model}. 
The limitation of PAH growth is bolstered by recent observations which indicate PAH accretion is only appreciable in dense environments. Notably, PAH mass increases by a factor of ${\sim}2$ in the translucent ISM of the MW \citep{zhang_2025:DustextinctioncurveVariationTranslucent}, while PAHs in low-metallicity galaxies are concentrated in dense molecular cores \citep{sandstrom_2010:SpitzerSurveySmall,tarantino_2025:JWSTCapturesGrowth}, indicating that PAH accretion is pushed to denser environments in low metallicity systems.
This, in conjunction with the conversion of gas-phase C into CO in molecular clouds \citep[e.g.][]{liszt_2007:FormationFractionationExcitation}, 
means PAH growth must also compete with CO formation in dense environments, further limiting the effective accretion of PAHs.
Therefore, ``top-down'' PAH formation could limit the growth of small C-based grains, and ultimately the total mass, suppressing the $2175 \; \angstrom$ bump strength. 
This could then explain our model's inability to produce the weakening bumps strengths of the LMC and SMC.

\section{Conclusions} \label{Conclusions}

In this work, we present a new discretized dust grain size evolution model integrated into the {\GIZMO} code base and coupled with the FIRE-3 stellar feedback and ISM physics model. 
All FIRE-3 cooling and heating processes, radiative transfer, and $\Hmol$ formation are integrated with the model, utilizing tracked dust abundances. 
This model includes the separate evolution of silicate, carbonaceous, and metallic iron dust species by accounting for their creation by SNe and AGB stars, growth via gas-dust accretion, destruction via thermal sputtering, SNe shocks, and astration, collisional shattering and coagulation, and sub-resolution turbulent dust diffusion in the same manner as the turbulent metal diffusion in FIRE-3 (see Fig.~\ref{fig:lifecycle_diagram} for a schematic representation).
We also present and utilize a novel prescription for SNR dust destruction which accounts for the shattering of large dust grains (Appendix~\ref{app:SNe_dust_processing}).

Using idealized simulations of MW, LMC, and SMC-mass galaxies, we test how variations in each dust process affect dust abundances and grains size distributions (Sec.~\ref{sec:model_variations}; Table~\ref{tab:sim_suite}), and investigate the origin of dust abundance and extinction variations seen in the Local Group (Sec.~\ref{sec:local_group_comparison}; Table~\ref{tab:local_group}).
We summarize our findings and their implications below.

\begin{enumerate}
    \item Our model uniquely predicts a bimodal grain size distribution for all dust species, with a population of small ($a\sim8$ nm) and large ($a\sim0.1\;\micron$) grains similar to observationally inferred size distributions (Fig.~\ref{fig:median_size_AV}). 
    Shattering in the WIM and SNRs creates small grains, accretion of these grains in the CNM produces the small grain population, and their succeeding coagulation in molecular clouds produces the large grain population.
    This bimodality is robust to order of magnitude changes in all dust processes (Fig.~\ref{fig:acc_dest_variations_DZ},~\ref{fig:shat_variations_DZ_GSD_AV}, and~\ref{fig:coag_variations_DZ_DSG_AV}).
    
    \item Dust abundances in the Local Group are determined by the relative strengths of dust growth by accretion and dust destruction by SNR (Fig.~\ref{fig:acc_dest_variations_DZ}), with variable accretion efficiencies explaining the D/Z and depletion trends seen in MW, LMC, and SMC (Fig.~\ref{fig:local_group_DZ}).
    The variation in depletion trends between elements is due to differing accretion rates for each dust species caused by element abundances, Coulomb enhancement, and CO formation in our model.
    Dust coagulation and shattering have minor effects on these trends as long as shattering occurs and coagulation is not too efficient as to compete with accretion (Fig.~\ref{fig:shat_variations_DZ_GSD_AV} and~\ref{fig:coag_variations_DZ_DSG_AV}).
    This explains why models which assume a constant MRN size distribution can reproduce Local Group dust abundance trends.
    
    \item The extinction curve slope and bump strength are determined by the efficiency of coagulation (Fig.~\ref{fig:shat_variations_DZ_GSD_AV} and~\ref{fig:coag_variations_DZ_DSG_AV}), which sets the small to large grain mass fraction for all dust species.
    Efficient accretion and decreasing coagulation efficiencies can explain the steepening slopes, caused by small silicate grains, as seen for LMC and SMC extinction curves (Fig.~\ref{fig:local_group_gsd_AV}).
    However, this also results in increasing 2175 $\angstrom$ bump strengths, caused by small carbonaceous grains, which are not seen in observations. 

    \item A population of very small ($a<1$ nm) carbonaceous grains, responsible for MIR emission features, does not exist in our simulations due to their rapid growth in the CNM, even when accretion is an overall inefficient producer of carbonaceous dust mass (Fig.~\ref{fig:local_group_DZ} and~\ref{fig:local_group_gsd_AV}). 
    The rapid conversion of small carbonaceous grains into PAHs via ``top-down'' formation in the WIM could remedy this issue, since PAH accretion growth may be severely limited compared to carbonaceous dust (Fig.~\ref{fig:PAH_model}).

    \item The observed cutoff of grains larger than $a\sim0.3\;\micron$ in the MW can be explained by either SNR efficiently shattering large grains or large grains not escaping molecular clouds.
    Detailed simulations of these environments incorporating dust evolution and dust-gas dynamics are needed to provide robust constraints and possible dependence on galactic environments.
    
    \item Other works predict an MRN-like power-law size distribution, likely caused by their low-resolution which requires sub-resolved treatment of shattering, accretion, and coagulation (Fig.~\ref{fig:gsd_sim_comparison}). 
    This also leads to a persistence of very small ($a<1$ nm) carbonaceous grains due to their continuous replenishment via shattering.
    Due to MW extinction curves being degenerate with the exact shape of the size distribution, an investigation into the dependence of dust emission spectra and polarization on size distributions are needed.

Our results show that while our current understanding of grain size evolution can generally explain dust abundances and extinction curves in the Local Group, uncertainties and variations between models impact the extreme ends of dust populations and could hamper predictions for dust in galaxies beyond the Local Group.
Notably, the dependence of small grain abundances on resolution has major implications for predicted extinction curves and the evolution and formation of PAHs.
Meanwhile, the disparate prescriptions for coagulation in dense gas and exclusion of SNR shattering by previous models affect the expected attenuation curves of massive high-z, galaxies whose dense environments likely coagulate grains to sizes larger than those seen in the Local Group.

\end{enumerate}

\section*{Acknowledgements}

We thank Massimiliano Parente and Desika Narayanan for providing simulation data for our comparisons with other works and Brandon Hensely for insightful discussions.
The authors acknowledge the Indiana University Pervasive Technology Institute \citep{stewart_2017:IndianaUniversityPervasive}, supported in part by Lilly Endowment Inc., for providing supercomputing, database, and storage resources that have contributed to the research results reported within this paper.
We ran simulations using: the Extreme Science and Engineering Discovery Environment
(XSEDE), supported by NSF grant ACI-1548562; Frontera allocations AST21010 and AST20016, supported by the NSF and TACC; Big Red 200 at the Indiana University Pervasive Technology Institute.
The data used in this work were, in part, hosted on facilities supported by the Scientific Computing Core at the Flatiron Institute, a division of the Simons Foundation. This work also made use of MATPLOTLIB \citep{hunter_2007:Matplotlib2DGraphics}, NUMPY \citep{harris_2020:ArrayProgrammingNumPy}, SCIPY \citep{virtanen_2020:SciPy10Fundamental}, and NASA’s Astrophysics Data System.

\datastatement{The data supporting the plots within this article are available on reasonable request to the corresponding author. A public version of the \GIZMO\ code is available at \gizmourl.}



\bibliographystyle{mnras}
\bibliography{references}


\appendix

\section{Accounting for Sub-Resolved Gas Clumping} \label{app:gas_clumping}

\begin{figure}
    \centering
    \includegraphics[width=\columnwidth]{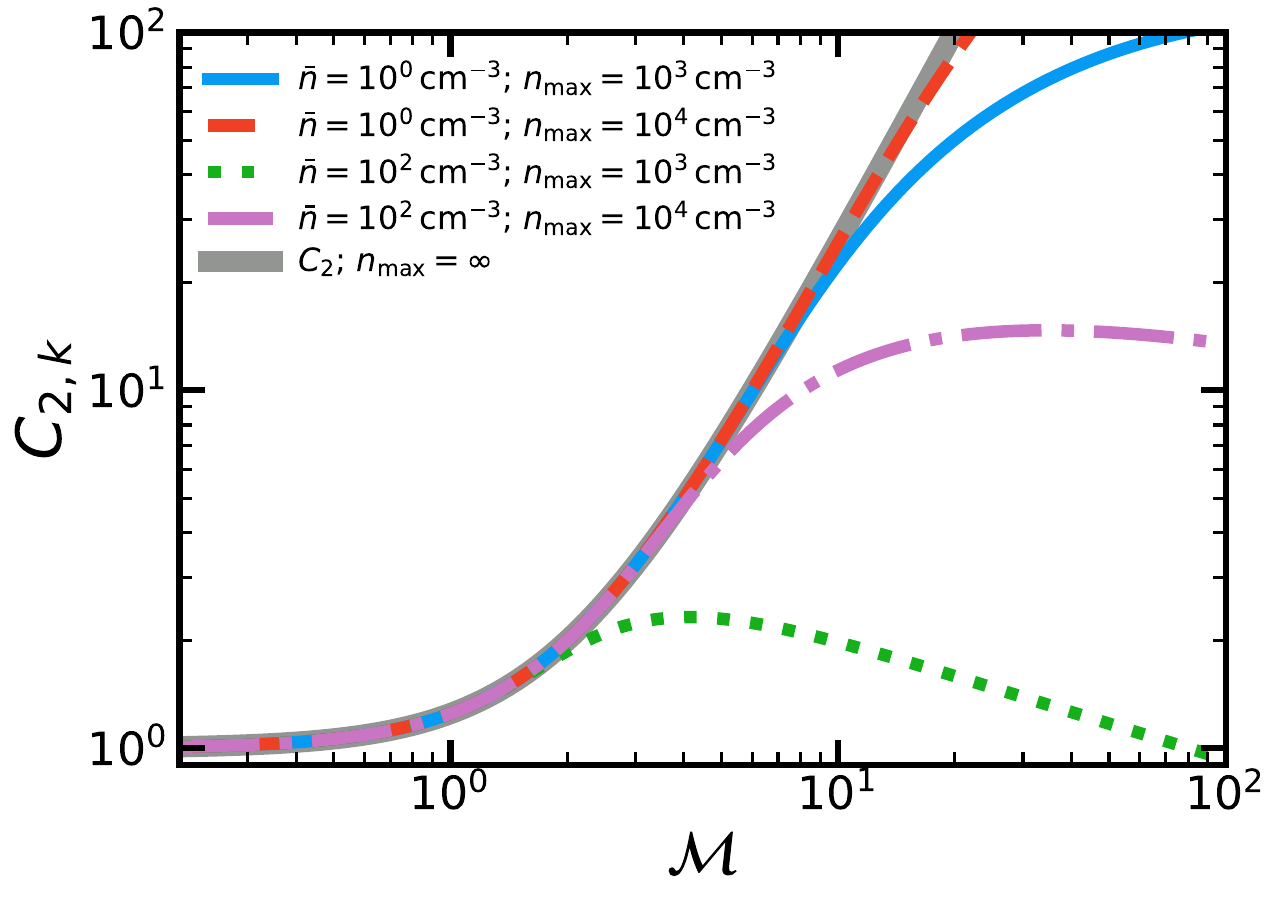}
    \caption{Relation between sub-resolved gas clumping factor and gas cell sonic Mach number assuming accretion turns off past a maximum density. We show the relation for the average gas densities of $\bar{n}=1\,{\rm and}  \,100\,\cmcubed$ and maximum densities of $n_{\rm max}=10^3\,{\rm and} \; 10^4\,\cmcubed$. We also include the clumping factor without the maximum density cutoff ({\it thick line}).}
    \label{fig:clumping_factor}
\end{figure}

To account for the enhancement of dust-gas and dust-dust interactions due to sub-resolution gas clumping we make the standard assumption that gas density in a given gas cell $i$ with volume averaged gas density $\bar{n}=\left< n\right>$ follows the volume-weighted log-normal probability density function (PDF) \citep[e.g.][]{scalo_1998:ProbabilityDensityFunction,federrath_2010:ComparingStatisticsInterstellar,hopkins_2013:ModelNonlognormalDensity}
\begin{equation} \label{eq:lognorm_pdf}
    p(s)= \frac{1}{\sqrt{2\pi}\sigma_s}\exp\left(-\frac{(s-s_0)^2}{2\sigma_s^2}\right),
\end{equation}
where $s=\ln(n/\bar{n})$, $s_0=-\sigma_s^2/2$, $\sigma_s^2=\ln(1+b^2 \mathcal{M}^2)$, $b$ is a geometric constant reflecting the compressive to solenoidal ratio (we set $b=1/2$ in line with the FIRE-3 $\Hmol$ chemical network; \citealt{hopkins_2023:FIRE3UpdatedStellar}), $\mathcal{M}$ is the cell sonic Mach number estimated as ($\mathcal{M})_{i} = \| \nabla \otimes {\bf v} \|_{i}\,\Delta x_{i}/ c_{s,\,i}$, $\nabla \otimes {\bf v}$ is the velocity gradient tensor, $\Delta x_{i}=(\rho_i/m_i)^{1/3}$ is the cell size in terms of the gas cell density and mass, and $c_{s,\,i}$ is the thermal sound speed.
We show the median Mach number relation with gas phase at the end of our {\bf m12\_lowres} simulation in Fig.~\ref{fig:gas_phase_clumping}.
Given this PDF, the gas-gas, dust-gas, and dust-dust interactions are enhanced by 2nd-order clumping factors $C_{\rm 2}^{\rm gg}$, $C_{\rm 2}^{\rm dg}$, and $C_{\rm 2}^{\rm dd}$ respectively \citep{li_2020:DustGrowthAccretion}.
For simplicity, we assume dust and gas are well coupled at all densities\footnote{Note that on scales smaller than our simulation resolution, such as within molecular clouds, dust and gas are not always well coupled \citep[e.g.][]{hopkins_2016:FundamentallyDifferentDynamics,soliman_2024:DustEvacuatedZonesMassive}.} and have the same PDFs such that $C_{\rm 2}^{\rm gg}=C_{\rm 2}^{\rm dg}=C_{\rm 2}^{\rm dd}$.
Thus, we define a general 2nd-order clumping factor $C_2=\langle n^2 \rangle/ \langle n \rangle^2 =  1+b^2\mathcal{M}^2$. 
For shattering and coagulation, accounting for sub-resolve clumping involves multiplying by $C_2$ as given in Eq.~\ref{eq:shat_analytic}. However, gas-dust accretion uniquely ``turns off'' past a certain density ($n_{\rm max}$) for each dust species. For carbonaceous dust, $n_{\rm max}$ is determined by the rapid conversion of gas-phase C to CO, which does not accrete onto dust. For silicates and metallic iron, $n_{\rm max}$ is determined by the freeze-out of molecules, such as water ice, onto dust grain surfaces. 
To account for this, we follow a procedure similar to \citet{dubois_2024:GalaxiesGrainsUnraveling}, deriving an effective clumping factor for each dust species $k$ 
\begin{equation} \label{eq:clumping_factor}
\begin{aligned}
    C_{\rm 2,k}^{\rm acc} &= \int_{-\infty}^{s_{{\rm max},k}} \frac{n^2}{\bar{n}^2} p(s) ds = \int_{-\infty}^{s_{{\rm max},k}} \exp(2 s) p(s) ds \\
    &= \frac{e^{\sigma_s^2}}{2} {\rm Erfc} \left( \frac{3 \sigma_s^2/2-s_{{\rm max},k}}{\sqrt{2}\sigma_s} \right)
\end{aligned}
\end{equation}
where $s_{{\rm max},k}=\ln(n_{{\rm max},k}/\bar{n})$ and Erfc is the complementary error function. 
For the C to CO transition we set $n_{\rm max}=10^3 \; \cmcubed$ and for molecular freeze out we set $n_{\rm max}=10^4 \; \cmcubed$  \citep{boogert_2015:ObservationsIcyUniverse}. 
The resulting relation between effective clumping factor and local Mach number is shown in Fig.~\ref{fig:clumping_factor} for differing gas cell number densities $\bar{n}$ and $n_{\rm max}$.

Additionally, dust processes such as accretion and coagulation depend on local gas temperature, which should decrease in sub-resolved clumped gas compared to the volume-averaged temperature $\bar{T}$ tracked by the gas cell. 
For these processes, we utilize an effective temperature $\Teff$ which is determined as follows.
Given the root-mean-squared density $\nH^{\rm rms} = (\langle n^2 \rangle)^{1/2} = (1+b^2\mathcal{M}^2)^{1/2} \langle n \rangle$, we assume an ideal gas with constant pressure throughout the cell volume $(\nH^{\rm rms}\Teff=\bar{n}_H\bar{T})$. The effective temperature is therefore $\Teff = \bar{T}/(1+b^2\mathcal{M}^2)^{1/2} = \bar{T} / (C_2)^{1/2} $.

\section{A New Sub-Resolution Prescription for SNe Dust Processing} \label{app:SNe_dust_processing}

To account for the synergistic effects of dust shattering and subsequent destruction via sputtering in SNe shocks, we utilize simplified (1) shattering and (2) sputtering routines applied in a three-step process of (1), (2), (1).
The numerical implementation of the sputtering and shattering routines are presented below, along with a comparison of this prescription with other works.

{\bf (1) Shattering:} To approximate large dust grains shattering into smaller grains, we define the dust shattering efficiency for a given grain size $a_j$ as
\begin{equation}
    \epsilon^{\rm SN}_{\rm shat}(a_j) = 1 - \exp\left( - \delta^{\rm SN}_{{\rm shat},i} \frac{a_j}{a_{\rm shat}} \right) ,
\end{equation}
where $a_{\rm shat}=0.05 \, \micron$ is the characteristic grain size above which shattering is efficient and $\delta^{\rm SN}_{{\rm shat},i}$ is the dust species shattering efficiency.
The total mass ejected from shattered grains is then 
\begin{equation}
     M_{\rm ej} = 
    \int_{\amin}^{\amax} \epsilon^{\rm SN}_{\rm shat}(a) \frac{4 \pi \rho_c}{3} a^3 \dnda da = \sum_i \epsilon^{\rm SN}_{\rm shat}(\aic) M_i^{\rm init}, 
\end{equation}
which we assume has a fragment size distribution following Eq.~\ref{eq:frag_dnda} with minimum and maximum fragments sizes of $a_{\rm frag,max}=0.3\;\micron$ and $a_{\rm frag,min}=a_{\rm min}$ respectively. The maximum fragment size is chosen to match the largest dust grains typically seen in the ISM \citep[e.g.][]{weingartner_2001:DustGrainSizeDistributions}.
The mass of grains injected into bin $j$ is then
\begin{equation}
    M_j^{\rm inj} =
\begin{cases}
\begin{aligned}
    & \frac{{\ajupper}^{0.7}-{\ajlower}^{0.7}}{a_{\rm frag,max}^{0.7}-{a_{\rm frag,min}^{0.7}}}  M_{\rm ej} & \ajupper < a_{\rm frag,max} ,\\
    & \frac{{a_{\rm frag,max}}^{0.7}-{\ajlower}^{0.7}}{a_{\rm frag,max}^{0.7}-{a_{\rm frag,min}^{0.7}}}  M_{\rm ej} &  \ajlower < a_{\rm frag,max} \leq \ajupper, \\

    & 0 & {\rm else}
\end{aligned}
\end{cases}
\end{equation}
and the net change in mass is $\Delta M_j = M_j^{\rm inj} - \epsilon^{\rm SN}_{\rm shat}(\ajc) M_j^{\rm init}.$
The number and mass of grains in each bin are then updated following  Appendix~\ref{app:update_mass_conserving}.

{\bf (2) Sputtering: } To approximate the destruction of small grains due to sputtering, we follow \citet{hirashita_2019:RemodellingEvolutionGrain} and \citet{dubois_2024:GalaxiesGrainsUnraveling}, defining a dust destruction efficiency for a grain of size $a_j$ as
\begin{equation}
    \epsilon^{\rm SN}_{\rm sput}(a_j) = 1 - \exp\left( - \frac{\delta^{\rm SN}_{{\rm sput},i}}{a_j/a_{\rm sput}} \right) ,
\end{equation}
where $a_{\rm sput}=0.05\,\micron$ is the characteristic grain size below which sputtering is efficient and $\delta^{\rm SN}_{{\rm sput},i}$ is the dust species sputtering destruction efficiency.
The change in mass of grains in bin $j$ is then $\Delta M_j = \epsilon^{\rm SN}_{\rm sput}(\ajc) M_j^{\rm init}$ and the final number and mass of grains in each bin are updated following Appendix~\ref{app:update_mass_conserving}. 

The two tunable parameters of this prescription are $\delta^{\rm SN}_{{\rm shat},i}$ and $\delta^{\rm SN}_{{\rm sput},i}$ which we set to match an expected mass fraction of grains destroyed $\epsilon_i$ for an initially MRN size distribution with $\amin=0.001\;\micron$ and $\amax=1\;\micron$ for dust species $i$.
We assume an expected $\epsilon_i\sim0.4$ for silicates, giving $\delta^{\rm SN}_{{\rm shat},i}=0.2$ and $\delta^{\rm SN}_{{\rm sput},i}=0.6$, based on detailed SNe dust processing simulations by \citet[][]{kirchschlager_2022:SupernovaInducedProcessing,kirchschlager_2024:SupernovaDustDestruction} which consider only silicate grains.

For carbonaceous and metallic iron, $\epsilon$ is not known, so we assume relative scalings of $\delta^{\rm SN}_{{\rm shat},i}$ and $\delta^{\rm SN}_{{\rm sput},i}$ between silicates, carbonaceous, and metallic iron accounting for differences in their sputtering and shattering rates.
For shattering, we assume a relative scaling of  1:1.3:1.5 based on our model's predicted shattering rates in WIM environments.
For sputtering, we assume a relative scaling of 1:0.66:0.8 given differences in predicted sputtering erosion rates for $10^6 \; {\rm K} \leq T \leq 10^7 \; {\rm K}$. This is also supported by detailed SNe dust destruction simulations by \citet{hu_2019:ThermalNonthermalDust}, which consider only sputtering for silicates and carbonaceous dust.
To summarize, we take $\delta^{\rm SN}_{{\rm shat},i}=0.2, 0.26, 0.3$ and $\delta^{\rm SN}_{{\rm sput},i}=0.6, 0.4, 0.48$ which gives $\epsilon_i\sim0.4, 0.34, 0.37$ for silicates, carbonaceous, and metallic iron, respectively.

\subsection{Comparison with Other Prescriptions}
\begin{figure}
    \centering
    \includegraphics[width=\columnwidth]{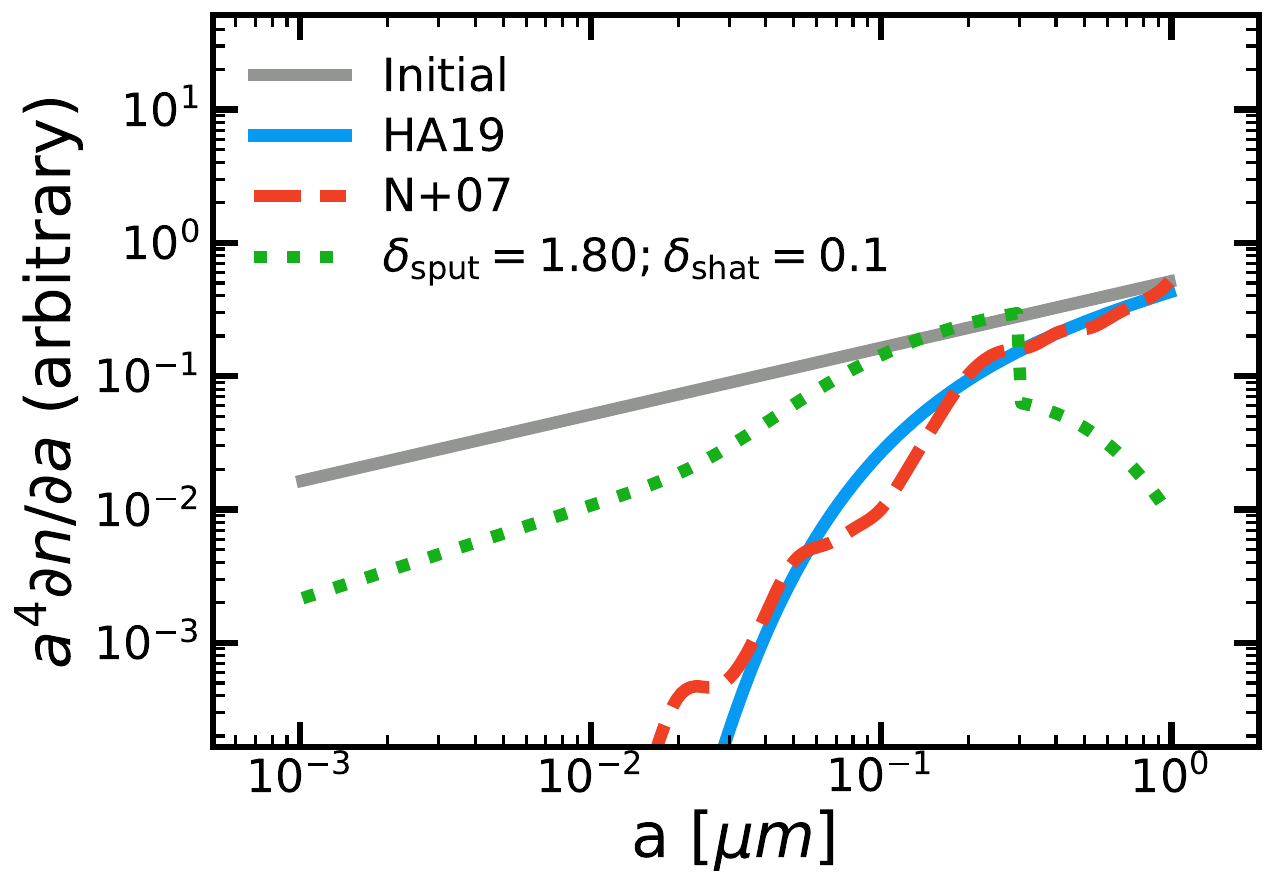}
    \caption{A comparison of how an initial MRN grain size distribution ({\it grey solid}) changes for different SNe dust processing prescriptions. We showcase the prescription presented in this work ({\it green dotted}). Our prescription decreases the number of large grains while still producing a large number of small grains due to its accounting of both grain shattering and sputtering. Other prescriptions presented in \citet{hirashita_2019:RemodellingEvolutionGrain}  ({\it blue solid}) and  \citet{nozawa_2007:EvolutionDustPrimordial} ({\it red dashed}) primarily reduce the number of small grains due to only accounting for sputtering.
    Note, all the prescriptions presented destroy ${\sim}55\%$ of the total dust mass, which is set by the fixed \citet{nozawa_2007:EvolutionDustPrimordial} prescription.}
    \label{fig:SNe_dust}
\end{figure}

Fig.~\ref{fig:SNe_dust} compares our SNe dust processing prescription with the prescriptions from \citet{hirashita_2019:RemodellingEvolutionGrain} and \citet{nozawa_2007:EvolutionDustPrimordial}, which are used in other grain size evolution models \citep[e.g.][]{mckinnon_2018:SimulatingGalacticDust,li_2021:OriginDustExtinction, dubois_2024:GalaxiesGrainsUnraveling}, by showcasing changes to an initial MRN grain size distribution. 
Note that the prescription from \citet{nozawa_2007:EvolutionDustPrimordial} is not tunable, so we present a tuned version of our prescription and the \citet{hirashita_2019:RemodellingEvolutionGrain} prescription to reproduce the total dust mass destruction fraction of ${\sim}55\%$ predicted by \citet{nozawa_2007:EvolutionDustPrimordial} for the given initial size distribution.
Due to its accounting for shattering, our prescription decreases the number of the largest grains while leaving behind a population of small grains in rough agreement with high-resolution SNe dust processing simulations from \citet{kirchschlager_2022:SupernovaInducedProcessing,kirchschlager_2024:SupernovaDustDestruction,scheffler_2025:DustDestructionSupernova}. 
In contrast, the \citet{hirashita_2019:RemodellingEvolutionGrain} and \citet{nozawa_2007:EvolutionDustPrimordial} prescriptions only account for dust sputtering and so heavily reduce the number of small grains while negligibly changing the number of the largest grains.

\section{Convergence Tests}
\label{app:convergence_tests}

\begin{figure}
    \centering
    \includegraphics[width=\columnwidth]{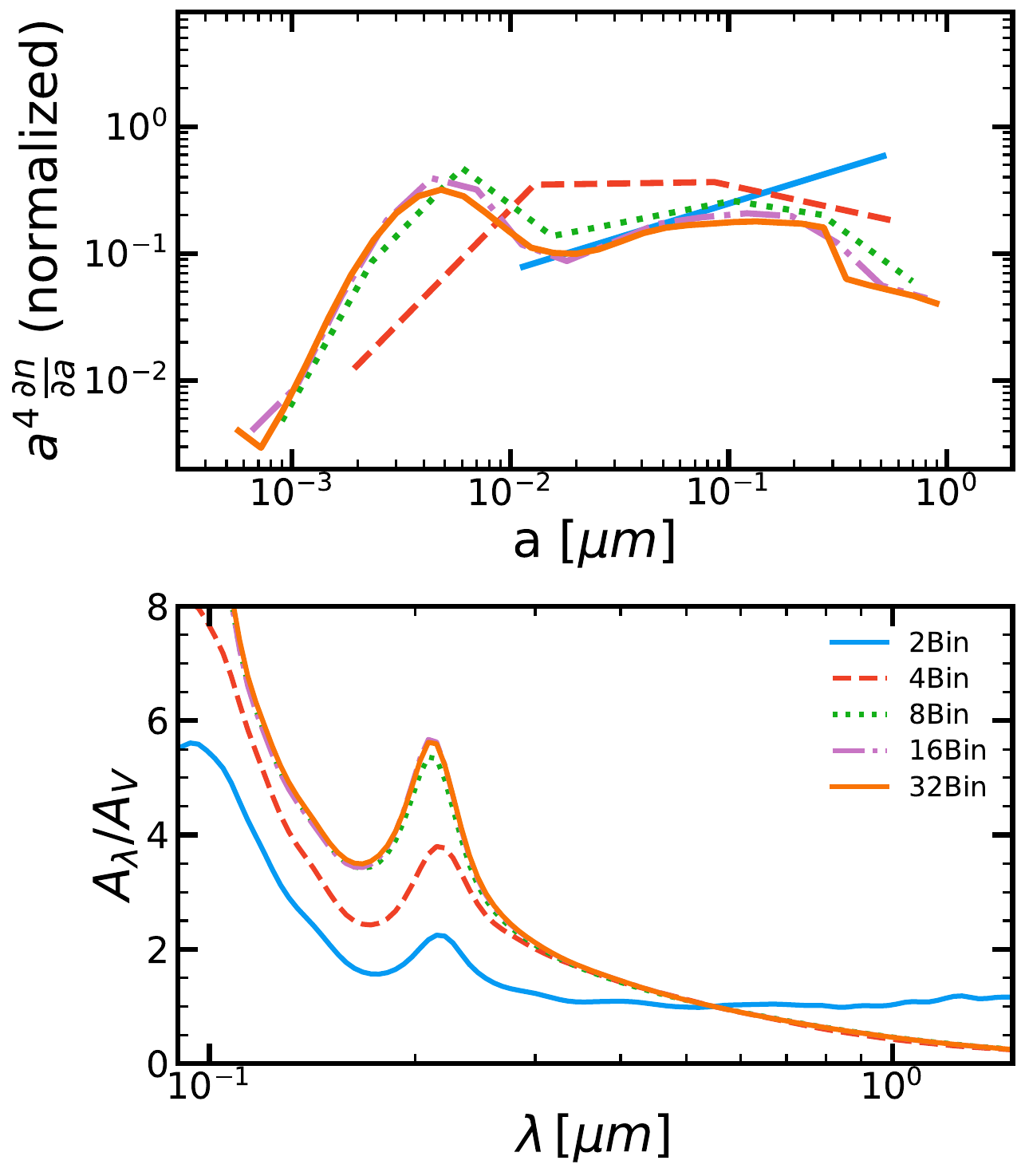}
    \caption{Resulting median grain size distribution and extinction curves for the {\bf Fiducial} model with 2, 4, 8, 16, and 32 grain size bins. Note we only show $a^4\dndaflat$ at the bin centers for ease of interpretation.}
    \label{fig:convergence_tests}
\end{figure}

To test the convergence of our model with the number of grain size bins, we reran the {\bf Fiducial} model with $N_{\rm bin} =2,4,6,8,16,$ and 32 and present the resulting median grain size distribution and derived extinction curves in Fig.~\ref{fig:convergence_tests}.
The predicted grain size distribution converges at $N_{\rm bin}=16$, although the derived extinction curves nears convergence at $N_{\rm bin}=8$.

\section{Time-Stepping Criteria} \label{app:time_stepping}

The time steps required to time-resolve grain processes can be shorter than the typical hydrodynamical time steps $\Delta t_{\rm hydro}$ set by gravity or fluid physical processes used in our simulations.
Shortening time steps for all physical processes in our simulation would be computationally wasteful, so we adopt a sub-cycling scheme by iterating each dust process over multiple smaller time steps as needed.
In particular, given a time step required for a dust process $\Delta t_{\rm proc}$ (where proc is either accretion, sputtering, coagulation, or shattering), if $\Delta t_{\rm hydro} > \Delta t_{\rm proc}$, we run $n_{\rm cycle}$ number of subcycles for the dust process with time step $\Delta t_{\rm cycle}$, where $n_{\rm cycle}={\rm ceil}(\Delta t_{\rm hydro}/\Delta t_{\rm proc})$ and $\Delta t_{\rm cycle}=t_{\rm hydro}/n_{\rm cycle}$. We describe how $\Delta t_{\rm proc}$ is determined for each process below.

For accretion and sputtering, $da/dt$ is typically constant across grain sizes,
and so the smallest grain size bin experiences the largest change and the shortest timescale. Therefore, we adopt a subcycling time step of 
\begin{equation}
   \Delta t_{\rm ac/sp}=\epsilon_{\rm cycle} \left( \frac{\delta a_1}{\dot{a}_{1, {\rm a/s}} }\right),  
\end{equation}
where $\delta a_1$ is the width of the smallest grain size bin, $\dot{a}_{1, {\rm a/s}}$ is the change in size for grains in the smallest bin given by the accretion or sputtering process, and $\epsilon_{\rm cycle}=0.3$ is a free parameter denoting the maximum size fraction grains can transverse across the smallest size bin in a single time step.

For coagulation and shattering, determining a timescale constraint is less straightforward. One possibility is using a time step criteria set by limiting the amount of grain mass moved between bins \citep[e.g.][]{mckinnon_2018:SimulatingGalacticDust}. For example, consider the effective mass timescale 
\begin{equation}
    \tau_{\rm mass} = \frac{\Mdust}{ \dot{M}_{\rm move}},
\end{equation}
where $\Mdust$ is the total dust mass in the gas cell and $\dot{M}_{\rm move}$ is the rate of mass moved between bins due to coagulation or shattering (such as the sum of $\dot{M}_i$ for all bins with $\dot{M}_i > 0$).
A subcycling timestep of $\Delta t_{\rm mass} = \epsilon_{\rm cycle, m} \tau_{\rm coll}$, where $\epsilon_{\rm cycle, m}<1$, is then adopted.
This translates to only allowing a fraction of $\epsilon_{\rm cycle}$ of the dust mass in the cell to move between bins in a time step.
This criterion is plausible for shattering since it predominantly affects large grains, where most of the dust mass resides. 
However, coagulation predominantly affects small grains, which constitute only a small fraction of the total dust mass. Therefore, $\Delta t_{\rm mass}$ is only feasible for coagulation if $\epsilon_{\rm cycle, m}\ll 1$. 
To avoid using extremely small $\epsilon_{\rm cycle, m}$, we also use a second time step criteria set by limiting the reduction in grain number due to coagulation. In particular, we consider an effective number timescale
\begin{equation}
    \tau_{\rm num} = \frac{N_{\rm dust}}{ \dot{N}_{\rm loss}},
\end{equation}
where $N_{\rm dust}$ is the total number of dust grains in the gas cell and $\dot{N}_{\rm loss}$ is the net number rate of grains lost over all bins due to coagulation.
A subcycling time step of $\Delta t_{\rm num} = \epsilon_{\rm cycle, n} \tau_{\rm num}$ is then used, where $\epsilon_{\rm cycle, n}<1$. 
This translates to only allowing the dust number fraction to be reduced by $\epsilon_{\rm cycle, n}$ in a time step.
For our purposes we set $\epsilon_{\rm cycle, m}=\epsilon_{\rm cycle, n}=0.1$. 
We tested running the {\bf Fiducial} model with and without time step subcycling and found that without subcycling, the abundance of large grains was heavily reduced due to coagulation not being time-resolved.

\section{Discretized Formulations} \label{app:discretized_forms}

For ease of reading and brevity, all discretized formulations used in our grain size evolution model have been collected here into sections which we briefly summarize.
Appendix~\ref{app:slope_limiting} details how bin slopes are limited after each bin update as needed.
Appendix~\ref{app:update_number_conserving} explains how bins are updated after the number-conserving processes of accretion and thermal sputtering, along with the treatment of edge cases when grains grow beyond $\amax$ or shrink below $\amin$.
Appendix~\ref{app:discretized_shat} outlines the discretization of the shattering and coagulation equations.
Appendix~\ref{app:update_mass_conserving} describes how bins are updated from the mass-conserving processes of shattering and coagulation.

\subsection{Slope Limiting} \label{app:slope_limiting}

When the number $N_i(t)$ and slope $s_i(t)$ of a given bin $i$ are updated, the resulting $s_i(t)$ may be large enough to cause $\dndaflat$ to be negative at one of the bin edges, which is unphysical. To avoid this, we include a slope limiting step that conserves the total mass within the bin $M_i(t)$ and recomputes $N_i(t)$ and $s_i(t)$ such that $\dndaflat=0$ at the bin edge. Assuming $\dndaflat$ is negative at the lower edge of the bin $\ailower$, we use the following linear set of equations to solve for the new number of grains, $\Tilde{N}_i(t)$, and slope, $\Tilde{s}_i(t)$, 
\begin{equation}
    M_i(\Tilde{N}_i(t),\Tilde{s}_i(t))=M_i(N_i(t),s_i(t))
\end{equation}
and
\begin{equation}
    \frac{\Tilde{N}_i(t)}{\aiupper-\ailower} + \Tilde{s}_i(t)(\ailower-\aic)=0.
\end{equation}
A similar procedure is used for the case that $\dndaflat$ is negative at the upper edge of the bin $\aiupper$.

\subsection{Updating Bins For Number Conserving Processes}  \label{app:update_number_conserving}

As shown in Sec.~\ref{sec:accretion} and~\ref{sec:sputtering}, the number-conserving processes of gas-dust and accretion and thermal sputtering alter the size of grains at a rate of $\dot{a}(a)$ over a given time step $\Delta t$.
To determine the number of grains that lie within a given bin $j$ after this time step, we use

\begin{equation} \label{eq:N_update}
\begin{aligned}
    N_j(t+\dt) &=\int_{\ajlower}^{\ajupper}\frac{\partial n (a, t+\dt)}{\partial a} \\
    &= \sum_{i=0}^{N-1} \mathbbm{1}_{x_1 \geq x_2}(i,j) N_{i \rightarrow j}(t,\dt) 
\end{aligned}
\end{equation}
where 
\begin{equation}
    \mathbbm{1}_{x_1 \geq x_2}(i,j) = 
    \begin{cases}
      1, \; {\rm if } \, x_1(i,j) \geq x_2(i,j)\\
      0, \; {\rm else}\\
    \end{cases}
\end{equation}
is the intersection function denoting whether grains in bin $i$ move to bin $j$ after the time step and $N_{i \rightarrow j}(t,\dt)$ is the number of grains moved. Given $x_1(i,j)=\max(\ailower,\ajlower-\dot{a}\dt)$ and $x_2(i,j)=\min(\aiupper,\ajupper-\dot{a}\dt)$, the bins intersect if $x_1(i,j)\ \geq x_2(i,j)$. The range of grain sizes in bin $i$ that move to bin $j$ (i.e. the intersection range) is $\left[ x_1(i,j) , x_2(i,j)\right]$.
The number of grains that move from bin $i$ to bin $j$ is then
\begin{equation}
\begin{aligned}
    N_{i \rightarrow j}(t,\dt) &= \int_{x_1}^{x_2} \frac{N_i(t)}{\aiupper-\ailower} + s_i(t)(a-\aic) \, da \\
    &= \left[ \frac{N_i(t) a}{\aiupper-\ailower} + s_i(t)\left(\frac{a^2}{2}-\aic a \right) \right]^{a=x_2}_{a=x_1}. \\
\end{aligned}
\end{equation}

Similarly, to determine the mass of grains in bin $j$ after timestep $\dt$ we use
\begin{equation} \label{eq:M_update}
     M_j(t+\dt) = \sum_{i=0}^{N-1} \mathbbm{1}_{x_1 \geq x_2}(i,j) M_{i \rightarrow j}(t,\dt)
\end{equation}
where
\begin{equation}
\begin{aligned}
    M_{i \rightarrow  j} &= \int_{x_1}^{x_2} m(a + \dot{a} \dt) \left( \frac{N_i(t)}{\aiupper - \ailower} + s_i(t)(a - \aic) \right) da \\
    &= \frac{4 \pi \rho_c}{3} \left[ \frac{N_i(t) (a+\dot{a}\dt)^4}{4 (\aiupper-\ailower)} + s_i(t) f_i^{M}(a,\dot{a},\dt) \right]^{a=x_2}_{a=x_1} \\
\end{aligned}
\end{equation}
denotes the mass transfer from bin $i$ to $j$ and we define
\begin{equation}
\begin{aligned}
f_i^M(a, \dot{a}, \dt) &= \frac{a^5}{5} + (3 \dot{a} \dt - \aic) \frac{a^4}{4} + \dot{a} \dt (\dot{a} \dt - \aic) a^3 \\
&+ (\dot{a} \dt)^2 (\dot{a} \dt - 3 \aic) \frac{a^2}{2} - \dot{a}^3 \dt^3 \aic a.    
\end{aligned}
\end{equation}
After $N_j(t+\dt)$ and $M_j(t+\dt)$ are calculated, we determine the new slope, $s_j(t+\dt)$, using Eq.~\ref{eq:slope_from_M}.

\subsubsection{Rebinning Edge Cases} \label{sec:rebin_edge}

Equations~\ref{eq:N_update} \&~\ref{eq:M_update} also determine the number and mass of grains that either (1) shrink below the minimum assumed grain size (move into bin $j=-1$) or (2) grow beyond the maximum assumed grain size (move into bin $j=N$). For case (1), we assume the grains are destroyed, reducing the total grain number and mass. For case (2), we rebin the grains into the $j=N-1$ bin in a mass-conserving manner by shrinking the rebinned grains to the maximum edge of bin $j=N-1$, $a_N^{\rm e}=a_{\rm max}$, calculating a new number, $\Tilde{N}_{N-1}(t+\dt)$, and slope $\Tilde{s}_{N-1}(t+\dt)$ for bin $N-1$. To do this, we first determine the average grain size of bin $j=N-1$ before rebinning as
\begin{equation} \label{eq:avg_grain_size}
\begin{aligned}
    \left< a \right>_{N-1} & (t+\dt) \\
    &=\frac{1}{N_{N-1}(t+\dt)} \int^{a_N^{\rm e}}_{a_{N-1}^{\rm e}} a \frac{\partial n (a, t+\dt)}{\partial a} \, da \\
    &= \left[ \frac{a^2/2}{a_N^{\rm e}-a_{N-1}^{\rm e}} + \frac{s_{N-1}}{N_{N-1}} \left( \frac{a^3}{3}-\frac{a_{N-1}^c a^2}{2}\right) \right]^{a=a_N^{\rm e}}_{a=a_{N-1}^{\rm e}}.
\end{aligned}
\end{equation}
Second, given the grain mass $M_N(t+\dt)$ to be added to bin $N-1$, we shrink these grains to $a_N^{\rm e}$ such that $N^{\rm rebin}(t+dt)=M_N(t+\dt)/(4 \pi \rho_c {a_N^{\rm e}}^3/3)$ gives the number of grains to be rebinned. These rebinned grains will then alter the average grain size in bin $N-1$ such that the new average grain size is
\begin{equation} \label{eq:size_rebin}
    \left< \Tilde{a} \right>_{N-1} (t+\dt) = \frac{N_{N-1} \left< a \right>_{N-1} + N^{\rm rebin}_{N-1} a^{\rm e}_N}{N_{N-1} + N^{\rm rebin}_{N-1}}.
\end{equation}
$ \left< \Tilde{a} \right>_{N-1} (t+\dt) $ can also be expressed in terms of $\Tilde{N}_{N-1}(t+\dt)$ and $\Tilde{s}_{N-1}(t+\dt)$ using Eq.~\ref{eq:avg_grain_size}.
Third, we enforce mass conservation such that
\begin{equation} \label{eq:mass_rebin}
\begin{aligned}
    &M_{N-1}(\Tilde{N}_{N-1}(t+\dt),\Tilde{s}_{N-1}(t+\dt))\\
    &=M_{N-1}(N_{N-1}(t+\dt),s_{N-1}(t+\dt))+M_N(t+\dt)
\end{aligned}
\end{equation}
where $M_{N-1}$ on the left hand side is determined from Eq.~\ref{eq:M_from_dnda} and $M_{N-1}$ and $M_N$ on the right hand side are determined from  Eq.~\ref{eq:M_update}.
We then solve for $\Tilde{N}_{N-1}(t+\dt)$ and $\Tilde{s}_{N-1}(t+\dt)$ by using Eq.~\ref{eq:mass_rebin} and the combination of Eq.~\ref{eq:avg_grain_size} and~\ref{eq:size_rebin} as a linear set of equations.

\subsection{Discretized Shattering/Coagulation Equation} \label{app:discretized_shat}

Following \citet[][see Appendix A]{mckinnon_2018:SimulatingGalacticDust}, we can discretize Eq.~\ref{eq:shat_analytic} to track the movement of mass between grain size bins by making three simplifying assumptions. 
First,  we assume all grains within the same bin have the same speed. Thus, we can approximate the relative velocity between two grains of size $a_1$ and $a_2$ (i.e in bins $k$ and $j$) as $v_{\rm rel}(a_1,a_2) \approx v_{\rm rel}(\akc,\ajc)$.
Second, we assume the mass of grains produced with size $a$ from shattering collisions between grains of size $a_1$ and $a_2$ (i.e. in bins $k$ and $j$) is the same for all grains within the same bin. Thus, we can approximate $m_{\rm shat}(a,a_1,a_2) \approx m_{\rm shat}(a,\akc,\ajc)$. Third, we assume all grains removed from bin $i$ have the same mass  $m_i(a) \approx m(\aic)$.
The discretized version of Eq.~\ref{eq:shat_analytic} is then
\begin{equation} \label{eq:shat_numeric}
\begin{aligned}
    \frac{V_{\rm cell}}{C_2} \frac{d M_i}{dt} &= -\pi  m(\aic) \sum^{N-1}_{k=0} v_{\rm rel}(\aic,\akc) \mathbbm{1}_{v_{\rm rel} > \vshat}(\aic,\akc) I^{i,k} \\
    &+ \pi \sum^{N-1}_{k=0} \sum^{N-1}_{j=0} v_{\rm rel}(\akc,\ajc)  \mathbbm{1}_{v_{\rm rel} > \vshat} (\akc,\ajc) m_{\rm shat}^{k,j}(i) I^{k,j}
\end{aligned}
\end{equation}
where 
\begin{equation}
\begin{aligned}
    m_{\rm shat}^{k,j}(i) &= \int_{\ailower}^{\aiupper} m_{\rm shat}(a,\akc,\ajc)da \\
    &= \int_{\ailower}^{\aiupper} \Bigg[m(a) \frac{\partial n_{\rm frag}}{\partial a} (a,\akc,\ajc) \\
    &+ (m(\akc)-m_{\rm ej}) \delta (a-a_{\rm rem}) \Bigg] da
\end{aligned}
\end{equation}
and
\begin{equation}
\begin{aligned}
    I^{k,j}(t) &= \int_{\aklower}^{\akupper} \int_{\ajlower}^{\ajlower} \Bigg[ \Bigg( \frac{N_k(t)}{\akupper - \aklower} + s_k(t) (a_1-\akc) \Bigg) \\
    &\times \Bigg( \frac{N_j(t)}{\ajupper - \ajlower} + s_j(t) (a_2-\ajc) \Bigg) (a_1+a_2)^2 \Bigg] da_2 da_1.
\end{aligned}
\end{equation}
For coagulation, $m_{\rm shat}^{k,j}(i)$ is replaced with 
\begin{equation}
    m_{\rm coag}^{k,j}(i) = 
    \begin{cases}
        (m_k+m_j)/2, \; & {\rm if} \; \ailower \leq \left( \frac{m_k+m_j}{4 \pi \rho_{\rm gr}/3} \right)^{1/3} < \aiupper, \\ 
        0, \; & {\rm else} 
    \end{cases}
\end{equation}
and $\mathbbm{1}_{v_{\rm rel} > \vshat} (\akc,\ajc)$ with $\mathbbm{1}_{v_{\rm rel} < v_{\rm coag}} (\akc,\ajc)$.

\subsection{Updating Bins from Mass Conserving Processes} \label{app:update_mass_conserving}

As is shown in Sec.~\ref{sec:mass_conserving_processes} and Appendix~\ref{app:discretized_shat}, over a given timestep $\dt$ the mass-conserving processes of shattering and coagulation will change the mass of grains in a given bin $i$ by $\Delta M_i = d M_i/dt \times \dt $. 
Here we describe the steps for how the grain size bin number, $N_i(t+\dt)$, and slope, $s_i(t+\dt)$, are updated from these processes.

{\bf (1) Rebinning Edge Cases:}
Grains can shatter below the minimum assumed grain size (move into bin $i=-1$) or coagulate beyond the maximum assumed grain size (move into bin $i=N$). 
In either case we rebin the grains in a mass-conserving manner such that $\Delta M_{-1}$ is added to $\Delta M_{0}$ and $\Delta M_{N}$ is added to $\Delta M_{N-1}$.
In practice, grains typically never coagulate beyond the maximum grain size since large grains have relative velocities above the coagulation velocity threshold. 

{\bf (2) Ensure Mass-Conservation:}
Due to approximations used in each process, mass may not be strictly conserved, $\Delta M_{\rm total} \equiv \sum^{N-1}_{i=0} \Delta M_i \neq 0$. To ensure mass-conservation we limit $\Delta M_i$ for certain bins.
In the case where $\Delta M_{\rm total}>0$, we limit the bins with $\Delta M_i > 0$, by taking the sum of all bins with $\Delta M_i > 0$, called $\Delta M_{\rm pos}$, and rescaling each bin $i$ with $\Delta M_i > 0$ such that  
\begin{equation}
    \Delta M_{i,{\rm limit}} = \left( 1 - \frac{\Delta M_{\rm total}}{\Delta M_{\rm pos}} \right) \Delta M_{i}.
\end{equation}
For the case where $\Delta M_{\rm total}<0$, we follow the same procedure, but limit the change in mass of bins with $\Delta M_i < 0$.

{\bf (3) Assume Average Grain Size:}
With the rescaled $\Delta M_i$ we update the mass in each bin $i$ with
\begin{equation} \label{eq:mass_injection_update}
    M_i(t + \dt) = M_i(t) + \Delta M_i,
\end{equation}
which can also be expressed in terms of $N_i(t+\dt)$ and $s_i(t+\dt)$, using Eq.~\ref{eq:M_from_dnda}.
To determine $N_i(t+\dt)$ and $s_i(t+\dt)$ we need an additional parameter, which we can construct via assumptions of the change in average grain size in each bin similar to the edge case rebinning procedure in Sec.~\ref{sec:rebin_edge}.

If bin $i$ loses grain mass ($\Delta M_i <0$), we assume the average grain size of the remaining grains is unchanged. 
If bin $i$ gains grain mass ($\Delta M_i > 0 $), we assume the injected grains have a set size distribution $\partial n_{\rm inj} / \partial a$. 
The size distribution of shattered grains is generally known ($\partial n / \partial a \propto a^{-3.3}$; \citealt{hellyer_1970:FragmentationAsteroids,jones_1996:GrainShatteringShocks}), but this is not the case for coagulated grains. For simplicity, we assume coagulated grains have the same size distribution as shattered grains, setting $\partial n_{\rm inj} / \partial a \propto a^{-3.3}$.
The change in average grain size in bin $i$ can then be generalized as
\begin{equation} \label{eq:average_size_change}
     \left< a \right>_i(t+\dt) = 
     \begin{cases} 
        {\left< a \right>_i}(t), \; & {\rm if} \; \Delta M_i < 0, \\ 
         \frac{N_{i}(t) {\left< a \right>_{i}}(t) + \Delta N_{i} \left< a \right>_i^{\rm inj}}{N_{i}(t) + \Delta N_{i}},  & {\rm if } \;  \Delta M_i \geq 0,
     \end{cases}
\end{equation}
where ${\left< a \right>_i}(t)$ is the average grain size in bin $i$ before injection as expressed by Eq.~\ref{eq:avg_grain_size}, 
\begin{equation} \label{eq:avg_inj_size}
    \left< a \right>_i^{\rm inj} = \frac{2.3 \left[ (\aiupper)^{-1.3} - (\ailower)^{-1.3} \right]}{1.3 \left[ (\aiupper)^{-2.3} - (\ailower)^{-2.3} \right]}
\end{equation}
is the average injected grain size, $\Delta N_i \equiv \Delta M_i /  \left< m \right>_i^{\rm inj}$ is the approximate number of injected grains, and 
\begin{equation} \label{eq:avg_inj_mass}
    \left< m \right>_i^{\rm inj} = \frac{4 \pi \rho_{\rm c}}{3} \frac{2.3 \left[ (\ailower)^{0.7} - (\aiupper)^{0.7} \right]}{0.7 \left[ (\aiupper)^{-2.3} - (\ailower)^{-2.3} \right]},
\end{equation}
is the average mass of injected grains.
Note that Eq.~\ref{eq:average_size_change} can also be expressed in terms of $N_i(t+\dt)$ and $s_i(t+\dt)$ using Eq.~\ref{eq:avg_grain_size}.
Therefore, using the system of equations from  Eq.~\ref{eq:mass_injection_update},~\ref{eq:M_from_dnda},~\ref{eq:average_size_change}, and~\ref{eq:avg_grain_size}, we can solve for $N_i(t+\dt)$ and $s_i(t+\dt)$. We then check the bin slopes and slope limit if needed.

\section{Calculating Extinction Curves} \label{app:calc_extinction}

Following \citep[e.g.][]{weingartner_2001:DustGrainSizeDistributions,draine_2011:PhysicsInterstellarIntergalactic}, for a given line of sight (LOS), the effective extinction curve is obtained by
\begin{equation}
    A_{\lambda} = \sum_k 2.5 \log e \int_{a_{\rm min}}^{a_{\rm max}} \pi a^2 Q_{{\rm ext},k} (a,\lambda) \int_{\rm LOS} \frac{\partial n_{d,k}}{\partial a} ({\bf r}, a) ds da,
\end{equation}
where $Q_{{\rm ext},k} (a,\lambda) = Q_{\rm abs} + Q_{\rm sca}$ is the ratio of extinction (due to scatter and absorption) to geometric cross section which is a function of grain size, wavelength, and grain species $k$ and $n_{d,k} (r, a)$ is the number density of grain species $k$ with sizes $[a, a + da]$ at position ${\bf r}$ along the line of sight.
For $Q_{{\rm ext},k}$, we create an interpolated function using the values from \citet{draine_1984:OpticalPropertiesInterstellar} and \citet{laor_1993:SpectroscopicConstraintsProperties} for astrosilicates and graphite.
No $Q_{{\rm ext},k}$ readily exist for metallic iron and so we adopt the function for astrosilicates, however metallic iron should have a relatively flat extinction curve \citep{draine_2013:MAGNETICNANOPARTICLESINTERStelLAR}.
Breaking the integral over $a$ into a summation of integrals over each bin of our grain size distribution and approximating the grain size for each bin with their midpoint gives the discretized form
\begin{equation}
\begin{aligned}
    {A_{\lambda}}
    &= 2.5 \pi \log e \sum_k \sum_i (\aic)^2 Q_{{\rm ext},k} (\aic,\lambda) (\aiupper-\ailower) \, \times \\
    & \int_{\rm LOS} \frac{\partial n_{d,k}}{\partial a} ({\bf r}, \aic) ds.
\end{aligned}
\end{equation}

We approximate each gas cell as an individual sight line, generalizing 
\begin{equation}
    \int_{\rm LOS} \frac{\partial n_{d,k}}{\partial a} ({\bf r}, a) ds \approx \frac{\partial n_{d,k}}{\partial a} (a) L,
\end{equation}
where $L$ is the kernel length of the gas cell. 
The discretized extinction curve\footnote{In the case of small $N_{\rm bin}$, we subsample the grain size distribution using Eq.~\ref{eq:dnda_discretized} within each bin to ensure the extinction curves are converged.} relative to the V band (5470 \r{A}) is then
\begin{equation} \label{eq:extinction_discretized}
\begin{aligned}
    \frac{A_{\lambda}}{A_V} 
    &= \frac{\sum_k \sum_i (\aic)^2 Q_{{\rm ext},k} (\aic,\lambda) (\aiupper-\ailower) \ \frac{\partial n_{d,k}}{\partial a} (\aic) }{\sum_k \sum_i (\aic)^2 Q_{{\rm ext},k} (\aic,V) (\aiupper-\ailower)  \ \frac{\partial n_{d,k}}{\partial a} (\aic)} \\
    &= \frac{\sum_k \sum_i (\aic)^2 Q_{{\rm ext},k} (\aic,\lambda)  N_{i,k}}{\sum_k \sum_i (\aic)^2 Q_{{\rm ext},k} (\aic,V)  N_{i,k}}.        
\end{aligned}
\end{equation}

\section{High-Temperature Dust Cooling} \label{app:dust_cooling}

Dust grains residing in hot plasmas collide with energetic electrons, heating the dust grains, which subsequently radiate in IR, cooling the gas.
Following \citet{dwek_1981:InfraredEmissionSupernova} and \citet{dwek_1987:InfraredDiagnosticDusty}, the heating rate via electron collisions for a single dust grain of size $a$ residing in a gas cell with temperature $T_{\rm gas}$ and electron density $n_{\rm e}$ is given by $H(a,T_{\rm gas},n_{\rm e}) = n_{\rm e} \, \Tilde{H}(a,T_{\rm gas})$ with
\begin{equation} 
   \frac{ \Tilde{H}(a,T_{\rm gas})}{\rm erg \; s^{-1} \; cm^3} =
    \begin{cases}
      5.38\times10^{-18} \left(\frac{a}{\rm \mu m}\right)^2 \left(\frac{T_{\rm gas}}{\rm K}\right)^{1.5} & {\rm if} \; x \geq 4.5 \\
      3.37\times 10^{-13}\left(\frac{a}{\rm \mu m}\right)^{2.41} \left(\frac{T_{\rm gas}}{\rm K}\right)^{0.88} & {\rm if} \; 1.5 \leq x < 4.5 \\
      6.48\times 10^{-6}\left(\frac{a}{\rm \mu m}\right)^3 & {\rm if} \; x < 1.5
    \end{cases}
\end{equation}
where $x=2.71\times10^8 (a/{\rm \mu m})^{2/3}(T_{\rm gas}/K)^{-1}$. The volumetric gas cooling rate due to dust heating, averaging over grain sizes, is then given by
\begin{equation} \label{eq:Lambda_dust}
\begin{split}
    \frac{\Lambda_{\rm dust}(a, T_{\rm gas}, n_{\rm e})}{\rm erg\;s^{-1}\;cm^{-3}} & = n_{\rm dust} \, n_{\rm e} \, \int^{a_{\rm max}}_{a_{\rm min}}\Tilde{H}(a,T_{\rm gas})  \frac{\partial n_{\rm gr}(a)}{\partial a} \; da \\
    & = n_{\rm dust} \, n_{\rm e} \, \left<\Tilde{H}(a,T_{\rm gas})\right>,
\end{split}
\end{equation}
where $\partial n_{\rm gr}/\partial a (a)$ is the grain size distribution with minimum and maximum grain sizes $a_{\rm min}$ and $a_{\rm max}$ respectively normalized to unity and 
\begin{equation}
   n_{\rm dust} = D \frac{\rho_{\rm gas}}{\left<m_{\rm dust}\right>}=\left(\frac{D m_{\rm p}}{X \left<m_{\rm dust}\right>} \right) \nH
\end{equation}
is the number density of dust grains for a given dust-to-gas ratio $D$, $X$ is the hydrogen mass fraction, 
\begin{equation}
\begin{split}
\left<m_{\rm dust}\right>= \frac{4 \pi \rho_{\rm gr}}{3} \int^{a_{\rm max}}_{a_{\rm min}} a^3 \frac{\partial n_{\rm gr}(a)}{\partial a} \; da = \frac{4 \pi \rho_{\rm c}}{3} \left< a^{3} \right>
\end{split}
\end{equation}
is the size-averaged grain mass, and $\rho_{\rm gr}$ is the mass density of the solid dust grain. 

Converting Eq.~\ref{eq:Lambda_dust} to a cooling function, we have
\begin{equation}
    \frac{\Lambda_{\rm dust}(a, T_{\rm gas}, n_{\rm e})}{\nH^2} = \left(
    \frac{3 D m_{\rm p}}{4 \pi \rho_{\rm c} X } \right) 
    \left( \frac{n_{\rm e}}{\nH} \right)
    \left( \frac{\left<\Tilde{H}(a,T_{\rm gas})\right>}{\left< a^{3} \right>} \right) \;  {\rm erg \; s^{-1} \; cm^{3}}. 
\end{equation}

For simplicity, we do not differentiate between dust species, taking $D$ to be the total dust-to-gas ratio from all dust species in the gas cell and $\rho_{\rm c}=3 \; {\rm g \,cm^{-3}}$ which is an intermediate density between silicate and carbonaceous dust. We also adopt a MRN grain size distribution $ \frac{\partial n_{\rm gr}(a)}{\partial a} \propto a^{-3.5}$ \citep{mathis_1977:SizeDistributionInterstellar} with $a_{\rm min}=4$ nm, and $a_{\rm max}=250$ nm. 

We note that while this routine averages over grain size, it is dominated by small grains. However, small grains are easily destroyed by thermal sputtering at nearly the same temperature ($T\gtrsim5\times10^5$ K) dust cooling becomes efficient, and so this routine will likely overpredict the efficiency of dust cooling.


\bsp	
\label{lastpage}
\end{document}